\PassOptionsToPackage{expansion=false}{microtype}
\PassOptionsToPackage{table}{xcolor}
\documentclass[manuscript,screen,nonacm]{acmart}

\usepackage{longtable}
\usepackage{xcolor}
\usepackage{array}
\usepackage{tabularx}

\definecolor{tabhead}{RGB}{239,244,250}   
\definecolor{tabband}{RGB}{247,250,253}   
\definecolor{tabkey}{RGB}{232,240,249}    
\definecolor{gradeS}{RGB}{27,108,44}      
\definecolor{gradeM}{RGB}{167,128,8}      
\definecolor{gradeW}{RGB}{196,90,18}      
\definecolor{gradeA}{RGB}{165,30,30}      
\definecolor{ctlpos}{RGB}{26,90,140}      
\definecolor{ctlneg}{RGB}{165,30,30}      
\definecolor{zonecyber}{RGB}{238,246,252} 
\definecolor{zoneevid}{RGB}{251,246,232}  
\definecolor{zonephys}{RGB}{252,239,235}  
\newcommand{\gS}{\textcolor{gradeS}{\textbf{\textsc{S}}}}

\newcommand{\gW}{\textcolor{gradeW}{\textbf{\textsc{W}}}}
\newcommand{\gA}{\textcolor{gradeA}{\textbf{\textsc{A}}}}
\newcommand{\spos}[1]{\textcolor{ctlpos}{\textbf{#1}}}
\newcommand{\sneg}[1]{\textcolor{ctlneg}{\textbf{#1}}}
\newcommand{\tband}[2]{\rowcolor{tabband}\multicolumn{#1}{l}{\textsc{#2}}\\}

\AtBeginDocument{%
  }

\setcopyright{none}
\copyrightyear{2026}
\acmYear{2026}
\acmDOI{}
\acmISBN{}

\setcitestyle{numbers,square}
\makeatletter
\def\NAT@sort{\z@}
\def\NAT@cmprs{\z@}
\def\@fnsymbol#1{\ensuremath{\ifcase#1\or \dagger\or *\or \ddagger\or
   \mathsection\or \mathparagraph\or \|\or **\or \dagger\dagger
   \or \ddagger\ddagger \else\@ctrerr\fi}}
\makeatother

\begin{document}

\title[AI Sandboxes]{AI Sandboxes: A Threat Model, Taxonomy, and Measurement Framework}

\author{Inderjeet Singh}
\authornote{Corresponding author.}
\email{inderjeet.singh@fujitsu.com}
\affiliation{%
  \institution{Fujitsu Research of Europe}
  \country{United Kingdom}
}

\author{Haitham Mahmoud}
\affiliation{%
  \institution{Fujitsu Research of Europe}
  \country{United Kingdom}
}

\author{Andr\'es Murillo}
\affiliation{%
  \institution{Fujitsu Research of Europe}
  \country{United Kingdom}
}

\renewcommand{\shortauthors}{Singh, Mahmoud, and Murillo}

\begin{abstract}
AI sandboxes produce evidence about systems that may act through software tools, networked devices, or physical actuators. The interpretation of that evidence depends on the boundary of the experiment. We define a sandbox apparatus, distinguish five overlapping archetypes, and specify fifteen claim-relative measurement dimensions. A weakest-link licensing rule caps a deployment claim at its weakest necessary evidence dimension. We connect this rule to a threat model covering both outward harm and corruption of the evaluation apparatus, then apply it to four worked examples spanning driving, industrial control, tool-using agents, and vision-language-action manipulation. We also map the required artifacts to standards and regulatory instruments. The framework identifies which claims a bounded experiment supports, which failures its controls contain, and which assumptions require further validation before deployment.
\end{abstract}



\maketitle

\section{Introduction}

AI systems are routinely evaluated in bounded environments before deployment. In systems security, a \textit{sandbox} restricts an untrusted program's permissions and access to resources \cite{nistGlossarySandbox}. AI researchers also use the term for simulators, digital twins, and supervised trials. These environments impose different experimental boundaries and support different inferences. A common name gives reviewers little basis for judging the evidence they produce.

Consider a vision-language-action (VLA) manipulation policy that passes a large GPU-simulated test suite, then makes unsafe contact on a real robot. Idealized gripper compliance and jitter-free simulation clocks can hide the failure, while an untested object trigger can activate a poisoned policy. This hypothetical failure exposes three distinct obligations: validate contact dynamics, measure control-loop timing, and test conditional adversarial behavior. Each requires its own apparatus and evidence.

Cyber-physical systems (CPS) couple computation, communication networks, and physical plants, historically under model-based control. Artificial Intelligence of Things (AIoT) embeds learning and inference across distributed, resource-constrained edge sensors and devices. Physical AI places high-capacity learned models, including VLA policies, diffusion controllers, and learned world models, inside closed-loop perception and actuation, challenging verification methods that assume analytically specified controllers.

SayCan, PaLM-E, RT-2, Open X-Embodiment, and OpenVLA illustrate the move from language-conditioned decisions to physical action \cite{ichter2023saycan,driess2023palme,zitkovich2023rt2,oneill2024openx,kim2024openvla}. Their evaluations must account for the consequences of learned control. Physical adversarial examples and LiDAR spoofing already show that sensor inputs can redirect a system \cite{eykholt2018physical,cao2019lidar}. BadVLA extends the concern to backdoored VLA policies whose triggered action deviations can coexist with retained clean-task performance \cite{zhou2025badvla}. Indirect prompt injection adds a separate route through tools and retrieved content \cite{greshake2023indirectprompt}.

\begin{figure}[htbp]
    \centering
    \includegraphics[width=\linewidth]{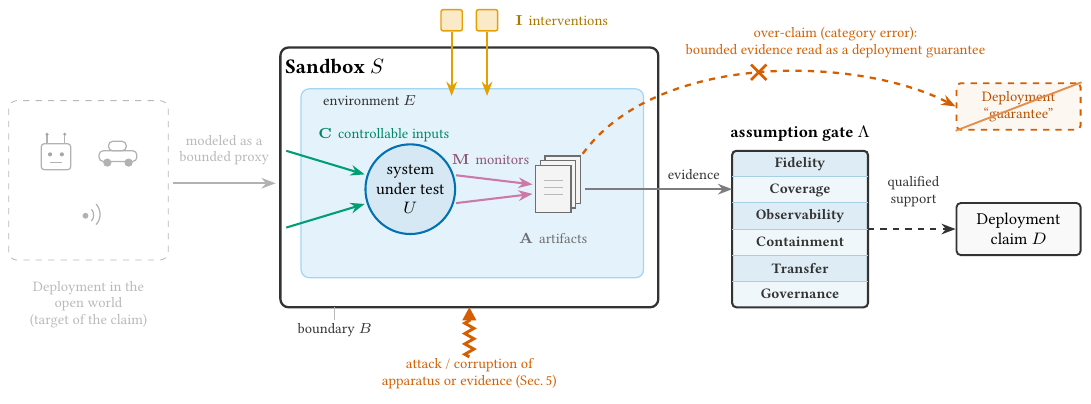}
    \Description{A left-to-right schematic. On the left, a faint deployment world (robot, vehicle, IoT node) is modeled as a bounded proxy inside a sandbox box. The sandbox contains the system under test in an environment field, with controllable inputs entering, monitors tapping the system, interventions acting on the boundary, and an evidence-artifact store. A solid arrow carries the evidence into an assumption gate drawn as a filter column with six named screens (fidelity, coverage, observability, containment, transfer, governance), then a dashed arrow supports a deployment claim. Two red failure annotations show over-claiming, a dashed red arc that leaps over the gate to a crossed-out deployment-guarantee box, and an attack that corrupts the apparatus or its evidence.}
    \caption{A sandbox is a bounded, instrumented experiment on a proxy of the deployed system. Its evidence artifacts license a deployment claim only through an explicit assumption gate, whose six filters (fidelity, coverage, observability, containment, transfer, and governance) are the assumption components $\Lambda$ of Section~\ref{sec:sandbox_claim}, under their stated qualifications. The two recurring failures, shown in red, are over-claiming (bypassing the gate to read bounded evidence as a deployment guarantee) and attacks that corrupt the apparatus or its evidence.}
    \label{fig:hero}
\end{figure}

\textbf{Thesis.} AI sandboxing should be treated as a system-level assurance discipline, not simply as a choice of simulator. Its boundary determines what an experiment represents, what it can contain, and which observations survive for review. We ask how those bounded observations license a deployment claim and what additional evidence that claim requires. Figure~\ref{fig:hero} gives the resulting apparatus-to-claim structure.

\subsection{Problem Gap and Contribution Scope}

Several mature literatures already study parts of this problem. Autonomous-driving testing work analyzes simulation and automated-driving-system evaluation in depth, including scenario generation, sensor modeling, and closed-loop testing \cite{kaur2021simulators,tang2023adstesting}. Embodied-AI and sim-to-real work organizes simulators, tasks, and transfer methods for robot learning \cite{duan2022embodiedaisurvey,muratore2022randomized}. Digital-twin research clarifies virtual-physical bindings, lifecycle monitoring, and security threats \cite{barricelli2019digitaltwin,alcaraz2022digitaltwinsecurity}. Cyber-range and security-testbed research systematizes adversarial experimentation, virtualized infrastructure, and exercise design \cite{yamin2020cyberranges}. Standards and governance frameworks now add further pressure: the NIST AI RMF frames AI risk management as a lifecycle activity, NIST TEVV emphasizes measurement discipline for AI evaluation, and the EU AI Act formalizes AI regulatory sandboxes as supervised settings for development, validation, testing, and documentation \cite{tabassi2023airmf,nistTEVV2026,euAIAct2024}.

These literatures organize their evidence around different units. Cyber-range taxonomies describe capabilities \cite{yamin2020cyberranges}, benchmark methodologies assess evaluation quality \cite{reuel2024betterbench,mohammadi2025agentevalsurvey}, and simulation-credibility standards qualify models for intended uses \cite{nasa2024std7009b}. Legal analyses examine supervised experimentation \cite{truby2022sandbox}, while recent work connects regulatory and technical sandboxes for EU AI Act compliance \cite{buscemi2025sandboxconfigurator,deckenbrunnen2026bathtubs}. We connect these units through a shared apparatus, a claim-relative evidence profile, and a threat model for the evidence chain. This lets a reviewer ask the same question across domains: which necessary support for the deployment claim remains missing? Table~\ref{tab:contribution_gap} states the scope of this synthesis.

\begin{table}[htbp]
\centering
\scriptsize
\renewcommand{\arraystretch}{1.10}
\setlength{\tabcolsep}{3pt}
\caption{Positioning the framework against adjacent technical literatures. The shaded final row states the position of this article.}
\label{tab:contribution_gap}
\begin{tabularx}{\textwidth}{>{\raggedright\arraybackslash}p{0.165\textwidth}>{\raggedright\arraybackslash}p{0.155\textwidth}>{\raggedright\arraybackslash}p{0.25\textwidth}>{\raggedright\arraybackslash}X}
\toprule
\rowcolor{tabhead}\textbf{Literature} & \textbf{Primary focus} & \textbf{What it explains well} & \textbf{Remaining gap and our position} \\
\midrule
Autonomous-driving testing and simulation \cite{kaur2021simulators,tang2023adstesting} & Vehicle-domain simulation and ADS testing & Scenario design, driving simulators, sensor models, test-generation methods & Single-domain emphasis. Security, AIoT, HiL evidence, and regulatory artifacts are not unified. We generalize the evidence question beyond road vehicles while retaining scenario and closed-loop testing insights. \\
\addlinespace
Embodied-AI and sim-to-real evaluation \cite{duan2022embodiedaisurvey,muratore2022randomized} & Robot tasks, simulators, transfer methods & Simulator capabilities, embodied benchmarks, domain randomization, reality-gap mitigation & Limited treatment of cyber containment, digital-twin attack surfaces, governance, and audit evidence. We connect embodiment and transfer to sandbox validity, containment, and assurance claims. \\
\addlinespace
Digital-twin and digital-twin security research \cite{barricelli2019digitaltwin,alcaraz2022digitaltwinsecurity} & Virtual-physical models and lifecycle monitoring & Twin definitions, application domains, synchronization, security threats & Centers on operational monitoring. Bounded pre-deployment evidence and TEVV staging remain outside its primary scope. We treat digital twins as one sandbox archetype within a broader assurance pipeline. \\
\addlinespace
Cyber-range and security-testbed research \cite{yamin2020cyberranges} & Cyber experimentation and adversarial exercises & Isolation, attack injection, range architecture, blue-team observability & Physical fidelity, actuation consequence, sim-to-real transfer, and safety-case linkage remain secondary. We couple cyber containment with physical consequence and evidence artifacts. \\
\midrule
\rowcolor{tabkey}This article & Physical AI, AIoT, and CPS sandboxes & Cross-domain taxonomy, threat model, measurement framework, standards alignment & An evidence-centered framework for sandbox validity across simulation, digital twins, HiL/SiL testbeds, cyber ranges, and regulatory sandboxes. \\
\bottomrule
\end{tabularx}
\end{table}

\subsection{Source Selection and Inclusion Criteria}
\label{subsec:section1-search}

We organize the literature as a structured cross-domain mapping, informed by Kitchenham and Charters \cite{kitchenham2007slr}. The source scope spans research accessible through ACM Digital Library, IEEE Xplore, Scopus, and arXiv, together with official standards and regulatory documents. The relevant registries include ISO, IEC, NIST, ETSI, ASAM, and UL, alongside active open-source simulation repositories. We focus on publications from 2016 to 2026 and trace their references back to foundational protection principles, assurance arguments, CPS semantics, and model verification and validation \cite{saltzer1975protection,kelly2004gsn,lee2008cps,oberkampf2010verification}.

The query vocabulary combines an AI-system term with an evaluation-environment term and a relevant assurance or deployment term. For example, a robotics query combines \texttt{robot learning}, \texttt{hardware-in-the-loop}, and \texttt{safety evaluation}. Parallel queries use \texttt{sandbox}, \texttt{testbed}, \texttt{cyber range}, \texttt{digital twin}, or \texttt{sim-to-real}, with domain terms for physical control or edge deployment. This domain-specific expansion avoids requiring foundational security and legal sources to contain modern AI terminology.

We include a source if it specifies or evaluates a bounded test environment, defines an assurance or standardization instrument, or analyzes an apparatus or CPS threat surface. We exclude model-architecture papers that report benchmark accuracy without analyzing the environment, boundary, transfer, or assurance inference. Foundational sources reached by backward snowballing supply definitions and methods. Research papers support technical and empirical claims, official instruments establish obligations, and platform documentation establishes version-specific capabilities. Representative platform sources cover Gazebo, CARLA, AirSim, Habitat, Isaac Sim/Lab, HELICS, and mosaik \cite{gazeboDocs2026,carlaDocs2026,airsimRepo2026,habitatSimRepo2026,isaacSimDocs2026,isaacLabRepo2026,helicsDocs2026,mosaikDocs2026}.

We extract the apparatus boundary and the claim each source can support, then code its relevant dimensions using Table~\ref{tab:section4-measurement-framework}. Appendix~\ref{app:heatmap-backing} records the family-level backing.
The final cited corpus contains 169 sources. Research literature accounts for 119, comprising 113 research papers and preprints plus two foundational books, two chapters, and two methodological or technical reports. Official standards, legal instruments, and regulatory or definitional guidance account for 29. Platform repositories and documentation account for 21. These are mutually exclusive source categories, with peer-reviewed and preprint versions distinguished by their bibliographic records. The bibliography file contains 172 records, including three uncited records retained from the original source file.

\subsection{Research Questions and Contributions}

The article is organized around five research questions:
\begin{description}
    \item[RQ1:] What constitutes an AI sandbox, and how does it differ from a benchmark, simulator, cyber range, digital twin, testbed, or regulatory sandbox?
    \item[RQ2:] Why do Physical AI, AIoT, and CPS require stronger sandbox assumptions than digital-only AI evaluation?
    \item[RQ3:] Which measurement dimensions make sandbox evidence comparable across domains?
    \item[RQ4:] What security threats, containment failures, and technical gaps limit current sandbox evidence?
    \item[RQ5:] How can standards and governance frameworks guide the design of next-generation AI sandboxes?
\end{description}

The article makes four contributions.
\begin{enumerate}
    \item \textbf{A cross-domain definition and taxonomy.} We formalize AI sandboxes as controlled and instrumented environments for test, evaluation, verification, and validation under bounded cyber-physical risk. The taxonomy separates simulation-based sandboxes, digital twins, adversarial/security sandboxes, regulatory sandboxes, and agent-based sandboxes while recognizing that real systems often combine them.
    \item \textbf{A cyber-physical threat model.} We connect sandbox infrastructure threats, such as isolation failure and cross-tenant leakage, with physical AI threats, such as adversarial perception, actuator command compromise, unsafe network bridges, and attacks on digital-twin state, and, distinctively from catalogs of attacks on ML systems, with attacks on the evaluation apparatus and the evidence it produces.
    \item \textbf{A measurement framework for sandbox evidence.} We define fifteen claim-relative dimensions and a weakest-link licensing cap. Four worked examples show how the same apparatus can support a narrow experimental claim while leaving a deployment claim weak or unsupported. The appendix records the family-level backing.
    \item \textbf{A standards-aligned research agenda.} We map sandbox design choices to AI risk management, TEVV, IoT cybersecurity, industrial cybersecurity, scenario-description, and model-exchange standards, then identify research problems in adaptive scenario generation, interoperable evidence artifacts, continuous assurance, and cyber-physical red teaming.
\end{enumerate}

Sections~\ref{sec:foundations} and~\ref{sec:physical-ai} define the apparatus and its physical-control requirements. Section~\ref{sec:measurement-framework} develops the measurement and composition rules. Sections~\ref{sec:key-challenges-gaps} and~\ref{sec:standards} analyze attacks on the evidence chain and the artifacts required for assurance review.

\section{Foundations, Definition, and Measurement Vocabulary}
\label{sec:foundations}

\subsection{From Isolation to Assurance Environments}
\label{sec:lineage}

Computer-security sandboxes restrict execution through permissions and resource boundaries \cite{nistGlossarySandbox}. Least privilege and complete mediation determine which accesses those boundaries must control \cite{saltzer1975protection}. Software fault isolation constrains a program's memory accesses, while virtual-machine introspection observes a guest from a privileged monitor \cite{wahbe1993sfi,garfinkel2003vmi}. Container deployment adds dependencies on host configuration and image provenance \cite{souppaya2017containers}. These mechanisms can establish an isolation claim, but other assurance claims require additional evidence.

Software testing and AI test, evaluation, verification, and validation (TEVV) add the obligation to produce reliable measurements. The ML Test Score treats production readiness as a property of the deployed pipeline \cite{breck2017mltest}. NIST TEVV connects measurements to their use context, and the AI Risk Management Framework (AI RMF) extends evaluation across the lifecycle \cite{nistTEVV2026,tabassi2023airmf}. Model cards, datasheets, and internal audits preserve the intended use and limitations needed to interpret a result \cite{mitchell2019modelcards,gebru2021datasheets,raji2020closing}.

Cyber ranges make adversarial behavior experimentally accessible \cite{yamin2020cyberranges}. They let analysts fix attacker access, observe defensive responses, and replay a compromise under containment. SCEPTRE and minimega extend this approach to operational-technology and network experiments \cite{hahn2021sceptre,crussell2016minimega}. For a robot or industrial controller, the remaining question is whether the emulated compromise reaches the physical consequence that the security claim concerns.

Benchmarks fix a comparison protocol. Interactive-agent benchmarks extend the measured object to state transitions and recovery across a trajectory \cite{liu2024agentbench,zhou2024webarena,lu2025toolsandbox,zhu2025abc}. Gazebo, MuJoCo, CARLA, AirSim, Habitat, and Isaac Gym make closed-loop behavior testable in modeled worlds \cite{koenig2004gazebo,todorov2012mujoco,dosovitskiy2017carla,shah2017airsim,savva2019habitat,makoviychuk2021isaacgym}. Domain randomization and direct simulation-versus-reality comparisons qualify which conclusions travel beyond those worlds \cite{tobin2017domainrandomization,peng2018simtoreal,muratore2022randomized,elmquist2025simtoreal}.

Digital twins and regulatory sandboxes add two further meanings that are often conflated with AI sandboxes. Digital-twin surveys characterize twins through virtual-physical binding, data flow, synchronization, lifecycle use, and operational monitoring \cite{barricelli2019digitaltwin,jones2020digitaltwin}. Security work also warns that twin connectivity can enlarge the attack surface across data pipelines, services, models, communication links, and physical counterparts \cite{alcaraz2022digitaltwinsecurity}. 

Regulatory sandboxes organize supervised experimentation. The FCA approach fixes process constraints, while the EU AI Act provides for testing and regulatory learning under competent-authority oversight \cite{fca2015regulatorysandbox,euAIAct2024}. Legal analyses examine accountability and consumer protection \cite{allen2019regulatorysandboxes,zetzsche2017regulating}. Governance-aware architectures can implement approval gates and audit records \cite{waseem2026governance}. Their technical effect depends on the experiment those records document.

We use \emph{AI sandbox} for a bounded, instrumented assurance environment. Its defining object is the experiment around a system, not merely the simulator or container that hosts it. Table~\ref{tab:conceptual_distinctions} separates these neighboring objects by the boundary they impose and the evidence they yield.

\begin{table}[htbp]
\centering
\footnotesize
\renewcommand{\arraystretch}{1.10}
\setlength{\tabcolsep}{3pt}
\caption{Conceptual distinctions among adjacent evaluation artifacts: each artifact's defining feature (italics), the boundary it fixes, the evidence it typically yields, what it does well (\spos{Does}), and what it cannot claim alone (\sneg{Cannot}). The shaded final row is the assurance-oriented object this article develops.}
\label{tab:conceptual_distinctions}
\begin{tabularx}{\textwidth}{>{\raggedright\arraybackslash}p{0.175\textwidth}>{\raggedright\arraybackslash}p{0.125\textwidth}>{\raggedright\arraybackslash}p{0.21\textwidth}>{\raggedright\arraybackslash}X}
\toprule
\rowcolor{tabhead}\textbf{Artifact} & \textbf{Primary boundary} & \textbf{Typical evidence} & \textbf{Strength and limit} \\
\midrule
\textbf{Benchmark}\newline\textit{fixed task, dataset, protocol, and metric for comparison} & Dataset/task & Scores, splits, error analyses, leaderboards, confidence intervals & \spos{Does:} comparable measurement of a narrow property. \sneg{Cannot:} deployment safety, interactive robustness, cyber-physical transfer, or behavior outside the protocol. \\
\addlinespace
\textbf{Simulator}\newline\textit{synthetic or emulated environment for executing agents, robots, controllers, or software} & Model-of-world & Scenario traces, simulator state, sensor streams, success/failure records & \spos{Does:} low-risk closed-loop exploration with repeatable variation. \sneg{Cannot:} real-world validity without calibration, uncertainty, and transfer evidence. \\
\addlinespace
\textbf{Testbed}\newline\textit{configured experimental infrastructure for a class of systems, protocols, or workflows} & Infrastructure & Logs, configurations, integration results, measurements & \spos{Does:} repeatable experimentation under explicit infrastructure assumptions. \sneg{Cannot:} generality beyond the tested hardware, software, workload, timing, and environment. \\
\addlinespace
\textbf{Cyber range}\newline\textit{contained environment for attack, defense, training, and security experimentation} & Network/security & Attack traces, packet captures, defensive telemetry, exercise records & \spos{Does:} adversarial experimentation with containment and observability. \sneg{Cannot:} physical safety or deployment assurance unless cyber effects couple to physical consequences. \\
\addlinespace
\textbf{Digital twin}\newline\textit{virtual representation linked to a physical or planned asset through models and data flows} & Virtual-physical and data-flow & Telemetry, synchronization state, replay logs, what-if analyses & \spos{Does:} lifecycle observability, operational replay, and asset reasoning. \sneg{Cannot:} deployment-safety assurance. Connectivity may also expand the attack surface. \\
\addlinespace
\textbf{HiL/SiL rig}\newline\textit{setup connecting production software or hardware to simulated, emulated, or real-time environments} & Interface and timing & Interface logs, latency/jitter traces, hardware profiles, fault responses & \spos{Does:} integration evidence for real code, devices, drivers, and timing. \sneg{Cannot:} field behavior without environmental coverage, containment, and transfer evidence. \\
\addlinespace
\textbf{Regulatory sandbox}\newline\textit{supervised legal or administrative framework for limited experimentation} & Process, legal, and governance & Approvals, sandbox plans, reports, guidance, risk documentation & \spos{Does:} regulator learning, supervised pilots, and documented process evidence. \sneg{Cannot:} technical validation, deployment-safety assurance, or compliance proof beyond the law and evidence. \\
\midrule
\rowcolor{tabkey}\textbf{AI sandbox}\newline\textit{bounded, instrumented TEVV environment with explicit controls, monitors, interventions, artifacts, and residual-risk assumptions} & Combined technical, physical, data, operational, and governance & Scenarios, seeds, versions, telemetry, traces, faults, attacks, interventions, governance artifacts & \spos{Does:} makes claims interpretable by binding results to boundary, evidence, containment, and assumptions. \sneg{Cannot:} unbounded deployment claims outside the modeled, measured, contained, and audited conditions. \\
\bottomrule
\end{tabularx}
\end{table}

\subsection{Definition and Boundary Model}
\label{sec:definition_boundary}

We define an AI sandbox as a controlled and instrumented environment (simulated, emulated, virtualized, or supervised in the real world) that supports TEVV of AI systems under bounded risk, with explicit mechanisms for isolation, monitoring, intervention, and evidence capture. The definition is broad enough to cover digital-only agents, robotics, AIoT, CPS, and regulated field trials, but narrow enough to exclude any environment that lacks a meaningful boundary or evidence function.

We model a sandbox as
\[
S = (U,E,B,C,M,I,A,R),
\]
where \(U\) is the system under test, \(E\) the environment model, and \(B\) the sandbox boundary. The controllable input and perturbation space is \(C\). The layer \(M\) measures execution, \(I\) provides intervention mechanisms, \(A\) preserves the evidence artifact bundle, and \(R\) records residual risk.

\begin{figure}[htbp]
    \centering
    \includegraphics[width=\linewidth]{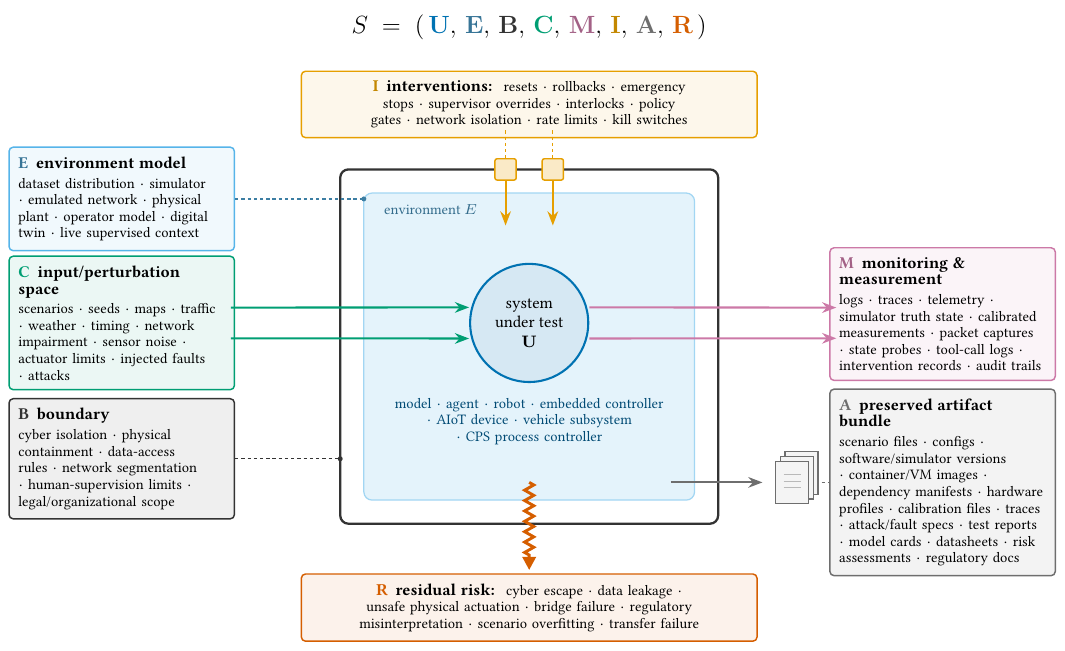}
    \Description{An annotated cutaway of a sandbox. A dark boundary rectangle B encloses a sky-colored environment field E with a central blue system-under-test U. Controllable inputs C enter horizontally through the left boundary into U. Monitors M tap U and exit horizontally to the right. Interventions I act as valves on the top boundary. An evidence-artifact store A receives the outflow on the right. Residual risk R is a red bolt piercing the bottom boundary outward, the mirror of the inward attack bolt in the hero figure. Each tuple letter is a color-keyed card listing its concrete examples.}
    \caption{Anatomy of an AI sandbox as the tuple $S=(U,E,B,C,M,I,A,R)$. Controllable inputs $C$ enter the boundary $B$ around the system under test $U$ in environment $E$. Monitors $M$ produce artifacts $A$. Interventions $I$ act on the boundary. Residual risk $R$ is what crosses $B$ outward despite the controls (the inward counterpart, an attack on the apparatus, appears in Figure~\ref{fig:hero}).}
    \label{fig:anatomy}
\end{figure}

Consider a ROS~2 warehouse mobile manipulator on a HiL stand. Its deployed controller is \(U\), coupled to a warehouse model \(E\) and a real gripper acting on a contained load. Network segmentation and a guarded workcell define \(B\). The experimenter draws object placements and injected delays from \(C\), while \(M\) records force and command timestamps. If contact exceeds the permitted envelope, \(I\) trips the safety relay. The resulting traces join the controller version and gripper calibration in \(A\). Unrepresented human contact and untested delay combinations remain in \(R\). Figure~\ref{fig:anatomy} gives the cross-domain catalog for each component.

The apparatus \(S\) contains and observes \(U\). Its output \(A\) records what happened under the configured boundary and interventions. A claim therefore needs both a description of the apparatus and the preserved record of the execution. A missing calibration file, for example, can prevent an otherwise complete force trace from supporting a contact-safety claim.

The tuple makes combinations of tools comparable. A robotics simulator may represent contact well yet omit the trace needed to diagnose a controller failure. A cyber range may record compromise in detail yet have no plant model to measure its consequence. These omissions become explicit when the claim is tied to the components of \(S\).

\subsection{Claim Semantics for Sandbox Evidence}
\label{sec:sandbox_claim}

We interpret evidence relative to a stated claim. The qualification must identify the assumptions under which the observed property holds and the uncertainty in extending it beyond the experiment.

A sandbox claim should therefore be read in the following form:
\begin{quote}
In sandbox \(S\), under assumptions \(\Lambda\), interventions \(I\), and measurement procedure \(M\), system \(U\) satisfies property \(P\) over an executed suite \(\mathcal{C}\subseteq C\), with uncertainty and validity qualification \(Q\). These observations support deployment claim \(D\) only to the extent justified by \(\Lambda\).
\end{quote}
Here \(P\) describes the tested in-sandbox property, while \(D\) states the intended deployment conclusion. The suite \(\mathcal{C}\) records the scenarios actually executed. It remains distinct from the controllable space \(C\).

The assumption gate \(\Lambda\) records six obligations. Fidelity assumptions identify represented physical channels, such as contact and sensor calibration, alongside cyber channels, such as scheduling and communication delay. Coverage assumptions delimit the executed events and attacker positions. Observability assumptions identify latent state that the monitors cannot recover. Containment assumptions bound the harm possible during execution. Transfer assumptions justify extending observations from \(\mathcal{C}\) to the target setting. Governance assumptions identify who may review \(A\) and how it enters an assurance decision.

An assumption can invalidate a conclusion whether it is explicit or hidden. Stating \(\Lambda\) makes each dependency reviewable and gives a counterexample a precise target. A missed force spike challenges observability. A collision that appears only with delayed inference challenges timing. Figure~\ref{fig:hero} places this gate between artifacts and the deployment claim. Section~\ref{subsec:worked-instantiations} assigns a licensing cap to the necessary evidence dimensions.

\subsection{Taxonomy of AI Sandbox Archetypes}
\label{sec:taxonomy}

We distinguish five overlapping archetypes in Table~\ref{tab:sandbox_archetypes}. A simulator can be coupled to a cyber range and a HiL rig, then used within a supervised pilot. Such combinations inherit the obligations of every interface they introduce.

The archetypes differ most in how they extend an experiment's reach. Simulation permits controlled variation. A twin connects the model to asset telemetry. A security range admits adversarial behavior, and a regulatory sandbox adds supervision. Agent-based sandboxes make stateful tool interaction testable. Combining them can close an evidence gap, but it can also create a new bridge through which commands or data cross the boundary. Agent-based evaluations matter here when tool authority reaches real accounts or physical control, as developed in Case~3 and the threat model.

\begin{table}[htbp]
\centering
\scriptsize
\renewcommand{\arraystretch}{1.10}
\setlength{\tabcolsep}{3pt}
\caption{Sandbox archetype taxonomy. Archetypes are non-exclusive. Real evaluation stacks may combine several rows. Each row gives the evaluation role with the typical evidence, the main strength (\spos{Strong}) and recurring failure mode (\sneg{Fails}), and representative examples.}
\label{tab:sandbox_archetypes}
\begin{tabularx}{\textwidth}{>{\raggedright\arraybackslash}p{0.10\textwidth}>{\raggedright\arraybackslash}X>{\raggedright\arraybackslash}p{0.265\textwidth}>{\raggedright\arraybackslash}p{0.21\textwidth}}
\toprule
\rowcolor{tabhead}\textbf{Archetype} & \textbf{Evaluation role and typical evidence} & \textbf{Strength and failure mode} & \textbf{Representative examples} \\
\midrule
Simulation-based & Synthetic, emulated, or physics-based environments for agents, controllers, robots, vehicles, policies, and perception-planning stacks. \emph{Evidence:} scenario traces, simulator truth state, sensor streams, success/failure records, sensitivity results. & \spos{Strong:} safe, repeatable closed-loop variation. \sneg{Fails:} when reality gap or transfer assumptions are overclaimed. & Gazebo, MuJoCo, CARLA, AirSim, Habitat, Isaac Gym/Isaac Sim \cite{koenig2004gazebo,todorov2012mujoco,dosovitskiy2017carla,shah2017airsim,savva2019habitat,makoviychuk2021isaacgym,isaacSimDocs2026} \\
\addlinespace
Digital-twin & Virtual representation linked to an asset, process, robot fleet, AIoT deployment, or operational telemetry stream. \emph{Evidence:} telemetry, replay logs, synchronization records, what-if results, drift indicators. & \spos{Strong:} lifecycle observability and virtual-physical analysis. \sneg{Fails:} when connectivity expands the attack surface or creates false confidence. & Manufacturing, CPS, AIoT, robotics operations, and lifecycle monitoring \cite{barricelli2019digitaltwin,jones2020digitaltwin,alcaraz2022digitaltwinsecurity} \\
\addlinespace
Adversarial/ security & Contained setting for attacks, faults, misuse, compromised components, red teaming, and defensive observation. \emph{Evidence:} attack traces, packet captures, fault logs, defensive telemetry, recovery/intervention records. & \spos{Strong:} controlled attack/fault injection. \sneg{Fails:} when the attacker model is too narrow, the target is abstracted, or physical consequences are omitted. & Cyber ranges, SCEPTRE, and minimega-style security testbeds \cite{yamin2020cyberranges,hahn2021sceptre,crussell2016minimega,garfinkel2003vmi,souppaya2017containers} \\
\addlinespace
Regulatory & Supervised legal or administrative setting for bounded testing, validation, documentation, and regulator-provider learning. \emph{Evidence:} approvals, sandbox plans, reports, written guidance, risk assessments, exit documentation. & \spos{Strong:} process legitimacy and regulatory learning. \sneg{Fails:} when participation is mistaken for technical validation or certification. & FCA regulatory sandbox. EU AI Act regulatory sandboxes \cite{fca2015regulatorysandbox,euAIAct2024,allen2019regulatorysandboxes,zetzsche2017regulating} \\
\addlinespace
Agent-based & Interactive environment for autonomous software or LLM agents over tools, websites, APIs, stateful tasks, or multi-agent interactions. \emph{Evidence:} tool-call logs, environment state, task success, dialogue trajectories, recovery events. & \spos{Strong:} long-horizon interaction and stateful decision making. \sneg{Fails:} when physical, legal, identity, privacy, and security consequences are abstracted away. & AgentBench, WebArena, ToolSandbox, and agentic benchmark checklists \cite{liu2024agentbench,zhou2024webarena,lu2025toolsandbox,zhu2025abc} \\
\bottomrule
\end{tabularx}
\end{table}

\subsection{Evaluation Stages}
\label{sec:evaluation_stages}

\begin{figure}[htbp]
    \centering
    \includegraphics[width=\linewidth]{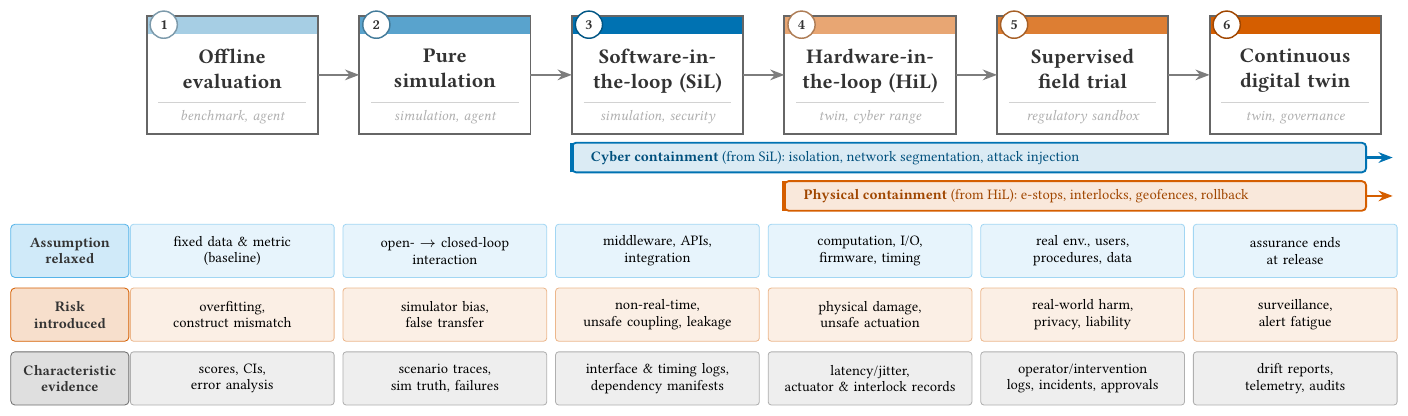}
    \Description{A six-stage evaluation pipeline reading left to right: offline evaluation, pure simulation, software-in-the-loop, hardware-in-the-loop, supervised field trial, and continuous digital twin. Stage headers are tinted blue for the three digital stages and warm orange from hardware-in-the-loop onward, where failures can reach the physical world. A cyber-containment band begins exactly at the software-in-the-loop stage and a physical-containment band exactly at the hardware-in-the-loop stage. Both persist to the end of the pipeline. Gray subtitles name each stage's dominant sandbox archetypes. Three annotation lanes below the stages give, per stage, the assumption relaxed, the risk introduced, and the characteristic evidence.}
    \caption{Staged evaluation pipeline for physical AI sandboxes. Each stage relaxes one assumption (lane~1), introduces a characteristic risk (lane~2), and yields specific evidence (lane~3). Cyber containment (sandboxed networks, attack-injection harnesses) is explicit from the SiL stage and persists through deployment. Earlier stages may also need isolation. Physical containment (e-stops, safety-rated interlocks, geofences) enters at the HiL stage, where the stage headers turn from blue (digital) to warm (physical reach). Gray subtitles name each stage's dominant sandbox archetypes.}
\label{fig:staged-pipeline}
\end{figure}

Each evaluation stage relaxes a different assumption and changes the required containment (Figure~\ref{fig:staged-pipeline}). An offline benchmark fixes the observations and scoring protocol, yielding error estimates while leaving feedback untested. Pure simulation admits closed-loop interaction, yielding state trajectories while introducing model error. Software-in-the-Loop (SiL) connects deployed software to the modeled environment, exposing integration faults and possible network leakage. Hardware-in-the-Loop (HiL) connects real devices or interfaces, yielding hardware timing and fault-response evidence while admitting physical damage. A supervised field trial exposes real operating conditions under restricted authority, yielding intervention and incident records at the cost of direct harm and privacy risks. Continuous monitoring relaxes the assumption of a fixed deployment distribution, producing drift evidence while adding persistent data exposure and dependence on timely remediation.

The useful archetype follows the question. Simulation supports controlled debugging, a cyber range exercises production interfaces under attack, and a twin can preserve field behavior for replay. A supervised pilot adds operating-procedure evidence. None of these choices makes the other stages redundant.

The ordering tracks which assumption each stage relaxes. A field trial cannot recover an unlogged software race, and a repeatable simulation cannot establish the response time of a hardware relay. A stage hand-off must state the new assumption being tested and the controls added to bound its consequences.

\subsection{Measurement Vocabulary}
\label{sec:measurement_vocabulary}

We use fifteen dimensions to describe what an apparatus can substantiate. Table~\ref{tab:section4-measurement-framework} defines their assessment procedures and required artifacts. Dimension~1, \emph{sensing and world fidelity}, concerns sensor channels and the surrounding scene. Timing fidelity, network realism, and actuator and plant realism isolate three other channels, so a strong result in one channel cannot silently count as validation of another.

Three principles govern the interpretation. \emph{Channel Specificity} requires validation of the channel carrying the claim, so a rendered grasp needs separate evidence for contact force. \emph{Reproducibility-Transfer Separation} distinguishes repeatability within \(S\) from correspondence with the target deployment. \emph{Process-Technical Separation} distinguishes supervised procedure from technical validation. Under the EU AI Act, applicable high-risk systems undergo \emph{conformity assessment}, and harmonized standards cited in the Official Journal confer a \emph{presumption of conformity} only for covered requirements \cite{euAIAct2024}. A \emph{safety case} is a structured argument connecting claims to evidence, commonly expressed using GSN or SACM \cite{kelly2004gsn,omgSACM2020}. We reserve \emph{certification} for formal third-party attestation.

These channel boundaries also guide artifact selection. Clock and deadline traces support timing claims \cite{lee2008cps,rajkumar2010cps}. Plant response and contact measurements support actuation claims. Network claims require protocol behavior under the relevant impairments. ASAM OpenSCENARIO and FMI help move scenarios or models across tools, but semantic equivalence still needs checking \cite{asamOpenScenarioDsl2026,fmiStandard3}. Governance records connect these measurements to risk-management and review processes \cite{tabassi2023airmf,iso23894,iso42001,fagan2020iotbaseline,etsi3036452024,mitchell2019modelcards,gebru2021datasheets}.

\subsection{Evidence Artifacts and Validity Threats}
\label{sec:evidence_validity}

Six artifact classes preserve the path from experimental design to claim (Figure~\ref{fig:evidence-chain}). Scenario records identify what was exercised, while platform records fix the apparatus that executed it. Execution traces preserve the observed behavior and interventions. Ground-truth records let reviewers test whether those observations measure the intended property. Perturbation records identify the fault or attacker that produced a deviation. Governance records bind the experiment to its approved scope and subsequent decision.

A detector evaluated only against its own labels can measure agreement while missing unsafe state, a construct-validity failure. Without the platform configuration and execution trace, a scheduler change can masquerade as a policy improvement, undermining internal validity. Missing field calibration leaves external validity unresolved. Missing scenario-level results and sampling records prevent a statistical assessment of uncertainty. An omitted perturbation script can conceal a weak attacker or an unsafe bridge, undermining security validity. A missing approval or claim-scope record can allow a bounded trial to justify an unsupported deployment decision, undermining governance validity. Preserving these records makes the cause of each failure independently examinable.

The next section applies these obligations to systems whose decisions change a physical state.

\section{Sandboxes in Physical AI, AIoT, and Cyber-Physical Systems}
\label{sec:physical-ai}

\subsection{Requirement Shifts Under Embodiment and Edge Deployment}

Physical control couples a learned decision to plant state. An evaluation must therefore represent the feedback path and measure where its assumptions fail. The following requirements identify the physical-control evidence that software-only evaluation leaves unresolved.

\subsubsection{Embodiment and Sensing}
\label{sec:embodiment-sensing}
 
A camera can blur during motion, while reflective material can distort a LiDAR return. If the sandbox preserves only clean inputs, its scenario distribution excludes the failure mechanism. Channel Specificity separates this sensing gap from contact and timing errors. Domain randomization tests a declared parameter family \cite{muratore2022randomized}. System identification and real-to-sim calibration fit the environment to measured behavior \cite{du2021autotuned}. Habitat and AI2-THOR support visual and interaction experiments whose validity still depends on the chosen task and embodiment \cite{savva2019habitat,szot2021habitat2}.

Generalist VLA policies increase the importance of this distinction. RT-2 and Open X-Embodiment connect broad visual-language representations to robot actions, while OpenVLA makes such a policy available as open weights \cite{zitkovich2023rt2,oneill2024openx,kim2024openvla}. A nominal task suite may leave a conditional backdoor untested. BadVLA studies trigger-induced control deviations alongside retained clean performance in simulated manipulation \cite{zhou2025badvla}. Its visual and object-trigger results leave real-hardware trigger transfer for separate validation. A physical object entering the camera view is therefore a distinct adversarial test condition. Its effect must be traced through the action stream to the robot's contact or motion envelope.

\subsubsection{Real-Time Control}
\label{sec:real-time-control}

A correct action delivered after its deadline can destabilize a control loop. Hard-real-time requirements make every deadline binding. Firm-real-time systems discard late results, while soft-real-time systems tolerate graded delay \cite{salamun2023weakly,erickson2022soft}. Unsynchronized simulators or emulators can introduce pauses and event-ordering errors that distort the measured response \cite{cosseron2024simulating,steinbrink2019cosim}. A timing claim needs end-to-end latency and deadline-miss distributions under the target load. SiL can expose software scheduling and interface behavior, but host timing does not establish the embedded device's preemption or throttling behavior. HiL adds measured controller and I/O timing, provided the coupled simulator respects the declared clock contract \cite{li2022hilbuilding,bompard2017realtime}.

\subsubsection{Safety Envelopes}
\label{sec:safety-envelopes}

Safety evaluation must test a declared operating envelope under faults as well as nominal operation. ISO~26262 and IEC~61508 organize functional-safety obligations, while ISO~21448 addresses hazards from limitations of intended functionality \cite{iso26262,iec61508,iso21448}. UL~4600 structures an argument supported by evidence for autonomous products \cite{ul4600}. SAE~J3016 defines automation roles, and ASAM scenario descriptions help express tests within the declared operating domain \cite{saej3016,asamOpenScenarioDsl2026}. We map these instruments to evidence obligations in Table~\ref{tab:section6-standards-map}, without treating the mapping as a substitute for their clauses.

\subsubsection{Network Effects}
\label{sec:network-in-the-loop}

A network fault can change the state seen by a controller before it changes the packet count seen by an analyst. FMI and HELICS let analysts couple plant and communication models \cite{fmiStandard3,palmintier2017helics}. SCEPTRE and minimega provide network and operational-technology emulation when protocol execution matters \cite{hahn2021sceptre,crussell2016minimega}. IEC~62443 supplies industrial segmentation concepts, and ISO/SAE~21434 supplies lifecycle cybersecurity obligations for road vehicles \cite{iec62443,isosae21434}. The experiment must connect a network impairment to its control effect and show which boundary prevented escalation.

\subsubsection{Edge and AIoT Resource Constraints}
\label{sec:edge-constraints}

In AIoT deployments models run on microcontrollers, embedded boards, mobile processors, or FPGAs under tight memory, power, and thermal budgets, so the evaluation question shifts from accuracy alone to whether the model meets the target device's resource budget. A model that is accurate on a GPU may be unusable on an embedded target. Benchmarks such as MLPerf Tiny make this measurable by reporting latency, accuracy, and energy on low-power platforms~\cite{banbury2021mlperftiny}, and FPGA-accelerated digital twins are being explored for latency-critical edge tasks such as collision avoidance where cloud round-trips are not an option~\cite{xu2025fpgatwin}. These are device-level evidence claims, strong for resource and timing behavior and silent on system-level safety.

\subsubsection{Containment Hand-offs Across SiL and HiL Staging}
\label{sec:hil-sil-staging}

The transition from SiL to HiL changes the authority of the same software command. In SiL, a delayed ROS~2 message may corrupt simulated state. In HiL, that message can reach a powered actuator. The hand-off must therefore carry an explicit command contract and a tested stop path, alongside the software and timing records established in Section~\ref{sec:evaluation_stages} \cite{li2022hilbuilding,bompard2017realtime,nguyen2023shiphil}.

Cyber containment becomes essential when SiL introduces production interfaces and reachable services. A host credential or middleware bridge can escape the modeled task even when the plant is simulated. At HiL, safety-rated interlocks and emergency stops must bound energy independently of the learned controller. A software stop that waits behind the same blocked middleware queue does not provide an independent physical boundary. The transition tests must therefore inject stale commands and delayed execution, record relay response, and verify that recovery cannot silently restore unsafe authority. Earlier stages can also require cyber isolation. Figure~\ref{fig:staged-pipeline} marks where these obligations become unavoidable in the depicted pipeline.

\subsection{Platform Patterns for Physical AI Sandboxes}

Physical-control evaluation commonly combines tools. A robotics simulator supplies sensor and plant models, middleware runs deployed code, and a HiL interface adds device timing. Infrastructure studies may instead couple energy and communication models through FMI, HELICS, or mosaik \cite{fmiStandard3,asamOpenScenarioDsl2026,palmintier2017helics,schutte2011mosaik}. Every connection needs an interface contract and an evidence record.

\subsubsection{Simulation and Testbed Platforms for Embodied AI Sandboxes}
\label{sec:simulation-testbed-platforms}

Explicit physics-engine sandboxes represent dynamics through specified equations and numerical solvers. MuJoCo, Gazebo, CARLA, and Isaac Lab differ in their sensor models and execution scale, but each exposes parameters whose error can be tested against measurements \cite{todorov2012mujoco,koenig2004gazebo,dosovitskiy2017carla,mittal2025isaaclab}. Gazebo and ROS support reuse of deployed robot code \cite{quigley2009ros,salimpour2025simtoreal}. Isaac Gym and Isaac Sim support parallel experimentation and synthetic data \cite{makoviychuk2021isaacgym,isaacSimDocs2026}. Habitat, AI2-THOR, and BEHAVIOR provide repeatable task environments \cite{savva2019habitat,szot2021habitat2,liu2022behaviorhabitat}. Channel Specificity determines which of these capabilities needs calibration for a given claim.

SIMPLER provides a concrete transfer measurement protocol. It compares manipulation policies in paired real and simulated tasks, examining policy-ranking agreement and sensitivity to changes in the evaluation conditions \cite{li2024simpler}. This supports a narrower inference than visual inspection: the harness can test whether a simulator preserves a policy comparison within the studied task distribution. It leaves human-contact safety and adversarial deployment conditions for separate validation.

Learned world foundation models change the source of environment error. Cosmos generates world rollouts for physical-AI development \cite{agarwal2025cosmos}. Such models offer photorealistic counterfactual generation at scale, but a generated contact sequence need not enforce the plant's conservation laws or actuator constraints. Plausible imagery can therefore coexist with hallucinated contact causality. Cosmos reports contact-rich physical inaccuracies. Each safety use still needs a separate test of the relevant dynamics and conservation constraints. In the tuple \(S\), the learned weights now form part of \(E\). Poisoning those weights can corrupt the evaluated world without changing the tested controller \(U\). A world-model sandbox consequently needs weight provenance and independent physical checks, alongside its visual evaluation.

Domain-specific tools concentrate model detail where it matters for their tasks. CARLA supports controlled urban-driving variation and scenario exchange \cite{dosovitskiy2017carla,asamOpenScenarioDsl2026}. AirSim illustrates why the source revision and maintenance state must accompany a historical platform citation \cite{shah2017airsim,airsimRepo2026}. Digital twins add operational replay, including edge and FPGA implementations for latency-sensitive applications \cite{das2022edgetwin,xu2025fpgatwin}. FMI, HELICS, and mosaik couple models at infrastructure scale, making solver and synchronization records necessary to interpret the joint experiment \cite{fmiStandard3,palmintier2017helics,schutte2011mosaik,ofenloch2024mosaik3}. Hybrid building and ship testbeds connect this software evidence to measured hardware responses \cite{li2022hilbuilding,nguyen2023shiphil}.

\section{Measurement Framework and Comparative Analysis}
\label{sec:measurement-framework}

\subsection{Cross-Domain Comparison Criteria}
\label{subsec:capability-validity}

A comparison must identify the claim a platform capability supports. A weather parameter becomes rainy-weather evidence only after a defined test measures the relevant sensor and control response. A packet capture becomes plant-resilience evidence only when it can be linked to process state. This is the capability-to-evidence inference that the three principles of Section~\ref{sec:measurement_vocabulary} constrain.

Verification checks whether an implementation solves its specified model. Validation checks whether the model adequately represents its referent for the intended use. Credibility also depends on the consequence of a wrong decision \cite{oberkampf2010verification,asme2018vv40,nasa2024std7009b}. A low-consequence debugging claim may tolerate coarse dynamics. A deployment claim needs evidence for every channel on which its safety or security argument depends.

We use a \emph{sandbox evidence profile} to make this relation explicit:
\[
  \mathrm{EvidenceProfile}(S,U,\mathcal{C},M,A,Q) \rightarrow
  \mathrm{Claim}(D \mid \Lambda),
\]
Here \(\mathcal{C}\subseteq C\) is the executed scenario or perturbation suite, \(A\) is its preserved artifact bundle, and \(Q\) qualifies measurement uncertainty and validity. The in-sandbox property \(P\) is the measured premise. The arrow denotes an inference to deployment claim \(D\) qualified by \(\Lambda\). CARLA can substantiate a recorded closed-loop comparison under specified maps and sensors \cite{dosovitskiy2017carla,carlaDocs2026}. HELICS and FMI can substantiate an executed coupling configuration \cite{palmintier2017helics,fmiStandard3}. Further claims require the corresponding transfer and model-validation evidence.

The evidence profile makes the unit of comparison explicit even when the substrates differ. It also identifies a repair: a missing timing trace calls for timing evidence, while a missing field comparison calls for a transfer study.

\subsection{Measurement Dimensions}
\label{subsec:measurement-dimensions}

The measurement dimensions in Table~\ref{tab:section4-measurement-framework} extend the vocabulary introduced in Section~\ref{sec:measurement_vocabulary} into an operational coding framework. Each dimension is scored relative to a particular assurance claim.

\begingroup
\footnotesize
\setlength{\LTcapwidth}{\textwidth}
\setlength{\tabcolsep}{3pt}
\renewcommand{\arraystretch}{1.10}
\begin{longtable}{>{\raggedright\arraybackslash}p{0.11\textwidth}>{\raggedright\arraybackslash}p{0.25\textwidth}>{\raggedright\arraybackslash}p{0.345\textwidth}>{\raggedright\arraybackslash}p{0.225\textwidth}}
\caption{Cross-domain measurement framework for AI sandbox evidence. Each dimension is coded relative to a stated assurance claim. The shaded bands group the fifteen dimensions into core evidence properties, cyber-physical realism and integration, and scale, openness, and accountability.}
\label{tab:section4-measurement-framework}\\
\toprule
\rowcolor{tabhead}\textbf{Dimension} & \textbf{Definition and measured object} & \textbf{Assessment procedure. Required artifacts \emph{(typical stages)}} & \textbf{Failure mode and validity threat}\\
\midrule
\endfirsthead
\multicolumn{4}{@{}l}{\small\textit{Table \ref{tab:section4-measurement-framework} continued.}}\\
\toprule
\rowcolor{tabhead}\textbf{Dimension} & \textbf{Definition and measured object} & \textbf{Assessment procedure. Required artifacts \emph{(typical stages)}} & \textbf{Failure mode and validity threat}\\
\midrule
\endhead
\bottomrule
\endlastfoot
\tband{4}{Core evidence properties}
Sensing and world fidelity & Claim-relevant match of camera, LiDAR, and radar observations, scene geometry, illumination, weather, and surrounding-agent or traffic behavior. Excludes execution timing, network transport, and actuator/plant response. & Compare sensor returns and scene/agent statistics with calibrated lab or field observations. Preserve sensor models, calibration residuals, scene provenance, and uncertainty ranges \emph{(simulation, SiL, HiL, twins)}. & Visual plausibility masks sensor error or unrealistic surrounding-agent behavior. \\
Controllability & Ability to vary scenarios, seeds, actors, faults, attacks, boundary cases, ODD parameters, and operating modes. & Inspect scenario grammar, parameter ranges, intervention APIs, attack/fault libraries, orchestration scripts. Scenario definitions, seed lists, generators, intervention scripts \emph{(simulation, ranges, regulatory trials)}. & Many generated scenarios over a weak grammar. Scenario bias and overfitting. \\
Observability & Ability to capture ground truth, hidden state, telemetry, logs, traces, interventions, timing, and outcomes. & Audit event schema, trace completeness, missingness, and alignment between logs and claim variables. Telemetry traces, ground truth, PCAP, controller and intervention logs, schemas \emph{(all stages)}. & Logs omit the variable needed to explain failure. False attribution. \\
Containment & Ability to bound cyber, physical, data, operational, and governance harm. & Review isolation, segmentation, access control, emergency stops, actuation gates, credential scope, and exit criteria. Network diagram, firewall rules, interlock tests, access records, regulatory plan \emph{(HiL, ranges, twins, regulatory)}. & Testbed remains connected to operations. Lateral movement or physical harm. \\
Reproducibility & Ability to rerun the experiment with stable versions, seeds, assets, configs, hardware, and scripts. & Independent rerun or artifact audit. Compare output deltas across replays and versions. Seeds, configs, versions, hashes, assets, model weights, replay scripts \emph{(benchmarks, simulation, SiL/HiL)}. & Cloud APIs, assets, or simulator defaults drift. Evidence decay. \\
\tband{4}{Cyber-physical realism and integration}
Scenario portability & Ability to express and replay scenarios across tools, stages, or domains. & Translate scenarios to another simulator or stage. Measure semantic equivalence and unsupported constructs. OpenSCENARIO/FMI files, translators, semantic notes, replay logs \emph{(AV, co-simulation, HiL, regulatory packages)}. & Portable syntax mistaken for identical execution. Semantic mismatch. \\
Timing fidelity & Realism and auditability of clocks, scheduling, latency, jitter, synchronization, deadlines, and real-time execution. & Compare clock traces, latency and jitter distributions, deadline misses, solver and scheduler settings. Clock logs, scheduler configs, real-time traces \emph{(robotics, AV, grid, ICS, AIoT, HiL)}. & Fixed-step simulation hides deadline failures. Invalid control or security conclusions. \\
Network realism & Representation of topology, protocol semantics, traffic, delay, packet loss, routing, radio conditions, and trust boundaries. & Compare network traces to target conditions. Inspect emulator topology and protocol coverage. PCAP, topology diagrams, emulator configs, protocol versions, mapping to plant effects \emph{(AIoT, ranges, CPS)}. & Packet traces omit PLC or control-loop semantics. Packet-level evidence overclaims physical resilience. \\
Actuator and plant realism & Representation of actuator, robot, vehicle, grid, building, industrial plant, or spacecraft dynamics relevant to the claim. & Compare response curves, saturation, delays, fault response, contact behavior, and operating envelope. Plant model, actuator calibration, HiL traces, field data, bill of materials \emph{(robotics, HiL, CPS, space)}. & Ideal actuators mask instability or unsafe commands. Missing physical consequence. \\
HiL/SiL integration & Inclusion of production software, firmware, controllers, devices, buses, sensors, actuators, or plant interfaces. & Interface coverage tests. Timing and I/O validation. Firmware and hardware inventory. Bill of materials, firmware versions, wiring diagrams, I/O traces, SiL build records \emph{(SiL, HiL, cyber-HiL)}. & Real controller coupled to unvalidated plant model. Interface evidence mistaken for transfer evidence. \\
Sim-to-real transfer evidence & Empirical correspondence between sandbox and deployment outcomes, including degradation and uncertainty. & Compare policy rankings, success rates, traces, safety events, perception errors, or attack impacts across sim and real. Real or lab traces, calibrated sim logs, transfer scripts, uncertainty treatment \emph{(simulation-to-field pipelines)}. & Simulation success treated as deployment proof. Unsupported external validity. \\
Attack/fault injection & Ability to represent adversarial and accidental disturbances with controlled scope, repeatability, observability, and safety. & Execute scripted attacks/faults. Document attacker model, access path, trigger, target, and consequence. Attack scripts, fault library, threat model, PCAP, process traces, containment logs \emph{(ranges, CPS, AI security, HiL)}. & Convenient faults substituted for credible adversaries. Weak attacker model. \\
\tband{4}{Scale, openness, and accountability}
Scalability & Ability to run enough experiments for the claim at acceptable cost and duration. & Measure throughput, parallelism, cost, failure rate, resource saturation, and statistical power. Hardware config, run logs, scheduler logs, cost records \emph{(GPU simulation, benchmarks, scenario sweeps)}. & Throughput amplifies biased scenarios. Quantity mistaken for coverage. \\
Openness and auditability & Reviewability of code, assets, configs, logs, licenses, data, models, and provenance. & Artifact inspection. License and provenance review. Reproduce the claim-to-evidence chain. Repository, assets, licenses, hashes, documentation, model cards \emph{(all technical sandboxes)}. & Proprietary assets or changing services hide assumptions. Unauditable evidence. \\
Governance artifacts & Process evidence: risk records, approvals, audit trails, scenario rationales, regulatory plans, exception handling, and traceability. & Review completeness and quality of claim-to-evidence mapping, oversight records, and residual-risk statements. Risk register, approvals, audit logs, documentation package, regulatory records \emph{(regulatory sandboxes, TEVV, assurance cases)}. & Paperwork substitutes for technical validation. Process evidence overclaimed as safety evidence. \\
\end{longtable}
\endgroup

Cross-tool comparison adds a separate semantic obligation. ASAM OpenSCENARIO exchanges scenario descriptions and FMI exchanges models \cite{asamOpenScenarioDsl2026,fmiStandard3}. A translator can preserve the file syntax while changing event ordering or contact behavior. Portability evidence must therefore include execution differences at the target interface.

\subsection{Coding Rubric and Heatmap}
\label{subsec:coding-rubric}

The framework uses five qualitative values: \textsc{Strong}, \textsc{Moderate}, \textsc{Weak}, \textsc{Absent}, and \textsc{Unclear}. We code qualitatively because a numerical score would imply precision the evidence rarely supports. The coding task makes the boundary of a claim auditable.

\textsc{Strong} requires documented measurements and reviewable artifacts for the stated dimension and claim. A calibrated sensing channel needs comparison traces. Reproducibility needs enough preserved configuration to rerun the experiment. Containment needs a tested boundary and intervention path. Governance needs a traceable decision record.

\textsc{Moderate} means that the capability is documented and plausible for the claim, but empirical validation, calibration, independent replication, or artifact completeness is incomplete. \textsc{Weak} means that the dimension is partial, proxy-level, dependent on undocumented assets or defaults, or disconnected from the dominant deployment dynamics or threat model. \textsc{Absent} means that the capability or artifact needed for the claim is not present. \textsc{Unclear} means public evidence is insufficient. Unknown evidence must not be upgraded or downgraded by intuition.

\begin{figure}[htbp]
    \centering
    \includegraphics[width=\linewidth]{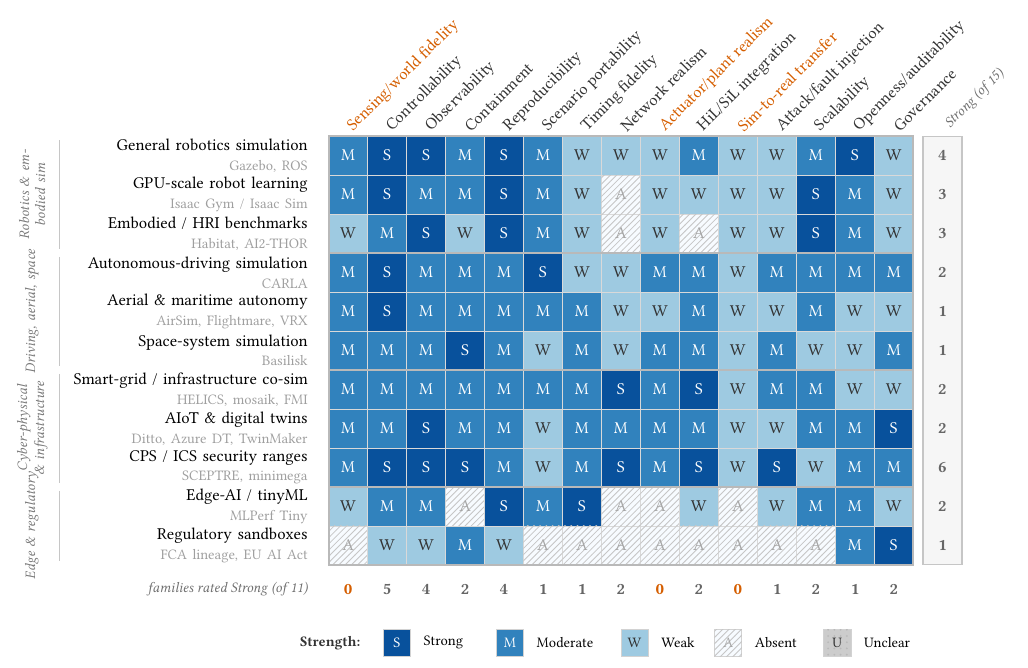}
    \Description{An eleven-by-fifteen color heatmap. Rows are sandbox families grouped by substrate (robotics and embodied simulation. Driving, aerial, and space. Cyber-physical and infrastructure. Edge and regulatory). Columns are the fifteen measurement dimensions. Each cell is shaded by strength and prints a redundant glyph: S strong, M moderate, W weak, A absent. A right margin gives, per family, the number of dimensions rated strong. A bottom row gives, per dimension, the number of families rated strong, with three columns reading zero. Those three zeros and their column headers (sensing/world fidelity, actuator/plant realism, sim-to-real transfer) are printed in red. A legend explains the strength scale.}
    \caption{Claim-relative coding of sandbox families across measurement dimensions (\textsc{S} strong, \textsc{M} moderate, \textsc{W} weak, \textsc{A} absent, \textsc{U} unclear). Each family $F_i$ is graded against its domain deployment reference claim $D_i^*$ in Column~3 of Table~\ref{tab:heatmap-backing}. Strength is dimension-specific, no family dominates, and the sensing/world-fidelity, actuator/plant-realism, and sim-to-real columns reach \textsc{Strong} for no family.}
    \label{fig:heatmap}
\end{figure}

The authors code each family \(F_i\) against a \emph{domain deployment assurance reference claim} \(D_i^*\). This is the deployment-level target in Column~3 of Table~\ref{tab:heatmap-backing}. The \textsc{Weak} and \textsc{Absent} cells in Figure~\ref{fig:heatmap} describe missing support for that target. Column~2 states a narrower \emph{in-sandbox claim} \(D_i^{\mathrm{local}}\). Dimensions such as field transfer or real-account containment can drop out of \(\mathrm{Rel}(D_i^{\mathrm{local}})\) when the local claim genuinely makes no assertion about them. If all remaining necessary dimensions are \textsc{Strong}, then \(\mathcal{L}(D_i^{\mathrm{local}})=\textsc{Strong}\). This conditional statement does not upgrade a \textsc{Moderate} reproducibility cell or justify removing a safety-relevant dimension.

The authors assigned the family-level grades using the operational criteria in Table~\ref{tab:section4-measurement-framework} and the source rationales in Appendix~\ref{app:heatmap-backing}. The comparison is a structured qualitative synthesis. Its audit trail consists of the stated reference claim, the criterion for each dimension, and the cited family-level evidence.

\textsc{Unclear} records insufficient evidence and maps to \textsc{Absent} only when licensing a claim. The current matrix contains no \textsc{Unclear} cells.

The heatmap rows are sandbox families organized around physical, cyber-physical, edge, infrastructure, and regulatory evidence. Agent-based sandboxes enter as a cross-cutting archetype: they appear where tool authority, real accounts, software effects, or physical consequence become the limiting evidence dimension, especially in the threat model and in Case~3 of Section~\ref{subsec:worked-instantiations}.

Figure~\ref{fig:heatmap} therefore compares evidence against stated domain targets. For example, an edge benchmark's lack of plant containment limits a system-safety inference while leaving its measured device-latency claim intact. Process-Technical Separation gives the corresponding interpretation for regulatory rows.

Appendix~\ref{app:heatmap-backing} records the local and deployment reference claims alongside the source rationale. Family membership summarizes an evaluation pattern. A particular configured campaign still needs its own evidence profile.

\subsection{Metrics and Evidence Artifacts for Physical AI and AIoT Sandboxes}
\label{subsec:metrics-artifacts}

Sandbox metrics characterize the apparatus and evaluation procedure. System metrics characterize \(U\). An instrumented simulator can faithfully report that a policy succeeded while leaving deployment-critical conditions unrepresented. Both kinds of metric must therefore identify the claim they measure.

For a physical-control campaign, instantiate the six artifact classes of Section~\ref{sec:evidence_validity} at the actual interfaces. Preserve controller and firmware revisions with the scenario manifest. Align physical ground truth with command and intervention timestamps. Keep attack specifications linked to the affected runs, and record the governance decision against that same manifest. These links let a reviewer reconstruct the result without relying on a platform's reputation.

\subsection{Validity Threats and Interpretive Cautions}
\label{subsec:validity-threats}

Transfer qualification must identify the dynamics on which the target claim depends. Under Reproducibility-Transfer Separation, randomization supports only the declared parameter family \cite{tobin2017domainrandomization,muratore2022randomized}. A HiL result also retains any unvalidated plant-model assumptions.

A scenario grammar can omit the event that causes failure. An autonomous-driving campaign may vary weather extensively while excluding a rare interaction between an occluded pedestrian and delayed braking \cite{tang2023adstesting,kaur2021simulators}. Report the declared event space and the executed subset \(\mathcal{C}\), with uncertainty over what remains untested.

Simulator-version dependence is a third threat. Physics engines, rendering pipelines, assets, middleware bridges, GPU kernels, solver settings, and concurrency behavior can change outcomes. Without version pinning and replay scripts, evidence decays. This is especially important for platforms whose maintenance status changes, including archived or discontinued projects. A historical simulator paper may establish scientific capability at publication time. Current evidence requires repository and documentation status checks.

Timing mismatch can invalidate a controller result when stale sensing or delayed actuation changes the feedback dynamics. Network simplifications can conceal the same failure by preserving packet statistics while omitting fieldbus semantics. The evidence must align communication events with the resulting controller and plant state.

Asset and data provenance are quieter but pervasive threats. Synthetic datasets and virtual environments may contain undocumented meshes, textures, object models, maps, traffic behaviors, human motion captures, or sensor models. These assets affect both reproducibility and external validity. For AIoT and digital twins, telemetry provenance and twin binding are equally important: a twin that silently changes data source, sampling rate, calibration, or actuation authority changes the evidence profile.

The security-specific threat is a weak attack model. A sandbox may support attack injection but test only benign faults, outdated exploits, or unrealistic attacker access. For CPS, the critical distinction is whether the attack produces a physical or operational consequence. For AI systems, prompt injection, model extraction, poisoning, adversarial perception, compromised plugins, simulator escape, and unsafe cloud bridges belong to different trust boundaries and require different controls~\cite{mitreATLAS,greshake2023indirectprompt,alcaraz2022digitaltwinsecurity}.

Under Process-Technical Separation, governance review must preserve the limits of the underlying measurement. An exit report that omits failed scenarios can change the apparent scope of a result even if every retained trace is authentic. The worked examples below make the resulting licensing limits explicit.

\subsection{Evidence Composition and Worked Instantiations}
\label{subsec:worked-instantiations}

The dimensions are assessed separately, while a deployment claim often depends on several jointly. We define a conservative serial composition rule and apply it to four literature-grounded examples. These examples specify conditional evidence profiles for named apparatus configurations. They report no new experimental success rates and make no blanket grade assignment to a product.

\paragraph{A weakest-link composition rule.}
Let \(L=\{\textsc{Absent}<\textsc{Weak}<\textsc{Moderate}<\textsc{Strong}\}\) be an ordered chain, hence a lattice. For licensing, map \textsc{Unclear} to \textsc{Absent}. For a claim \(D\), let the nonempty set \(\mathrm{Rel}(D)\) contain the dimensions whose failure would individually invalidate its argument. With coded strength \(e(d)\in L\), define
\[
  \mathcal{L}(D)=\min_{d\in\mathrm{Rel}(D)}e(d).
\]
The rule caps the licensed claim at each necessary support: \(\mathcal{L}(D)\leq e(d)\) for every relevant \(d\). Raising a non-minimal dimension leaves the cap unchanged. If several dimensions tie at the minimum, all must improve before the cap rises. The rule is an ordinal licensing policy. It supplies neither a probability of safety nor a lower bound on actual reliability, and \textsc{Strong} grades do not prove joint sufficiency. Dependencies and interactions must remain in the claim argument.

Across stages, include each necessary transfer link in the serial argument. A weak SiL-to-HiL link caps a field claim even when the recorded simulation is reproducible. Assurance-case dependency reasoning motivates this conservative cap \cite{kelly2004gsn}. Prior simulation-credibility work instead retains separate factor assessments \cite{blattnig2008credibility}. We retain that profile as well, using the minimum only to limit the inference a reader may draw.

\paragraph{Selecting \(\mathrm{Rel}(D)\).}
Derive the necessary dimensions from the claim's hazard and threat argument, constrained by applicable standards. The mapping requires a claim-specific rationale. Vehicle safety can depend on sensing, contact dynamics, and timing, while ISO/SAE~21434 adds cybersecurity obligations. IEC~62443 anchors industrial network and attack evidence. Tables~\ref{tab:section6-standards-map} and~\ref{tab:section6-dimension-to-artifact} record these anchors. Where no instrument settles the mapping, the threat model must state the remaining judgment. A dimension cannot be removed merely because its grade is inconvenient.

\paragraph{Compensating controls and runtime assurance.}
A verified runtime envelope changes the architecture and the claim being licensed. A Simplex-style runtime-assurance wrapper, control barrier function, or hardware interlock can enforce a safety constraint even when the learned policy remains unverified \cite{schierman2020rta,ames2017cbf,alshiekh2018shielding,iec61508}. For a guarded claim \(D_{\mathrm{safety}}\), \(\mathrm{Rel}(D_{\mathrm{safety}})\) centers on monitor-envelope validity, timing fidelity, containment, and fail-safe actuation authority. Within the fifteen dimensions, envelope validity needs observable state and validated monitor sensing, while actuation authority needs tested intervention and plant response. Policy-side sensing/world fidelity can cease to be necessary only if the safety argument is independent of that policy's perception. Shared spoofable sensors or a delayed stop path preserve the dependency. The wrapper earns no automatic grade increase. It requires its own evidence profile, exactly the monitor-validity problem in Challenge~R13 of Table~\ref{tab:section5-ranked-challenges}.

\begin{table}[htbp]
\centering
\scriptsize
\renewcommand{\arraystretch}{1.12}
\setlength{\tabcolsep}{3pt}
\caption{Four conditional worked profiles. Grades apply to the configurations described in the text. The local claim records a bounded result or comparison and requires preserved evidence for its remaining necessary dimensions.}
\label{tab:worked-comparison}
\begin{tabularx}{\textwidth}{>{\raggedright\arraybackslash}p{0.12\textwidth}>{\raggedright\arraybackslash}p{0.17\textwidth}>{\raggedright\arraybackslash}p{0.15\textwidth}>{\raggedright\arraybackslash}p{0.22\textwidth}>{\raggedright\arraybackslash}p{0.075\textwidth}>{\raggedright\arraybackslash}X}
\toprule
\rowcolor{tabhead}\textbf{Sandbox \(S\)} & \textbf{Over-read claim \(D\)} & \textbf{Key \textsc{Strong} dimensions} & \textbf{Bottlenecks in \(\mathrm{Rel}(D)\)} & \textbf{\(\mathcal{L}(D)\)} & \textbf{Narrowed claim \(D^{\mathrm{local}}\)} \\
\midrule
\(D_1\) CARLA driving & High simulated success establishes safe urban deployment & Controllability and scenario portability & Timing and field transfer \textsc{Weak}. Sensing/world fidelity \textsc{Moderate} & \textsc{Weak} & The recorded stack execution handles the modeled suite \\
\addlinespace
\(D_2\) SWaT detector & Detection on the testbed establishes protection of a target water plant & Observability, containment, and network realism & Target-attack representation and cross-plant transfer \textsc{Weak} & \textsc{Weak} & Detector behavior on the recorded plant and attacks \\
\addlinespace
\(D_3\) WebArena/ ToolSandbox agent & Task success establishes safe real-account tool use & Observability and reproducibility & Real-account containment evidence \textsc{Absent}. Adversarial injection and consequence transfer \textsc{Weak} & \textsc{Absent} & Task completion under the sandboxed tool authority \\
\addlinespace
\(D_4\) Embodied VLA manipulation & High Isaac Lab/SIMPLER task success establishes safe collaborative deployment & Observability, controllability, and scalability & Contact/plant response, timing, and attack/fault coverage \textsc{Weak}. Nominal paired transfer \textsc{Moderate} & \textsc{Weak} & Nominal task comparison within the paired evaluation conditions \\
\bottomrule
\end{tabularx}
\end{table}

\paragraph{Case 1: CARLA driving.}
Let \(D_1\) claim safe urban deployment from high closed-loop success in CARLA. Here \(U\) is the planning stack, \(E\) fixes maps and sensor models, and \(\mathcal{C}\) fixes scripted traffic and weather. Resets provide \(I\), collision and success measurements provide \(M\), and retained traces provide \(A\). Controllability and scenario portability are \textsc{Strong} for this configured suite \cite{dosovitskiy2017carla,asamOpenScenarioDsl2026}. Reproducibility remains \textsc{Moderate} if assets or versions are incomplete \cite{carlaDocs2026}. Sensing/world fidelity is \textsc{Moderate}, while unmeasured hardware timing and absent field correlation leave timing and transfer \textsc{Weak} \cite{lee2008cps,tang2023adstesting}. Thus \(\mathcal{L}(D_1)=\textsc{Weak}\). A descriptive claim about the preserved simulated executions can be \textsc{Strong} if its necessary observation and scenario records are complete. A claim about reproducible future regression would also retain reproducibility in its relevant set. Road-safety inference still needs an argument about target-domain coverage and exposure \cite{kalra2016driving,koopman2017avsafety}.

\paragraph{Case 2: SWaT industrial-control detector.}
Let \(D_2\) infer protection of a target water plant from detection on SWaT. The physical testbed includes real controllers and process instrumentation \cite{mathur2016swat}. In the conditional profile, archived process traces and contained injection establish \textsc{Strong} observability, containment, and network realism, alongside the range's control and injection capabilities \cite{giraldo2018physics}. Those capabilities do not establish representation of the target plant's attacker access or topology. For \(D_2\), attack/fault injection is \textsc{Weak} on target-campaign representativeness, and transfer to the target plant is \textsc{Weak}. WADI illustrates that another water-system testbed introduces a different process and attack context \cite{ahmed2017wadi}. The cap is \(\mathcal{L}(D_2)=\textsc{Weak}\). Narrowing the claim to recorded detector behavior on the exercised plant removes cross-plant generalization, while retaining measurement and labeling obligations.

\paragraph{Case 3: WebArena and ToolSandbox agents.}
Let \(D_3\) infer safe operation on real accounts from high task success with sandboxed tools. WebArena supplies self-hosted applications, and ToolSandbox supplies stateful tool interactions \cite{zhou2024webarena,lu2025toolsandbox}. A versioned task suite with complete trajectories can provide \textsc{Strong} observability and reproducibility \cite{zhu2025abc}. For the target claim, however, evidence that permission gates contain real-account authority is \textsc{Absent}. If the campaign omits indirect prompt injection and irreversible API effects, adversarial testing and consequence transfer remain \textsc{Weak} \cite{greshake2023indirectprompt,mitreATLAS}. The missing evidence belongs to the containment dimension. Physical consequence is irrelevant to this software-only claim. Hence \(\mathcal{L}(D_3)=\textsc{Absent}\). The local task-completion claim is licensed only within the recorded tool authority.

\paragraph{Case 4: embodied VLA manipulation.}
Let \(D_4\) claim that an OpenVLA-style policy with high task success in GPU simulation and a real-to-sim harness is safe for collaborative human-shared workspace deployment. Consider an Isaac Lab campaign with preserved simulator state and scripted interventions, coupled to a SIMPLER-style paired task comparison \cite{kim2024openvla,mittal2025isaaclab,li2024simpler}. These are complementary apparatuses, without an assumed plug-compatible implementation. In the specified profile, state capture, controlled variation, and parallel execution provide \textsc{Strong} observability, controllability, and scalability. Paired policy-ranking evidence raises nominal-task transfer to \textsc{Moderate}. This grade is our conditional rubric judgment. SIMPLER supplies the paired-evaluation method.

Collaborative safety still depends on contact-force compliance and end-to-end action latency under ROS~2 load. If those channels are unvalidated, actuator/plant realism and timing remain \textsc{Weak}. Omitting object-triggered VLA backdoors or depth spoofing leaves attack/fault injection \textsc{Weak} \cite{zhou2025badvla,cao2019lidar}. Thus \(\mathcal{L}(D_4)=\textsc{Weak}\), with nominal-task transfer alone insufficient to lift it. HiL force-limiting interlocks and adversarial trigger suites address these bottlenecks only when their response traces and operating envelopes are validated. A guarded safety claim must then be reassessed using the runtime-assurance decomposition above.

Table~\ref{tab:worked-comparison} exposes the practical value of the cap. Increasing scenario throughput cannot repair missing device timing, target-attack coverage, or real-account containment. Each bottleneck identifies a different next measurement.

\section{Key Challenges and Research Gaps}
\label{sec:key-challenges-gaps}

\subsection{From Validity Threats to Research Gaps}
\label{subsec:section5-validity-to-gaps}

A research gap remains when a recurring failure prevents an experiment from supporting its intended claim. Table~\ref{tab:section5-ranked-challenges} separates these gaps from missing product features by stating the question each raises and the evidence that would settle it.

Composition is the common difficulty. A calibrated plant model can still be coupled to the wrong clock, and a correctly logged attack can still omit the relevant physical effect. Joint assurance requires the interface assumptions to survive these combinations \cite{muratore2022randomized,tang2023adstesting,yamin2020cyberranges,alcaraz2022digitaltwinsecurity,nistTEVV2026,tabassi2023airmf}.

We group gaps by where the claim loses support. Representation gaps omit a necessary channel or operating condition \cite{oberkampf2010verification,asme2018vv40,nasa2024std7009b}. Security and containment gaps permit harm or corruption of the apparatus \cite{souppaya2017containers,alcaraz2022digitaltwinsecurity,yamin2020cyberranges,greshake2023indirectprompt}. Assurance gaps break the trace from preserved evidence to a justified decision \cite{kelly2004gsn,raji2020closing}.

One failure can cross these groups. If a twin accepts a spoofed telemetry source, its calibration can become invalid and its replay can hide an unsafe state. An intact audit log then documents the wrong experiment. A useful research result must identify which dependency failed and what additional evidence would repair the claim.

\subsection{Ranked Technical Challenges}
\label{subsec:section5-ranked-challenges}

Table~\ref{tab:section5-ranked-challenges} ranks gaps by their breadth across families, their ability to invalidate a deployment claim, and their invisibility in ordinary aggregate logs. The prioritization follows these stated criteria. It has no empirical ranking interpretation. A shared cloud evaluation can put isolation first, while spacecraft testing can put mission-environment mismatch first. The stated target claim determines the local order.

\begingroup
\scriptsize
\setlength{\LTcapwidth}{\textwidth}
\setlength{\tabcolsep}{3pt}
\renewcommand{\arraystretch}{1.10}
\begin{longtable}{>{\raggedright\arraybackslash}p{0.02\textwidth}>{\raggedright\arraybackslash}p{0.155\textwidth}>{\raggedright\arraybackslash}p{0.25\textwidth}>{\raggedright\arraybackslash}p{0.27\textwidth}>{\raggedright\arraybackslash}p{0.225\textwidth}}
\caption{Ranked technical challenges limiting AI-sandbox evidence, each paired with the concrete research question it raises and the evidence that would demonstrate progress. The ordering reflects cross-domain assurance risk: breadth across sandbox families, claim-invalidating potential, and the difficulty of detecting the failure from ordinary benchmark evidence. Affected domains and stages appear in italics under each challenge.}
\label{tab:section5-ranked-challenges}\\
\toprule
\rowcolor{tabhead}\textbf{R} & \textbf{Challenge and gap} & \textbf{Why current evidence falls short} & \textbf{Concrete research question and method} & \textbf{Evidence needed to show progress} \\
\midrule
\endfirsthead
\multicolumn{5}{@{}l}{\small\textit{Table \ref{tab:section5-ranked-challenges} continued.}}\\
\toprule
\rowcolor{tabhead}\textbf{R} & \textbf{Challenge and gap} & \textbf{Why current evidence falls short} & \textbf{Concrete research question and method} & \textbf{Evidence needed to show progress} \\
\midrule
\endhead
\bottomrule
\endlastfoot
\textbf{1} & Compositional sim-to-real transfer \emph{(robotics, AVs, aerial, maritime, space, industrial CPS. Simulation to field)} & Domain and dynamics randomization reduce selected transfer gaps, but evidence for one channel or parameter family does not compose across perception, planning, control, timing, networking, and plant dynamics~\cite{tobin2017domainrandomization,peng2018simtoreal,muratore2022randomized,elmquist2025simtoreal}. & How can evidence across perception, planning, control, timing, networking, attack/fault behavior, and plant dynamics be composed into a bounded assurance claim? \emph{Method:} claim graphs, typed assumptions, uncertainty propagation, counterexample replay. & Paired sim/HiL/field traces, calibration residuals, uncertainty intervals, sensitivity analysis, and failure attribution across coupled layers, traceable by independent reviewers. \\
\addlinespace
\textbf{2} & Real-time semantics, latency, jitter, and synchronization \emph{(CPS, AIoT, AVs, robotics, grids, HiL, edge AI)} & Fixed-step simulation and mean latency hide deadline-miss patterns, stale sensors, bus contention, clock skew, and scheduler interactions~\cite{lee2008cps,rajkumar2010cps,salamun2023weakly,zhang2013timesync}. & How should sandboxes reproduce and report a deployment's timing class under realistic load? \emph{Method:} real-time tracing, co-simulation clock contracts, load and contention profiles. & End-to-end timing traces, jitter tails, deadline-miss distributions, clock-source assumptions, and load profiles. \\
\addlinespace
\textbf{3} & Heterogeneous co-simulation and interface semantics \emph{(smart grids, buildings, infrastructure, twins, aerospace. Co-simulation, SiL, HiL)} & FMI, HELICS, mosaik, and middleware bridges improve exchange and orchestration, but syntactic coupling does not prove unit, solver, clock, event-ordering, or plant-model equivalence~\cite{fmiStandard3,palmintier2017helics,schutte2011mosaik,steinbrink2019cosim,bompard2017realtime}. & Which interface contracts make coupled federations semantically equivalent across engines and time-step policies? \emph{Method:} typed interface contracts, unit and solver metadata, cross-engine replay. & Interface contracts, synchronization logs, step-size sensitivity, and cross-engine replay deltas. \\
\addlinespace
\textbf{4} & Long-tail scenario coverage and rare-event confidence \emph{(AVs, robots, drones, agents, infrastructure. Benchmark, simulation, field)} & A scenario generator can omit rare hazards, degraded sensing, adversarial adaptation, or unusual human intervention despite a large campaign~\cite{tang2023adstesting,kaur2021simulators,kalra2016driving,koopman2017avsafety}. & How can generators measure coverage over operationally meaningful event spaces, and how should finite campaigns report uncertainty without implying proof of safety? \emph{Method:} hazard-linked grammars, ODD bins, coverage-guided fuzzing, rare-event simulation, dependence-aware intervals, hidden holdouts. & Declared event space, coverage maps, failure clustering, uncertainty over uncovered regions, fresh-holdout performance, and negative-result interpretation. \\
\addlinespace
\textbf{5} & Sensor, actuator, contact, plant, and environment mismatch \emph{(robotics, AVs, AIoT, industrial CPS, aerial and maritime autonomy)} & Platform-level fidelity claims can hide the channel that carries the safety or security property: contact, tire-road interaction, LiDAR returns, illumination, wear, saturation, vibration, weather, or aging~\cite{duan2022embodiedaisurvey,todorov2012mujoco,shah2017airsim,savva2019habitat,szot2021habitat2,cao2019lidar}. & Which channel-specific calibration and validation evidence is necessary for the channel that carries the claim? \emph{Method:} channel-wise system identification, noise and saturation models, contact validation. & Channel-specific calibration, response curves, noise and saturation models, contact validation, and lab/field trace comparison. \\
\addlinespace
\textbf{6} & Network realism and cyber-physical coupling \emph{(AIoT, ICS, smart grids, vehicles, robots, tool-using agents. Ranges, twins, HiL)} & Packet-level realism does not establish physical consequence, process deviation, mission impact, privacy exposure, or recovery behavior~\cite{yamin2020cyberranges,mathur2016swat,ahmed2017wadi,liu2009fdi,giraldo2018physics}. & How can authorized attack and fault injection be coupled to physical consequence without unsafe operational bridges? \emph{Method:} lab-contained plant/twin/HiL coupling, consequence metrics, interlocks. & Network topology, PCAPs, impairment profiles, attacker access path, plant traces, safety margins, response latency, final state, and containment evidence. \\
\addlinespace
\textbf{7} & HiL/SiL staging and lab-to-field transfer \emph{(robotics, vehicles, buildings, ships, space, industrial CPS)} & HiL exposes real code and interfaces, but an unvalidated plant model, simplified I/O, or nonrepresentative network remains a simulation assumption~\cite{li2022hilbuilding,nguyen2023shiphil,nasa2024std7009b}. & What evidence package makes HiL/SiL results reviewable outside the original laboratory? \emph{Method:} hardware BOM, firmware records, bus-timing schemas, digital thread, lab protocol. & Hardware/firmware IDs, bus timing, relay configuration, plant validation, intervention logs, field correlation. A second-lab rerun or emulator replay with bounded trace delta. \\
\addlinespace
\textbf{8} & Reproducibility under dependency and asset graphs \emph{(all sandbox families)} & Containers and open repositories do not preserve changing simulator defaults, meshes, maps, services, model endpoints, GPU kernels, hardware revisions, or governance context~\cite{mitchell2019modelcards,gebru2021datasheets,ntia2021sbom,nist2022ssdf,souppaya2017containers}. & How can drifting assets, services, and defaults be pinned or declared so that evidence does not decay? \emph{Method:} artifact manifests, hashes, SBOMs, seed registries, nondeterminism notes. & Artifact manifests, hashes, licenses, SBOMs, seed records, replay scripts, nondeterminism notes, and independent reruns. \\
\addlinespace
\textbf{9} & Digital-twin drift and telemetry provenance \emph{(AIoT, industrial CPS, buildings, infrastructure, vehicles. Twin, replay, monitoring)} & Connectivity can make a twin appear current while telemetry sources, sampling, calibration, topology, or actuation authority drift~\cite{barricelli2019digitaltwin,jones2020digitaltwin,alcaraz2022digitaltwinsecurity}. & How can twin drift and telemetry-binding changes be detected and propagated into claim validity? \emph{Method:} drift monitors, provenance-aware twin state, recalibration triggers. & Telemetry lineage, source authentication, drift metrics and history, recalibration records, replay deltas, and bridge access logs. \\
\addlinespace
\textbf{10} & Human, operator, and supervisor behavior \emph{(AVs, plants, aviation, robotics, regulatory sandboxes. HiL, field, governance)} & Humans can mask failures, introduce delays, create unmodeled recovery paths, or become part of the mitigation without being represented as system components~\cite{koopman2017avsafety,euAIAct2024,fca2017regulatorysandboxlessons}. & How should human intervention be represented when it affects safety outcomes? \emph{Method:} human-in-the-loop protocols, workload models, intervention-trace analysis. & Operator protocols, alert timing, missed or late interventions, authority maps, training state, and separation of autonomous from operator-assisted success. \\
\addlinespace
\textbf{11} & Benchmark and scenario overfitting \emph{(ML benchmarks, embodied agents, AVs, robotics, tool-use environments)} & Repeated exposure to public tasks and assets can reward familiarity with the benchmark~\cite{recht2019imagenet,zhu2025abc,liu2024agentbench,zhou2024webarena,lu2025toolsandbox}. & How can public leaderboards reduce overfitting and asset leakage while retaining reproducibility? \emph{Method:} rotating holdouts, seed registries, benchmark cards, leakage audits, independent retests. & Private or rotating scenarios, public/private performance deltas, contamination tests, negative-result publication, and governance audit. \\
\addlinespace
\textbf{12} & Evidence composition across safety, security, and governance \emph{(full assurance pipeline)} & Separate logs, dashboards, risk registers, safety cases, and metrics rarely encode dependency, uncertainty, and residual-risk relations~\cite{kelly2004gsn,raji2020closing,tabassi2023airmf,nistTEVV2026}. & What evidence package lets a safety engineer, regulator, or reviewer audit scenarios, seeds, traces, faults, monitor states, operator actions, and residual risk, and attribute failures causally across layers? \emph{Method:} evidence schemas, hash manifests, SACM/GSN-style linkage, counterfactual replay, causal graphs over sandbox variables. & Claim-to-evidence graphs, assumption dependencies, artifact hashes, reviewer replay, residual-risk records, and stable root-cause ranking on seeded failures. \\
\addlinespace
\textbf{13} & Runtime assurance and monitor validity \emph{(robotics, aviation, AVs, industrial CPS. SiL, HiL, runtime)} & Shields, monitors, and barrier functions depend on dynamics, sensing, timing, actuation authority, and implementation correctness~\cite{alshiekh2018shielding,ames2017cbf,schierman2020rta}. & Under which dynamics, sensing, timing, and actuation-authority assumptions is a runtime monitor valid, and how is its implementation verified? \emph{Method:} formal monitor specification, injected violations, envelope validation. & Proof assumptions, activation traces, injected-violation results, response latency, and envelope validation. \\
\addlinespace
\textbf{14} & Agentic tool-use containment \emph{(LLM agents, software agents, tool-using embodied systems)} & Task success can hide unauthorized file, API, browser, memory, credential, or robot-control effects~\cite{greshake2023indirectprompt,zhou2024webarena,lu2025toolsandbox,mitreATLAS}. & How can tool-using agents be evaluated with realistic state while preserving least privilege and evidence integrity? \emph{Method:} capability-based tool sandboxing, stateful replay, taint and provenance labels, indirect-prompt suites. & Capability-scoped tools, permission manifests, transcripts, state snapshots, canary and secret-handling tests, and utility-security tradeoff curves. \\
\addlinespace
\textbf{15} & Space and aerospace mission-environment mismatch \emph{(spacecraft, rovers, aviation, high-altitude and remote autonomy)} & Mission conditions make communication delay, hardware faults, and limited recovery central evaluation requirements~\cite{nasa2024std7009b,basiliskDocs2026,kenneally2020basilisk,biesiadecki2007mer}. & How can mission-specific hazards be represented without overclaiming benign laboratory evidence? \emph{Method:} mission-phase scenario grammars, delayed-command HiL, fault and radiation assumptions, GN\&C traces. & Delay/loss traces, mission-phase coverage, fault assumptions, onboard-autonomy traces, operations constraints, and M\&S credibility review. \\
\end{longtable}
\endgroup

The ranking favors evidence that reveals a claim-invalidating dependency. A campaign with a declared event space and preserved failures can be more informative than a larger campaign reporting only mean success. Uncertainty about omitted conditions and operator interventions belongs in that comparison, alongside sampling uncertainty.

\subsection{Security and Containment Threat Model}
\label{subsec:section5-threat-model}

We represent a threat instance using components that map directly to the apparatus \(S\):
\[
\tau=(\alpha,\pi,B_{\mathrm{viol}},X_{\mathrm{target}},Y_{\mathrm{impact}},V_{\mathrm{ctrl}}).
\]
Here \(\alpha\) states the actor's goal, knowledge, and capability. The path \(\pi\) reaches a violated boundary \(B_{\mathrm{viol}}\) within \(B\), and \(X_{\mathrm{target}}\) identifies the targeted component of \(S\). The consequence \(Y_{\mathrm{impact}}\) can be cyber, physical, or governance-related. The record \(V_{\mathrm{ctrl}}\) identifies the control, its validation artifacts, and the residual risk it leaves in \(R\). Table~\ref{tab:section5-threat-model} uses these same fields, combining path with boundary and target with consequence to keep the table readable.

Two failure modes connect \(\tau\) to the claim-licensing rule.
\emph{Containment breach} permits outward harm. The threat compromises \(B\) or disables \(I\), so execution of \(U\) in \(E\) realizes damage, lateral movement, or leakage among the residual risks recorded in \(R\). T1, T3, T6b, and T9 illustrate these routes.
\emph{Evidentiary corruption} creates false assurance by changing the apparatus or its record:
\[
(E,\mathcal{C},M,A)\mapsto
(\widetilde E,\widetilde{\mathcal{C}},\widetilde M,\widetilde A).
\]
For a fixed claim \(D\) and relevant set \(\mathrm{Rel}(D)\), let \(e(d)\) denote the warranted grade and \(\hat e(d)\) the grade inferred from the corrupted record. False assurance requires inflated support at a relevant dimension and an increase in the minimum:
\[
\exists d\in\mathrm{Rel}(D):\hat e(d)>e(d),\qquad
\min_{d\in\mathrm{Rel}(D)}\hat e(d)>
\min_{d\in\mathrm{Rel}(D)}e(d).
\]
Writing \(\mathcal{L}(D\mid A)\) for the licensing decision from a bundle gives
\[
\mathcal{L}(D\mid\widetilde A)>\mathcal{L}(D\mid A).
\]
Inflating one non-bottleneck dimension alone cannot satisfy this inequality. The attack must conceal or raise every remaining minimum. An altered claim scope or omitted relevant dimension is a separate form of argument corruption that must also be caught by scope review.

Poisoned environment assets (T5), spoofed bridge telemetry (T6a), and unrecorded sensor spoofing during testing (T7) can corrupt what the experiment appears to have exercised. Benchmark gaming (T8), selective reporting (T4), and governance over-claiming (T10) can then change the decision without touching \(U\) in production. For a learned world model, T5 can target weights inside \(E\). For a VLA controller, a compromised model or physical trigger can produce conditional deviations while nominal tests remain satisfactory \cite{agarwal2025cosmos,zhou2025badvla}.

MITRE ATLAS catalogs adversarial tactics and techniques against AI-enabled systems, including attacks beyond the classifier itself \cite{mitreATLAS}. Our distinction is the assurance objective: we trace an attack through a component of \(S\) to outward harm or an inflated \(\mathcal{L}(D)\). We make no claim that apparatus attacks fall outside every ATLAS technique. Security analysis of externally run GPAI evaluations already treats evaluation infrastructure as an attack target \cite{tlaie2025gpaievals}. The extension here includes physical interventions and twin/HiL bridges, together with the evidence dimensions they can invalidate.

A control therefore needs its own evidence. Its configuration and activation record must let a reviewer establish which boundary it defended during the relevant run. Controls that share the compromised sensor, clock, or authority path retain a common failure dependency.

\begingroup
\scriptsize
\setlength{\LTcapwidth}{\textwidth}
\setlength{\tabcolsep}{3pt}
\renewcommand{\arraystretch}{1.10}
\begin{longtable}{>{\raggedright\arraybackslash}p{0.15\textwidth}>{\raggedright\arraybackslash}p{0.19\textwidth}>{\raggedright\arraybackslash}p{0.19\textwidth}>{\raggedright\arraybackslash}p{0.19\textwidth}>{\raggedright\arraybackslash}p{0.195\textwidth}}
\caption{Cyber-physical threat instances aligned with \(\tau\). Each row states the actor \(\alpha\), path \(\pi\) and violated boundary \(B_{\mathrm{viol}}\), targeted component \(X_{\mathrm{target}}\) and impact \(Y_{\mathrm{impact}}\), then the control-validation and residual-risk record \(V_{\mathrm{ctrl}}\). Blue, amber, and red tints identify infrastructure, evidence/governance, and cyber-physical paths in Figure~\ref{fig:threat}.}
\label{tab:section5-threat-model}\\
\toprule
\rowcolor{tabhead}\textbf{Threat class} & \textbf{Actor \(\alpha\)} & \textbf{Path \(\pi\), boundary \(B_{\mathrm{viol}}\)} & \textbf{Target \(X_{\mathrm{target}}\), impact \(Y_{\mathrm{impact}}\)} & \textbf{Control record \(V_{\mathrm{ctrl}}\)} \\
\midrule
\endfirsthead
\multicolumn{5}{@{}l}{\small\textit{Table \ref{tab:section5-threat-model} continued.}}\\
\toprule
\rowcolor{tabhead}\textbf{Threat class} & \textbf{Actor \(\alpha\)} & \textbf{Path \(\pi\), boundary \(B_{\mathrm{viol}}\)} & \textbf{Target \(X_{\mathrm{target}}\), impact \(Y_{\mathrm{impact}}\)} & \textbf{Control record \(V_{\mathrm{ctrl}}\)} \\
\midrule
\endhead
\bottomrule
\endlastfoot
\cellcolor{zonephys}\textbf{T1}\;Misaligned or malicious SUT~\cite{greshake2023indirectprompt,lu2025toolsandbox} & Task-aware actor with task/API access can maximize score, hide unsafe behavior, misuse tools, or escape constraints using task, API, code, or environment access. & Tool APIs, browser, files, shell, simulator API, middleware, robot-control API. SUT-to-sandbox boundary. & \(U\), intervention authority \(I\), and tool/action records in \(A\). misleading score, data exposure, unsafe action. & Capability-scoped authorized red-team tasks, egress tests, tool-call replay. Residual long-horizon policy bypass. \\
\addlinespace
\cellcolor{zonecyber}\textbf{T2}\;External infrastructure adversary~\cite{souppaya2017containers,nist2022ssdf} & Infrastructure-aware actor can steal models or data, disrupt evaluation, or tamper evidence through exposed services, credentials, or supply-chain footholds. & Public API, CI/CD, artifact store, cloud API. Network, identity, orchestration, and artifact-store boundary. & \(U\), monitors \(M\), and artifact store \(A\). exfiltration, denial of evaluation, invalid evidence. & Authorized penetration test, IAM review, network-exposure review, audit-log review. Residual unknown vulnerabilities. \\
\addlinespace
\cellcolor{zonecyber}\textbf{T3}\;Compromised tenant or shared substrate~\cite{garfinkel2003vmi,naghibijouybari2018gpu} & Co-resident actor with substrate access can leak data or move laterally through shared host, GPU, scheduler, storage, or network. & Container/VM/GPU/cloud substrate. Tenant isolation boundary. & Tenant boundary \(B\) and other tenants' \(U\) and \(A\). confidentiality breach and polluted results. & Isolation tests, cloud-configuration audit, side-channel risk review. Residual provider opacity. \\
\addlinespace
\cellcolor{zoneevid}\textbf{T4}\;Insider or artifact manipulator~\cite{raji2020closing,kelly2004gsn} & Insider with campaign knowledge can selectively report, alter logs, leak proprietary data, or overstate a result using legitimate access. & Data store, report repository, evidence package, safety case, regulatory submission. Artifact and governance boundary. & Scenario records and artifacts \(A\). false assurance or privacy breach. & Role separation, immutable logs, hash manifests, independent review. Residual biased interpretation. \\
\addlinespace
\cellcolor{zonecyber}\textbf{T5}\;Compromised plugin, asset, package, or dependency~\cite{ntia2021sbom,nist2022ssdf,biggio2012poisoning} & Supplier with component knowledge can poison simulator behavior, leak data, or gain execution through trusted components. & Simulator plugin, container image, package registry, map, mesh, texture, model, or scenario asset. Supply-chain boundary. & Environment \(E\) including world-model weights, and artifacts \(A\). invalid evidence or code execution. & Signed assets, dependency pinning, SBOM review, sandboxed plugin tests. Residual semantically wrong assets. \\
\addlinespace
\cellcolor{zonephys}\textbf{T6}\;Twin, HiL, fieldbus, or telemetry attacker~\cite{alcaraz2022digitaltwinsecurity,zhang2013timesync,giraldo2018physics} & Bridge-aware actor with channel access can spoof telemetry, replay state, manipulate clocks, or inject commands into a cyber-physical bridge. & MQTT, OPC UA, ROS, fieldbus, PMU/GNSS/PTP, cloud-edge bridge. Operational, timing, and physical boundary. & Twin environment \(E\), measurement \(M\), and intervention \(I\). unsafe plant state or false replay. & Replay-protection tests, signed telemetry, delay injection, plant-consequence traces. Residual bridge changes after test. \\
\addlinespace
\cellcolor{zonephys}\textbf{T7}\;Adversarial perception or sensor attacker~\cite{cao2019lidar,giraldo2018physics,zhou2025badvla} & Sensor-aware actor with physical-channel access can corrupt the perceived world during HiL or field evaluation by spoofing, jamming, or physically perturbing sensor channels. & LiDAR, camera, radar, GNSS, or acoustic channel via emitters or physical adversarial objects. Sensor-to-plant boundary. & Sensing in \(U\) and ground-truth measurement \(M\). wrong perceived state and invalid evaluation evidence. & Sensor-redundancy consistency checks, physical-plausibility monitors, authorized spoofing-injection tests. Residual novel physical channels. \\
\addlinespace
\cellcolor{zoneevid}\textbf{T8}\;Benchmark-gaming actor~\cite{recht2019imagenet,zhu2025abc} & Task-aware submitter can inflate performance or hide failures by exploiting public tasks, seeds, assets, or scoring rules. & Leaderboard, public scenario grammar, task suite, repeated submissions. Benchmark governance boundary. & Executed suite \(\mathcal{C}\subseteq C\) and reported artifacts \(A\). unsupported robustness or safety claim. & Private holdouts, leakage audits, rotating scenarios, independent retest. Residual reproducibility-secrecy tension. \\
\addlinespace
\cellcolor{zoneevid}\textbf{T9}\;Privacy adversary or careless handler~\cite{nist2020privacy,gebru2021datasheets} & Handler with trace access and contextual knowledge can identify people, recover proprietary operations, expose credentials, or leak sensitive field context. & Field-log replay, telemetry store, notebooks, trace releases, reports. Data-governance boundary. & Preserved telemetry and credentials in \(A\), crossing \(B\). privacy harm and legal exposure. & Privacy review, minimization, de-identification test, access audit, secret scan. Residual linkage risk. \\
\addlinespace
\cellcolor{zoneevid}\textbf{T10}\;Governance over-interpretation actor~\cite{euAIAct2024,allen2019regulatorysandboxes,zetzsche2017regulating} & Decision-maker with report access can treat bounded sandbox participation or a narrow experiment as certification or deployment proof. & Regulatory sandbox, safety case, risk register, executive report. Process/evidence interpretation boundary. & Governance evidence in \(A\) and residual-risk record \(R\). unsafe deployment decision. & Claim-to-evidence review, scope statement, residual-risk sign-off. Residual institutional pressure. \\
\end{longtable}
\endgroup

\begin{figure}[htbp]
    \centering
    \includegraphics[width=\linewidth]{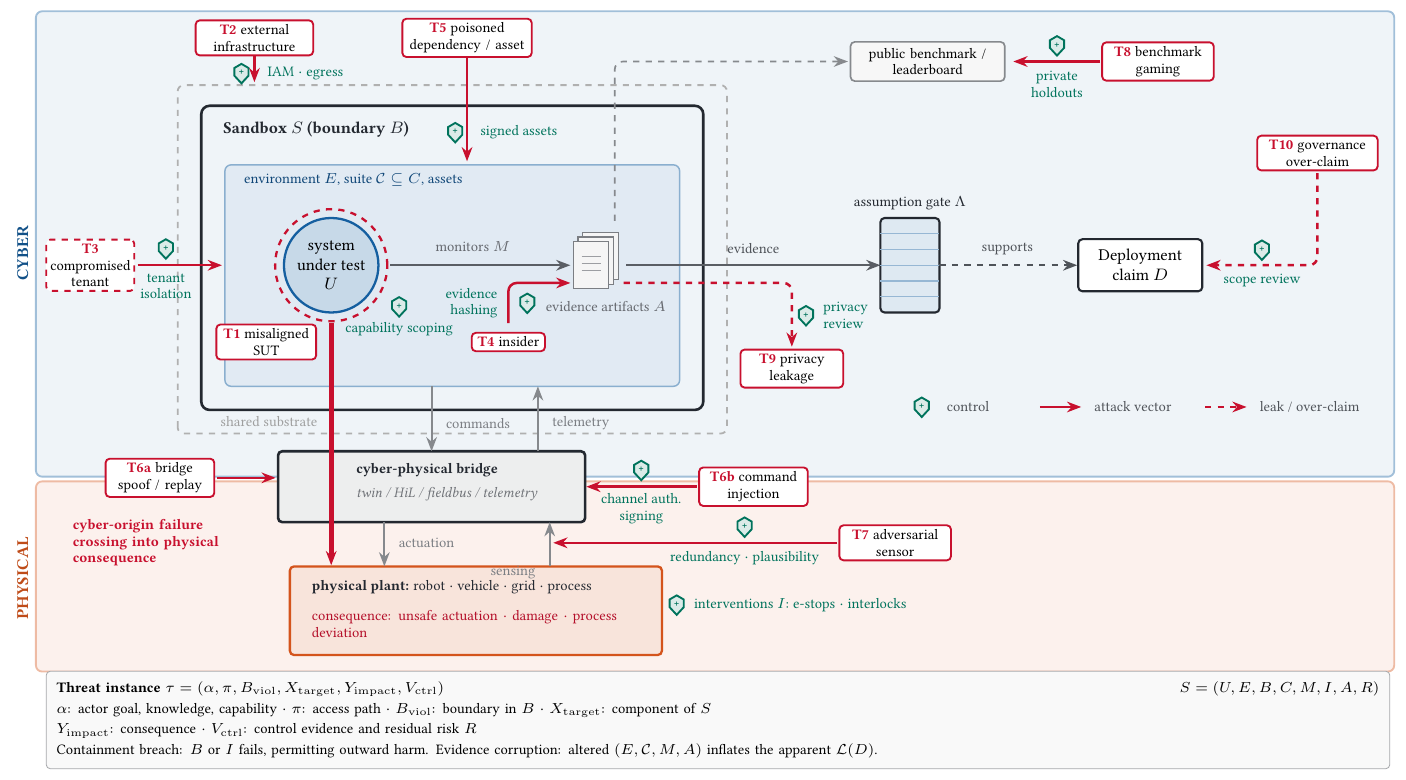}
    \Description{The sandbox apparatus of the earlier anatomy figure redrawn under attack, split into a cyber zone above and a physical zone below. Inside the cyber zone, a shared substrate encloses the sandbox boundary, environment field, system under test, and evidence artifacts, which flow through the assumption gate to the deployment claim. A public benchmark hangs off the artifact store. A cyber-physical bridge straddles the zone border and connects to a physical plant below via actuation and sensing channels. Ten red attack vectors, labeled T1 to T10 after the threat-model table, each strike the asset their table row names: the system under test, the substrate, a neighboring tenant, the artifacts, the simulator assets, the bridge (two vectors, spoof/replay and command injection), the sensing channel, the benchmark, an outward privacy leak, and the deployment claim. Green shields sit on the attacked paths and carry short control names. A bold red vertical descends from the system under test through the bridge into the plant, labeled cyber-origin failure crossing into physical consequence. A bottom strip unpacks one threat instance as the threat instance tau.}
    \caption{Threat surface of an AI sandbox, drawn on the apparatus of Figures~\ref{fig:hero} and~\ref{fig:anatomy}. Red vectors are the threat classes T1 to T10 of Table~\ref{tab:section5-threat-model}, each striking the asset its row names. The bridge class T6 appears as its two manifestations, spoof/replay (T6a) and command injection (T6b), and dashed red marks outward leakage (T9) and over-claim pressure (T10). Green shields are controls from Table~\ref{tab:section5-threat-control}, placed on the paths they guard. The bold red vertical is the central crossing: a cyber-origin failure descending from the system under test through the twin/HiL bridge into physical consequence. The bottom strip defines \(\tau\), aligned with the table fields.}
    \label{fig:threat}
\end{figure}

The same connection can support observation and enable harm. A bidirectional twin bridge may record the plant accurately while permitting an injected command to cross \(B\). Conversely, an isolated experiment may contain every action yet preserve a selectively edited \(A\). Figure~\ref{fig:threat} places these two failure modes on the apparatus.

\subsection{Threats, Controls, and Residual Risk}
\label{subsec:section5-threat-control}

A containment claim requires a record that the relevant control was active during execution. For software isolation, protection principles and container guidance identify permissions and mediation paths that must be checked \cite{saltzer1975protection,wahbe1993sfi,garfinkel2003vmi,souppaya2017containers}. A physical control adds a measurable response deadline and an energy path. Table~\ref{tab:section5-threat-control} states the artifacts needed to test these claims.

\begingroup
\scriptsize
\setlength{\LTcapwidth}{\textwidth}
\setlength{\tabcolsep}{3pt}
\renewcommand{\arraystretch}{1.10}
\begin{longtable}{>{\raggedright\arraybackslash}p{0.185\textwidth}>{\raggedright\arraybackslash}p{0.36\textwidth}>{\raggedright\arraybackslash}p{0.21\textwidth}>{\raggedright\arraybackslash}p{0.175\textwidth}}
\caption{Threat-to-control matrix, keyed to the threat classes T1 to T10 of Table~\ref{tab:section5-threat-model}. Controls reduce risk only when their configuration, activation, monitoring, and review are included in the evidence package. Each row pairs preventive controls with detection and intervention (\spos{Prevent}, \spos{Detect/respond}), the artifacts that evidence them, and the residual risk.}
\label{tab:section5-threat-control}\\
\toprule
\rowcolor{tabhead}\textbf{Threat (classes)} & \textbf{Controls} & \textbf{Evidence artifacts} & \textbf{Residual risk} \\
\midrule
\endfirsthead
\multicolumn{4}{@{}l}{\small\textit{Table \ref{tab:section5-threat-control} continued.}}\\
\toprule
\rowcolor{tabhead}\textbf{Threat (classes)} & \textbf{Controls} & \textbf{Evidence artifacts} & \textbf{Residual risk} \\
\midrule
\endhead
\bottomrule
\endlastfoot
Sandbox escape, container/VM breakout, or host compromise (T1 to T3)~\cite{souppaya2017containers,saltzer1975protection} & \spos{Prevent:} VM/container isolation, seccomp/AppArmor, no privileged containers, patching, least privilege, egress filtering. \spos{Detect/respond:} system-call logs, EDR, egress monitoring. Kill job, revoke tokens, rotate secrets, rebuild image. & Image digest, runtime policy, host hardening record, network diagram, scan report. & Kernel or hypervisor zero-days and misconfiguration can bypass intended isolation. \\
\addlinespace
Cross-tenant or GPU leakage (T3)~\cite{naghibijouybari2018gpu} & \spos{Prevent:} dedicated hardware for sensitive runs, tenant isolation, storage/network segmentation, no secrets in shared memory. \spos{Detect/respond:} side-channel risk review, resource anomaly alerts. Evacuate tenant, invalidate affected outputs. & Tenant map, GPU allocation log, access logs, sensitivity classification. & Hardware side channels and provider internals may be opaque. \\
\addlinespace
Credential, API-key, signing-key, or robot-token exposure (T1, T2)~\cite{nist2022ssdf,greshake2023indirectprompt} & \spos{Prevent:} capability-based access, short-lived tokens, secret vaults, no secrets in logs, prompts, traces, or screenshots. \spos{Detect/respond:} secret scanners and audit logs. Revoke or rotate keys, disable bridge, incident review. & IAM policy, token scope, secret-scan report, key-rotation record. & Legitimate tool access can still leak through outputs or indirect prompt paths. \\
\addlinespace
Scenario, data, model, asset, or dependency poisoning (T5)~\cite{biggio2012poisoning,gebru2021datasheets,ntia2021sbom,nist2022ssdf} & \spos{Prevent:} signed datasets and scenarios, review gates, dependency pinning, SBOMs, asset manifests, sandboxed plugins. \spos{Detect/respond:} distribution-shift and integrity checks, regression tests. Quarantine artifact and rerun affected claims. & Dataset hash, datasheet, scenario manifest, SBOM, lockfile, signature log. & Signed artifacts may still be scientifically invalid or malicious before signing. \\
\addlinespace
Telemetry spoofing, stale twin state, or time manipulation (T6)~\cite{alcaraz2022digitaltwinsecurity,zhang2013timesync,giraldo2018physics} & \spos{Prevent:} source authentication, replay protection, timestamp validation, redundant clocks, one-way gateways where appropriate. \spos{Detect/respond:} drift and clock-skew alarms, cross-sensor consistency, delay-injection tests. Freeze twin authority or switch to safe mode. & Telemetry lineage, bridge configuration, key records, sync logs, delay profile. & Drift, spoofing, and sensor degradation can be hard to distinguish without plant truth. \\
\addlinespace
Command injection through HiL, fieldbus, or twin bridge (T6)~\cite{mathur2016swat,ahmed2017wadi,nguyen2023shiphil} & \spos{Prevent:} network segmentation, allow-listed commands, command signing, safety relays, physical interlocks, safe loads. \spos{Detect/respond:} command audit, process anomaly detection, actuator-state monitor. Emergency stop, relay trip, supervisor override. & Bridge configuration, command schema, relay wiring, interlock test, safe-load record. & Unsafe energy paths or delayed effects may remain outside the interlock. \\
\addlinespace
Adversarial perception or sensor attack (T7)~\cite{eykholt2018physical,cao2019lidar} & \spos{Prevent:} sensor redundancy, physical plausibility checks, constrained test setup, validated robust training where available. \spos{Detect/respond:} cross-sensor residuals, detector alerts, planner-state checks. Slow, stop, degrade mode, human intervention. & Sensor model, calibration record, attack/fault script, physical setup description. & Detection can be attack-specific and environment-specific. \\
\addlinespace
Benchmark gaming and leaderboard overfitting (T8)~\cite{recht2019imagenet,zhu2025abc} & \spos{Prevent:} private or rotating holdouts, seed registries, submission limits, benchmark cards. \spos{Detect/respond:} leakage and contamination audits, public/private performance deltas, independent retests. Invalidate or requalify affected scores. & Benchmark card, holdout registry, leakage-audit report, retest record. & Reproducibility and secrecy remain in tension. Familiarity effects may persist. \\
\addlinespace
Evidence tampering, selective reporting, or governance overclaiming (T4, T10)~\cite{raji2020closing,kelly2004gsn,tabassi2023airmf} & \spos{Prevent:} immutable logs, hash-linked evidence packages, role separation, preregistered protocol, claim-scope review. \spos{Detect/respond:} completeness audit and independent replay. Invalidate report, rerun, qualify claim. & Hash manifest, signed report, protocol, exception log, risk register, residual-risk statement. & Honest but biased interpretation can survive technical integrity checks. \\
\addlinespace
Privacy leakage from field logs or operational telemetry (T9)~\cite{nist2020privacy,gebru2021datasheets} & \spos{Prevent:} minimization, de-identification, access control, retention limits, licensing/consent review, secret scanning. \spos{Detect/respond:} DLP, re-identification checks, secret scans. Remove or segregate data, notify owners, rotate exposed credentials. & Privacy review, data inventory, de-identification record, access log, retention decision. & Linkage attacks and context can defeat de-identification. \\
\addlinespace
Runtime monitor or safety-envelope invalidity (cross-cutting: T1, T6, T7)~\cite{alshiekh2018shielding,ames2017cbf,schierman2020rta} & \spos{Prevent:} independently checked envelope, shield or barrier assumptions, RTA wrapper, fail-safe actuation authority. \spos{Detect/respond:} violation-rate logs, response latency, monitor-health checks. Safe controller, shield, emergency stop. & Monitor specification, proof assumptions, envelope configuration, intervention trace. & A monitor may be correct for the wrong model or delayed beyond usefulness. \\
\end{longtable}
\endgroup

\begin{figure}[htbp]
    \centering
    \includegraphics[width=\linewidth]{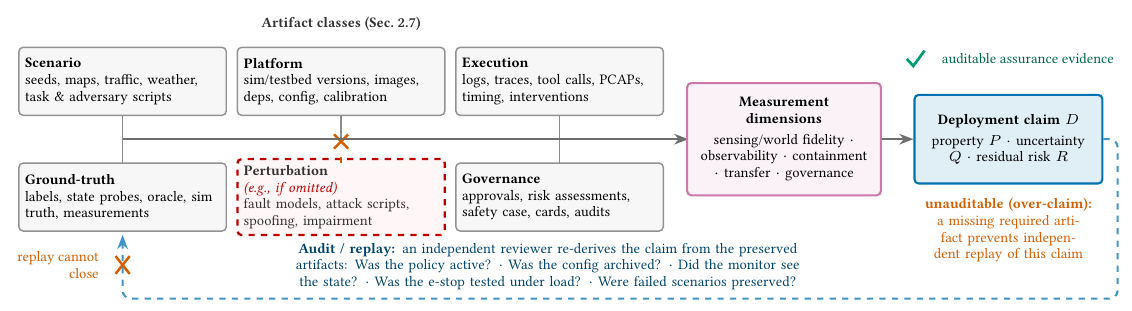}
    \Description{A left-to-right provenance pipeline. On the left, six artifact-class boxes (scenario, platform, execution, ground-truth, perturbation, governance) arranged in a three-column grid join a horizontal evidence trunk through short vertical elbows. The trunk feeds a measurement-dimensions box, which feeds a deployment-claim box. A dashed audit-and-replay loop returns from the claim to the artifacts, carrying a question set: was the policy active, was the config archived, did the monitor see the state, was the e-stop tested under load, were failed scenarios preserved. The perturbation artifact has a dashed red border and reads Perturbation (e.g., if omitted). Its feed arrow is broken with a red cross, the replay loop is marked cannot close, and the claim is labeled unauditable over-claim. When the chain is complete the claim is labeled auditable assurance evidence.}
    \caption{Sandbox capability becomes assurance evidence only when the artifact chain (scenario, platform, execution, ground-truth, perturbation, governance) supports the measured dimensions and the claim, and an independent reviewer can replay it. A missing required perturbation artifact (illustrated by the dashed red box) severs the chain and the replay loop, rendering the downstream claim unauditable.}
    \label{fig:evidence-chain}
\end{figure}

A control record must connect the tested configuration to an observed response. For an emergency stop, this includes the initiating event, measured relay latency, and final actuator state under load. For evidence integrity, it includes the archived manifest and rejected runs. A failed control remains useful scientific evidence when the record identifies the broken assumption. Figure~\ref{fig:evidence-chain} shows how an omitted perturbation record blocks that reconstruction.

\subsection{Research Gaps Specific to Physical AI and AIoT Sandboxes}
\label{subsec:section5-research-gaps}

Physical-control assurance requires composition at the interfaces where failures propagate. A delayed VLA action and a stale twin measurement can each cross from software into contact. A useful campaign aligns the initiating perturbation with plant state and the intervention that bounded it. Without that alignment, a reviewer cannot decide whether the policy recovered, a supervisor rescued it, or the apparatus suppressed the failure. The challenge table specifies the corresponding research questions and validation artifacts.

Progress should be demonstrated by an independently reviewable artifact. A coverage method can be tested against withheld field failures. A drift detector can be tested on telemetry with known source changes. A runtime monitor can be tested for both missed interventions and disruptive false interventions under degraded sensing. Negative results identify the operating assumptions that the next claim must retain.

Standards can specify which records make these results usable in an assurance decision. The next section maps those obligations to the measurement framework and identifies where a portable artifact schema remains missing.

\section{Standardization, Regulatory Alignment, and Next-Generation Sandboxes}
\label{sec:standards}

\subsection{Why Standards Belong in the Sandbox Evidence Chain}
\label{subsec:section6-rationale}

Standards bodies and regulators define the review obligations under which sandbox evidence is used. Their instruments determine which records an assessor expects and how the records enter a decision. The experiment must still establish the claim-relative measurements.

AI evaluation research supplies the methodological basis for this mapping. Evaluation surveys and multi-construct frameworks separate the measured task from the wider conclusion \cite{chang2024survey,liang2023helm}. Measurement theory and benchmark-validity studies examine whether a metric represents its stated construct \cite{jacobs2021measurement,raji2021everything}. Ecological validity concerns correspondence to the deployed interface \cite{devries2020ecological}. Extreme-risk evaluations and reporting standards add requirements for interpreting bounded tests and reproducing their results \cite{shevlane2023extreme,phuong2024frontier,dodge2019showyourwork}. Our framework applies these obligations to physical channels and their evidence chain.
 
The AI RMF and NIST TEVV organize lifecycle evaluation and measurement \cite{tabassi2023airmf,nistTEVV2026}. Dioptra and ARIA give concrete institutional examples of evaluation infrastructure and staged testing \cite{nistDioptra2024,nistARIA2024}. ISO/IEC~23894 guides risk management, while ISO/IEC~42001 defines an auditable organizational management system \cite{iso23894,iso42001}. The EU AI Act adds supervised sandbox plans and exit records \cite{euAIAct2024}. Each addresses a different part of the decision process.

Process-Technical Separation matters when these records disagree with the experiment. An assurance argument can conceal weak evidence \cite{leveson2011safetycase,rushby2015safetycase}. An evaluation file can omit relevant operating conditions \cite{tang2023adstesting}, and an organizational audit need not cover a particular product's untested behavior \cite{raji2020closing}. The reverse problem also matters: a well-run experiment can be unusable to an assessor if its configuration and provenance are missing \cite{breck2017mltest,gebru2021datasheets}.

We separate process, engineering, and interchange roles. Process instruments organize review and accountability \cite{iso42001,tabassi2023airmf}. Engineering standards constrain domain-specific safety and security arguments \cite{iec61508,iso26262,iso21448,ul4600,isosae21434,iec62443,fagan2020iotbaseline,etsi3036452024}. Interchange specifications move scenarios and models between tools \cite{asamOpenScenarioDsl2026,fmiStandard3,ssp2019spec}. Regulatory instruments determine how the resulting evidence contributes to conformity assessment or deployment authorisation \cite{euAIAct2024,avact2024}. Structured assurance arguments record the dependencies between these roles \cite{kelly2004gsn,denney2018acas,bloomfield2010safetycase}.

\subsection{Mapping Sandbox Design to Standards and Regulation}
\label{subsec:section6-standards-map}
 
Table~\ref{tab:section6-standards-map} maps each instrument to the apparatus choices and artifacts it can require or guide. Requirements depend on the applicable clause and legal route. The table gives an assurance interpretation. Applicability must be checked for the named system and decision.

\begingroup
\scriptsize
\setlength{\LTcapwidth}{\textwidth}
\setlength{\tabcolsep}{3pt}
\renewcommand{\arraystretch}{1.10}
\begin{longtable}{>{\raggedright\arraybackslash}p{0.215\textwidth}>{\raggedright\arraybackslash}p{0.30\textwidth}>{\raggedright\arraybackslash}p{0.255\textwidth}>{\raggedright\arraybackslash}p{0.165\textwidth}}
\caption{Mapping standards and regulatory instruments to AI-sandbox design choices and evidence obligations. Rows are grouped by role (process, engineering, interchange, regulatory) in the shaded bands that mirror Figure~\ref{fig:standards}. Short scope descriptors appear in italics under each instrument. Each row identifies what a sandbox must provide for its evidence to be usable under that instrument.}\label{tab:section6-standards-map}\\
\toprule
\rowcolor{tabhead}\textbf{Instrument and scope} & \textbf{Sandbox design implication} & \textbf{Evidence artifacts expected} & \textbf{Strongly relevant domains} \\
\midrule
\endfirsthead
\multicolumn{4}{@{}l}{\small\textit{Table \ref{tab:section6-standards-map} continued.}}\\
\toprule
\rowcolor{tabhead}\textbf{Instrument and scope} & \textbf{Sandbox design implication} & \textbf{Evidence artifacts expected} & \textbf{Strongly relevant domains} \\
\midrule
\endhead
\bottomrule
\endlastfoot
\tband{4}{Process standards}
\textbf{NIST AI RMF 1.0}~\cite{tabassi2023airmf,nistTEVV2026}\newline\textit{lifecycle AI risk management} & Sandbox must support Map, Measure, Manage, and Govern activities across pre-deployment evaluation, monitoring, and incident response. & Risk register, measurement plan, evaluation reports, monitoring outputs, governance roles. & All AI domains. Common reference in U.S. federal AI guidance. \\
\addlinespace
\textbf{ISO/IEC 42001:2023}~\cite{iso42001}\newline\textit{AI management system} & Sandbox scope and outputs must be preserved as auditable management-system records. & AI policy, scope statement, sandbox procedure, internal-audit and management-review records. & Organizations seeking certifiable AI governance. \\
\addlinespace
\textbf{ISO/IEC 23894:2023}~\cite{iso23894}\newline\textit{AI risk-management guidance} & Campaign records must link measured outcomes to assumptions and risk-treatment decisions. & Risk identification log, treatment plan, residual-risk statement, scope justification. & Any AI system requiring documented risk management. \\
\addlinespace
\textbf{NIST AI TEVV}~\cite{nistTEVV2026,raji2020closing}\newline\textit{measurement science for AI test, evaluation, verification, and validation} & Sandbox metrics must be tied to defined constructs with explicit validity, uncertainty, and use context. & Metric definitions, validity analysis, dataset and scenario provenance, evaluator-independence notes. & High-stakes evaluation, benchmarks, regulatory testing. \\
\tband{4}{Engineering standards}
\textbf{IEC 61508}~\cite{iec61508}\newline\textit{functional safety for electrical, electronic, and programmable systems} & Sandbox must support hazard analysis, Safety Integrity Level allocation, V\&V, and proof-test design. & Hazard log, Safety Integrity Level allocation, V\&V plan, proof-test records, FMEDA. & Industrial CPS, infrastructure, generic safety-critical systems. \\
\addlinespace
\textbf{ISO 26262}~\cite{iso26262}\newline\textit{functional safety for road vehicles} & Sandbox must trace from item definition through ASIL allocation, V\&V, and confirmation measures. & ASIL allocation, safety case, test specifications, traceability matrix. & Automotive electrical and electronic systems. \\
\addlinespace
\textbf{ISO 21448 (SOTIF)}~\cite{iso21448,koopman2017avsafety}\newline\textit{safety of the intended functionality} & Sandbox must test triggering conditions for hazards from sensing and algorithmic limitations, including unknown-unsafe scenarios. & Triggering-condition catalog, scenario coverage, residual-risk argument. & ADAS/AD, robotics with learned perception. \\
\addlinespace
\textbf{ANSI/UL 4600}~\cite{ul4600,kelly2004gsn,ratiu2024spi}\newline\textit{safety case for autonomous products} & Sandbox must produce evidence that maps to a claims-and-arguments structure with explicit assumptions. & GSN-style safety case, hazard analysis, scenario coverage, runtime-monitor records. & Autonomous vehicles, mobile robots, autonomous machinery. \\
\addlinespace
\textbf{ISO/SAE 21434}~\cite{isosae21434}\newline\textit{road-vehicle cybersecurity} & Sandbox must include threat assessment, attack and fault injection, and cybersecurity validation across the vehicle lifecycle. & Threat analysis and risk assessment (TARA), cybersecurity case, attack and test logs, post-production monitoring. & Automotive cybersecurity. \\
\addlinespace
\textbf{IEC 62443}~\cite{iec62443}\newline\textit{industrial cybersecurity} & Sandbox must reflect the zone-and-conduit architecture, security-level allocation, and component requirements. & Zone and conduit diagram, security-level rationale, component requirements, test reports. & ICS, OT, and AIoT integrated with industrial processes. \\
\addlinespace
\textbf{NISTIR 8259A. ETSI EN 303 645}~\cite{fagan2020iotbaseline,etsi3036452024}\newline\textit{IoT and consumer-IoT baseline security} & AIoT sandboxes must test baseline device capabilities such as identification, secure update, secure defaults, and data protection. & Device-capability inventory, update logs, secret-handling tests, privacy review. & AIoT devices and edge-AI products. \\
\tband{4}{Interchange standards}
\textbf{ASAM OpenSCENARIO. OpenDRIVE. OpenODD}~\cite{asamOpenScenarioDsl2026,scholtes2021odd}\newline\textit{scenario, map, and ODD interchange} & Sandbox should encode scenarios and operating envelopes in portable form with explicit semantic notes. SAE J3016 automation levels fix the driver's role and the operational design domain the scenarios must cover~\cite{saej3016}. & Scenario files, map files, ODD specification, conversion logs, semantic-difference notes. & ADAS/AD, mobile robotics, large-scale scenario campaigns. \\
\addlinespace
\textbf{FMI / SSP}~\cite{fmiStandard3,ssp2019spec,palmintier2017helics}\newline\textit{model exchange and co-simulation} & Sandbox should treat coupled models as versioned FMUs or SSP components with declared interfaces, units, and step semantics. & FMU/SSP files, interface contracts, unit and solver metadata, synchronization logs. & CPS, smart-grid, infrastructure co-simulation. \\
\addlinespace
\textbf{SBOM and SSDF}~\cite{ntia2021sbom,nist2022ssdf}\newline\textit{software supply-chain transparency} & Sandbox artifacts must include manifests for software, models, and assets used in each experiment. & SBOM, signature records, dependency lockfiles, build-provenance attestations. & All sandbox stacks. \\
\tband{4}{Regulatory instruments and programs}
\textbf{EU AI Act}~\cite{euAIAct2024,allen2019regulatorysandboxes,zetzsche2017regulating}\newline\textit{EU AI regulation, including regulatory sandboxes} & Sandbox plans must define scope, data use, supervision, risk controls, and exit. Applicable high-risk systems require conformity assessment under Article~43. Article~40 links cited harmonized standards to presumption of conformity for covered requirements. & Sandbox plan, technical documentation, conformity-assessment records, exit report, post-market plan. & EU-deployed AI systems, especially high-risk. \\
\addlinespace
\textbf{UK Automated Vehicles Act 2024}~\cite{avact2024}\newline\textit{authorisation and liability for self-driving vehicles} & Sandbox outputs must be traceable and legally defensible as evidence for authorisation decisions. & Authorisation dossier, ODD declaration, statement of safety principles, post-market data. & UK road vehicles. \\
\addlinespace
\textbf{Program-level instruments: UK AISI sandboxing guidance. MITRE Federal AI Sandbox. HSE AI guidance}~\cite{ukaisandbox2024,mitreFederalAISandbox2026,hseAI2024}\newline\textit{supervised evaluation programs and workplace-safety duties} & Sandbox should support government-supervised red-team protocols for dangerous capabilities and systemic risks, controlled compute with governance and transition pathways from prototype to mission deployment, and evidence that foreseeable workplace risks are identified and controlled. & Evaluation protocols, capability reports, mitigation plans, access controls, environment baselines, transition records, workplace risk assessments, incident reporting. & Frontier AI systems, U.S. federal AI use cases, UK workplaces deploying AI-enabled machinery. \\
\end{longtable}
\endgroup

The common requirement is traceability from the intended use to the reviewed evidence \cite{tabassi2023airmf,iso23894,iso42001,nistTEVV2026,euAIAct2024,ul4600}. The necessary technical measurements then depend on the domain.

For road vehicles, the safety argument must address functional faults and limitations of intended functionality \cite{iso26262,iso21448}. Vehicle and industrial cybersecurity add credible attack paths and their operational effects \cite{isosae21434,iec62443}. IoT guidance adds device and update-security obligations \cite{fagan2020iotbaseline,etsi3036452024}. These obligations constrain \(\mathrm{Rel}(D)\) through the specific claim argument.

Existing instruments leave methodological work for learned control. ISO~26262 and SOTIF address complementary sources of vehicle hazards \cite{iso26262,iso21448}. IEC~62443 provides industrial cybersecurity structure but requires an application-specific account of learned-component threats \cite{iec62443}. The EU AI Act combines national sandbox implementation with coordination provisions \cite{euAIAct2024}. Across the instruments mapped here, the remaining gap is a dedicated portable schema for paired sim-to-real evidence. General model-validation and documentation obligations already provide partial anchors.
The mapping therefore identifies both an applicable obligation and the measurement method still needed to discharge it.
 
\subsection{From Measurement Dimensions to Standardized Evidence Artifacts}
\label{subsec:section6-evidence-mapping}
 
Table~\ref{tab:section6-dimension-to-artifact} connects each dimension to the artifact a reviewer can inspect.

An artifact records evidence for a dimension. An audit log, for example, supports observability only if it records the state and timing needed by the claim. The review question in each row tests that connection.

\begingroup
\footnotesize
\setlength{\LTcapwidth}{\textwidth}
\setlength{\tabcolsep}{3pt}
\renewcommand{\arraystretch}{1.10}
\begin{longtable}{>{\raggedright\arraybackslash}p{0.135\textwidth}>{\raggedright\arraybackslash}p{0.295\textwidth}>{\raggedright\arraybackslash}p{0.50\textwidth}}
\caption{Mapping between sandbox measurement dimensions and standardized evidence artifacts. Each row names the instruments that anchor the dimension (full citations in Table~\ref{tab:section6-standards-map}), the standardized artifact family a sandbox should generate (italics), and the review question those artifacts must answer. The shaded bands repeat the dimension groups of Table~\ref{tab:section4-measurement-framework}.}\label{tab:section6-dimension-to-artifact}\\
\toprule
\rowcolor{tabhead}\textbf{Dimension} & \textbf{Anchoring instruments} & \textbf{Standardized artifact family and review question} \\
\midrule
\endfirsthead
\multicolumn{3}{@{}l}{\small\textit{Table \ref{tab:section6-dimension-to-artifact} continued.}}\\
\toprule
\rowcolor{tabhead}\textbf{Dimension} & \textbf{Anchoring instruments} & \textbf{Standardized artifact family and review question} \\
\midrule
\endhead
\bottomrule
\endlastfoot
\tband{3}{Core evidence properties}
Sensing and world fidelity & ASME VV40, NASA-STD-7009B, ISO~26262, SOTIF \cite{asme2018vv40,nasa2024std7009b,iso26262,iso21448} & \textit{Artifacts: sensor and scene model, calibration report, credibility record \cite{oberkampf2010verification}.} Do observed sensor returns and surrounding-agent behavior match the claim's target conditions, with stated uncertainty? \\
Controllability & OpenSCENARIO/OpenDRIVE/ OpenODD. UL~4600. SOTIF \cite{asamOpenScenarioDsl2026,ul4600,iso21448} & \textit{Artifacts: scenario catalog, ODD declaration, parameter-sweep specification~\cite{scholtes2021odd}.} Which scenarios were exercised, and what was deliberately excluded? \\
Observability & AI RMF. ISO/IEC~42001. UL~4600. HSE guidance \cite{tabassi2023airmf,iso42001,ul4600,hseAI2024} & \textit{Artifacts: telemetry schema, audit log, intervention log~\cite{breck2017mltest,raji2020closing}.} Which states were monitored, and can a reviewer replay a failure? \\
Containment & IEC~62443. ISO/SAE~21434. SBOM/SSDF guidance \cite{iec62443,isosae21434,nist2022ssdf} & \textit{Artifacts: architecture diagram, network and IAM policy, interlock-test record~\cite{souppaya2017containers,saltzer1975protection}.} Which boundaries existed, and were they tested under load? \\
Reproducibility & SSDF. SBOM minimum elements. ISO/IEC~42001 \cite{nist2022ssdf,ntia2021sbom,iso42001} & \textit{Artifacts: SBOM, dependency lockfile, asset hash manifest, seed registry~\cite{gebru2021datasheets}.} Can an independent reviewer rerun the experiment from the artifacts? \\
\tband{3}{Cyber-physical realism and integration}
Scenario portability & ASAM OpenSCENARIO. FMI/SSP \cite{asamOpenScenarioDsl2026,fmiStandard3,ssp2019spec} & \textit{Artifacts: OpenSCENARIO/OpenDRIVE bundles, FMU and SSP packages.} Can a scenario or model move across tools without silently changing semantics? \\
Timing fidelity & IEC~61508. ISO~26262. FMI for clocked systems \cite{iec61508,iso26262,fmiStandard3} & \textit{Artifacts: real-time trace, scheduler configuration, clock-source declaration~\cite{lee2008cps,rajkumar2010cps}.} Does the sandbox reproduce the deployment's timing class? \\
Actuator and plant realism & ISO~26262. SOTIF. IEC~61508 for safety-related actuation \cite{iso26262,iso21448,iec61508} & \textit{Artifacts: actuator-limit declaration, plant and contact model, actuator calibration data~\cite{shah2017airsim,muratore2022randomized}.} Are the channels carrying the safety or security property modeled and bounded? \\
Network realism & IEC~62443. ISO/SAE~21434. ETSI EN~303~645 \cite{iec62443,isosae21434,etsi3036452024} & \textit{Artifacts: topology and impairment profile, PCAP archive, protocol-version log~\cite{yamin2020cyberranges,giraldo2018physics}.} Are the relevant network conditions and adversary positions represented? \\
HiL/SiL integration & IEC~61508. ISO~26262. FMI/SSP. UL~4600 \cite{iec61508,iso26262,fmiStandard3,ssp2019spec,ul4600} & \textit{Artifacts: interface-control document, SiL build record, HiL wiring and I/O trace, firmware inventory~\cite{li2022hilbuilding,nguyen2023shiphil}.} Which production interfaces were exercised, and which were still simulated or stubbed? \\
Attack and fault injection & ISO/SAE~21434. IEC~62443. UL~4600. MITRE ATLAS \cite{isosae21434,iec62443,ul4600,mitreATLAS} & \textit{Artifacts: TARA, attack scripts, fault library, consequence trace~\cite{greshake2023indirectprompt}.} Was a credible attacker exercised, and what was the physical or operational consequence? \\
Sim-to-real transfer & NASA-STD-7009B. ASME VV40. UL~4600 \cite{nasa2024std7009b,asme2018vv40,ul4600} & \textit{Artifacts: paired sim/HiL/field traces, calibration deltas, uncertainty intervals~\cite{muratore2022randomized,elmquist2025simtoreal}.} What is the transfer gap, and how does it qualify the deployment claim? \\
\tband{3}{Accountability}
Scalability & AI RMF. Benchmark and TEVV reporting practice \cite{tabassi2023airmf,nistTEVV2026,dodge2019showyourwork} & \textit{Artifacts: run manifest, scheduler log, throughput and cost record, statistical-power statement.} Were enough runs completed for the claim, and did scaling change the scenario distribution or failure modes? \\
Openness and auditability & ISO/IEC~42001. SBOM/SSDF. EU AI Act documentation duties \cite{iso42001,ntia2021sbom,nist2022ssdf,euAIAct2024} & \textit{Artifacts: artifact index, license and provenance record, hash manifest, access log.} Can an authorized reviewer inspect the materials needed to reproduce the claim-to-evidence chain? \\
Governance and audit & AI RMF. ISO/IEC~23894 and~42001. EU AI Act. UL~4600 \cite{tabassi2023airmf,iso23894,iso42001,euAIAct2024,ul4600} & \textit{Artifacts: risk register, GSN-style safety case, management-system records, regulator file~\cite{kelly2004gsn,denney2018acas}.} Can a reviewer trace from claim to evidence to residual-risk decision? \\
\end{longtable}
\endgroup
 
\begin{figure}[htbp]
    \centering
    \includegraphics[width=0.95\linewidth]{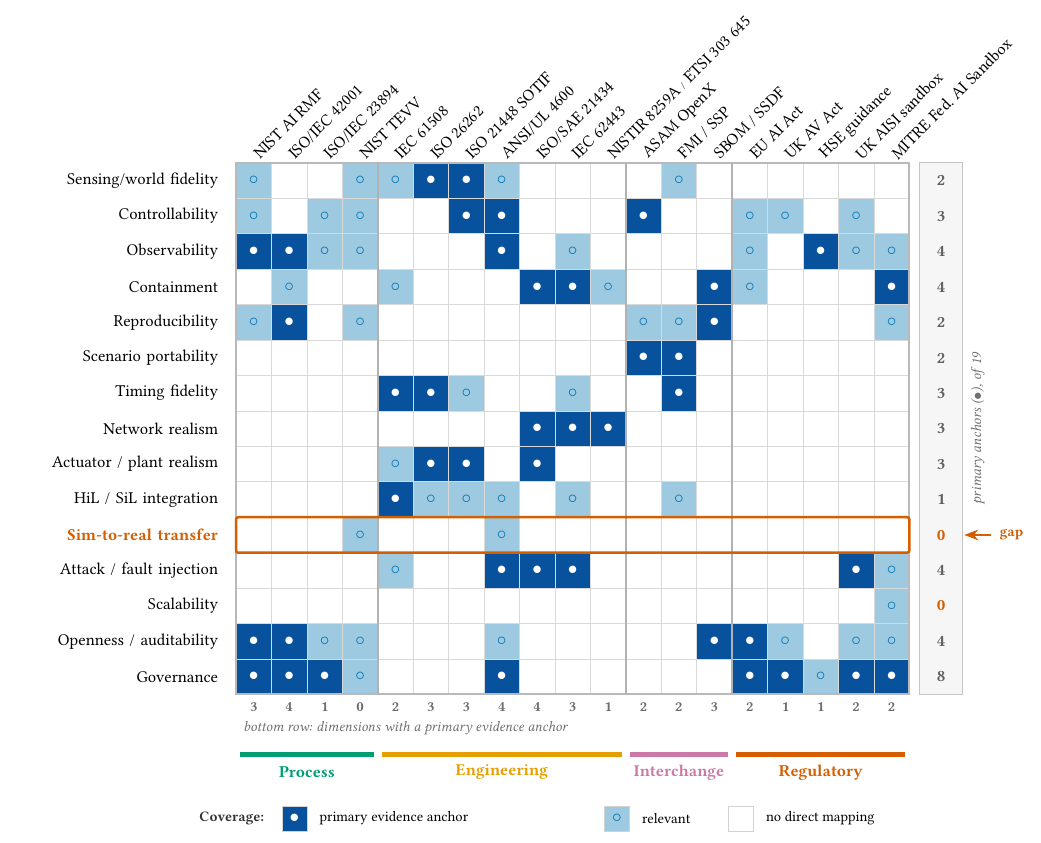}
    \Description{A coverage matrix. Rows are the measurement dimensions. Columns are standards and regulatory instruments grouped by role into process, engineering, interchange, and regulatory bands. Dark cells mark dimensions an instrument makes non-negotiable. Light cells with a dot mark relevant coverage. Empty cells mark no coverage. The sim-to-real transfer row is empty across all instruments and is outlined and labeled as the gap. Scalability is likewise unanchored. Right-margin and bottom-row counts summarize coverage per dimension and per instrument.}
    \caption{Evidence anchors in the mapped standards and regulatory instruments. Dark cells identify primary anchors assigned by this mapping, light cells identify related guidance, and empty cells indicate no direct mapping in this comparison. Transfer has related TEVV and safety-case guidance but no primary anchor for a dedicated sim-to-real evidence package. The cells express the authors' analytical mapping. A conformity determination requires clause-level assessment.}
    \label{fig:standards}
\end{figure}

Figure~\ref{fig:standards} distinguishes primary anchors from related guidance. Its transfer row contains related TEVV and UL~4600 coverage. The missing element is the dedicated schema described above. At the campaign level, a telemetry log without a schema can block review \cite{breck2017mltest,raji2020closing}, and an isolation test without a zone-and-conduit model can leave an IEC~62443 argument incomplete \cite{iec62443}. A replayable scenario package addresses a different obligation \cite{asamOpenScenarioDsl2026}.

Preserving the applicable artifact lets an assessor check whether the experiment supports the proposed decision. Certification, conformity assessment, and authorisation remain distinct uses of that evidence.
A transfer schema would need to record the paired conditions and calibration procedure as well as their observed differences \cite{tobin2017domainrandomization,peng2018simtoreal,muratore2022randomized,elmquist2025simtoreal}. This gives the next-generation agenda a concrete target.

\subsection{Next-Generation Sandbox Directions}
\label{subsec:section6-next-gen}
 
The directions below connect open challenges to a reviewable artifact. Evaluation infrastructure is also an open problem in technical AI governance \cite{reuel2024openproblems}. Here the test of progress is whether the proposed artifact can change or qualify a deployment claim.

\paragraph{Adaptive and coverage-guided scenario generation.}
Scenario generation should report the declared event space and which regions remain untested.
Coverage-guided fuzzing and falsification provide search methods for this task \cite{li2020avfuzzer,tian2018deeptest,annpureddy2011staliro,donze2010breach}. Driving-scenario research adds hazard and actor models \cite{ding2023survey,tang2023adstesting,kaur2021simulators}. Rare-event analysis constrains the confidence that finite campaigns can support \cite{kalra2016driving,koopman2017avsafety}. A useful artifact pairs a hazard-linked coverage map with withheld scenarios and reports where coverage failed to predict their outcomes.

\paragraph{Continuous assurance pipelines.}
Field change can invalidate the assumptions that licensed release.
Lifecycle monitoring in the AI RMF, ISO/IEC~42001, and EU AI Act motivates a deployment-to-sandbox feedback loop \cite{tabassi2023airmf,iso42001,euAIAct2024}. A twin can preserve changed behavior for replay, while drift methods identify when the original distribution no longer fits \cite{barricelli2019digitaltwin,jones2020digitaltwin,gama2014conceptdrift,lu2019driftsurvey}. The artifact should record the replay trigger, integrity checks, and the claims retained or withdrawn after re-evaluation.

\paragraph{Interoperable evidence and scenario portability.}
Scenario and model interchange supplies part of the portability mechanism \cite{asamOpenScenarioDsl2026,fmiStandard3,ssp2019spec,scholtes2021odd}. The remaining task is to exchange the full evidence chain, with a manifest linking each scenario to execution records and the resulting claim. SACM and assurance-case tooling provide structures for these dependencies \cite{omgSACM2020,denney2018acas,wei2019assurance}. Demonstrating replay in independently developed review environments would test whether the schema preserves meaning.

\paragraph{Cyber-physical red teaming with consequence metrics.}
A cyber-physical red-team artifact must link the perturbation to its consequence. Existing range and AI threat frameworks supply attack descriptions \cite{yamin2020cyberranges,mitreATLAS}. CPS security studies identify the process state and intervention response needed to interpret them \cite{giraldo2018physics,alcaraz2022digitaltwinsecurity,cardenas2008secure}. Water-system testbeds and SCEPTRE show settings in which such traces can be collected \cite{mathur2016swat,ahmed2017wadi,hahn2021sceptre}. The deliverable is a contained campaign with an explicit attacker model and time-aligned operational effects.

\paragraph{Regulatory-technical interoperability.}
Article~57 of the EU AI Act requires national AI regulatory sandboxes across Member States and permits joint provision. National implementation can create divergent technical evidence expectations and incentives for cross-border \emph{sandbox arbitrage}, in which applicants seek the least demanding interpretation. This prospective governance risk needs comparative evidence from national practice. Article~57's coordination provisions and Article~58's common arrangements seek consistent practice \cite{euAIAct2024}.
A portable minimum evidence-package schema would let authorities compare the same campaign without changing its technical claim. It should preserve scope and target context, link measurements to the executed suite, and carry the residual-risk decision. EU conformity assessment, UK AV Act authorisation, and workplace-safety review would still apply their own legal tests \cite{euAIAct2024,avact2024,hseAI2024,ukaisandbox2024}. Regulatory-sandbox scholarship supplies design principles for comparing these processes \cite{ranchordas2021sandboxes,allen2019regulatorysandboxes,zetzsche2017regulating}. A multi-instrument mapping of one preserved campaign would test the schema's portability.

\paragraph{Shared and federated sandbox infrastructure.}
Shared evaluation infrastructure introduces tenant boundaries into the evidence chain \cite{souppaya2017containers,naghibijouybari2018gpu}. Federated evaluation could let organizations contribute measurements while preserving their local data and model boundaries. Existing shared infrastructure and secure-aggregation methods supply partial mechanisms \cite{mitreFederalAISandbox2026,kairouz2021federated,bonawitz2017secureagg}. A valid demonstration must show both confidentiality under the stated threat model and consistency between the joint result and preserved local evidence.

\begin{figure}[htbp]
\centering
\includegraphics[width=0.72\linewidth]{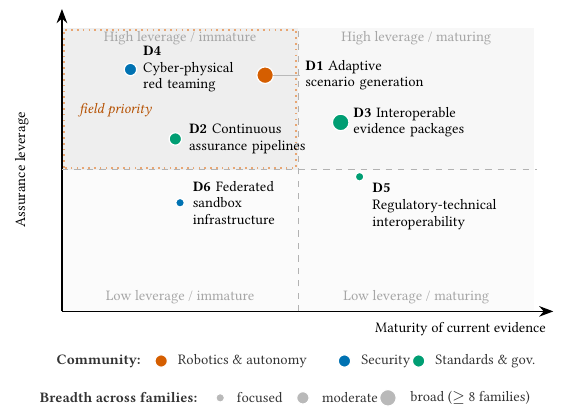}
\Description{A two-axis roadmap for next-generation AI sandboxes. The horizontal axis shows maturity of current evidence, and the vertical axis shows assurance leverage. Six research directions are placed across the four quadrants.}
\caption{Research-and-engineering roadmap for next-generation AI sandboxes. Each direction is scored on the maturity of current evidence (does a reviewable artifact exist today) and assurance leverage (how much a credible artifact would change a deployment claim). Color denotes the primary research community and dot size denotes breadth across sandbox families. Positions are indicative and domain-dependent.}
\label{fig:section6-roadmap}
\end{figure}

Figure~\ref{fig:section6-roadmap} gives a qualitative research-planning view of these directions. Position expresses the authors' assessment of current artifact maturity and potential effect on an assurance decision. These domain-dependent positions have no empirical score interpretation.
 
\subsection{Synthesis: What a Standards-Aligned Sandbox Looks Like}
\label{subsec:section6-synthesis}
 
A standards-aligned sandbox starts with the deployment decision and identifies its necessary evidence. It preserves the measurements and control records that the applicable instruments require, linking them through a reviewable assurance argument \cite{kelly2004gsn,denney2018acas,bloomfield2010safetycase}. The relevant domain standards constrain that argument \cite{iso26262,iso21448,iec62443,isosae21434,fagan2020iotbaseline}, while scenario and provenance specifications make its records reusable \cite{asamOpenScenarioDsl2026,fmiStandard3,ssp2019spec,ntia2021sbom,gebru2021datasheets}. Tested boundaries protect both the experiment and its artifacts \cite{souppaya2017containers,saltzer1975protection,alcaraz2022digitaltwinsecurity}. A risk or regulatory record then states which decision those artifacts support \cite{tabassi2023airmf,iso23894,iso42001,euAIAct2024,ul4600}.
 
The unresolved task is to preserve these dependencies across tool and organizational boundaries. The licensing cap identifies which missing support blocks a claim. The evidence package gives the reviewer a way to check that judgment.

\section{Conclusion}
\label{sec:conclusion}

A sandbox defines the conditions under which an AI-system claim can be tested. The apparatus tuple separates the tested system from its environment and evidence, while the taxonomy distinguishes the roles of the tools around it. For physical control, a claim can fail at a timing or contact interface even when task success remains high. The fifteen dimensions make those necessary channels explicit, answering RQ1 through RQ3.

The weakest-link rule limits inference from the resulting profile. The four cases identify different bottlenecks: field transfer in driving, target-attack representation in industrial control, real-account containment for tool use, and contact/timing evidence for VLA manipulation. A runtime safety envelope can change the claim structure, but it must establish its own sensing and intervention assumptions. The threat model answers RQ4 by tracking both outward harm and evidence corruption through the same apparatus. A forged record can license a dangerous decision without any attack on the deployed controller.

Standards alignment answers RQ5 by specifying how these measurements enter review. Conformity assessment, a structured safety case, and third-party certification use evidence for different decisions. The immediate research target is not simply a richer simulated world, but a portable record that exposes each claim's necessary assumptions and survives independent replay. An assurance sandbox earns its deployment relevance when a reviewer can trace the conclusion to the experiment and identify exactly where that conclusion stops.

\bibliographystyle{ACM-Reference-Format}
\begingroup
\sloppy
\bibliography{references}


\begin{thebibliography}{169}


\ifx \showCODEN    \undefined \def \showCODEN     #1{\unskip}     \fi
\ifx \showISBNx    \undefined \def \showISBNx     #1{\unskip}     \fi
\ifx \showISBNxiii \undefined \def \showISBNxiii  #1{\unskip}     \fi
\ifx \showISSN     \undefined \def \showISSN      #1{\unskip}     \fi
\ifx \showLCCN     \undefined \def \showLCCN      #1{\unskip}     \fi
\ifx \shownote     \undefined \def \shownote      #1{#1}          \fi
\ifx \showarticletitle \undefined \def \showarticletitle #1{#1}   \fi
\ifx \showURL      \undefined \def \showURL       {\relax}        \fi
\providecommand\bibfield[2]{#2}
\providecommand\bibinfo[2]{#2}
\providecommand\natexlab[1]{#1}
\providecommand\showeprint[2][]{arXiv:#2}

\bibitem[Agarwal et~al\mbox{.}(2025)]%
        {agarwal2025cosmos}
\bibfield{author}{\bibinfo{person}{Niket Agarwal}, \bibinfo{person}{Arslan
  Ali}, \bibinfo{person}{Maciej Bala}, \bibinfo{person}{Yogesh Balaji},
  \bibinfo{person}{Erik Barker}, \bibinfo{person}{Tiffany Cai},
  \bibinfo{person}{Prithvijit Chattopadhyay}, \bibinfo{person}{Yongxin Chen},
  \bibinfo{person}{Yin Cui}, \bibinfo{person}{Yifan Ding}, {et~al\mbox{.}}}
  \bibinfo{year}{2025}\natexlab{}.
\newblock \showarticletitle{Cosmos World Foundation Model Platform for Physical
  {AI}}.
\newblock \bibinfo{journal}{\emph{arXiv preprint arXiv:2501.03575}}
  (\bibinfo{year}{2025}).
\newblock
\urldef\tempurl%
\url{https://arxiv.org/abs/2501.03575}
\showURL{%
\tempurl}


\bibitem[Ahmed et~al\mbox{.}(2017)]%
        {ahmed2017wadi}
\bibfield{author}{\bibinfo{person}{Chuadhry~Mujeeb Ahmed},
  \bibinfo{person}{Venkata~Reddy Palleti}, {and} \bibinfo{person}{Aditya~P.
  Mathur}.} \bibinfo{year}{2017}\natexlab{}.
\newblock \showarticletitle{{WADI}: A Water Distribution Testbed for Research
  in the Design of Secure Cyber Physical Systems}. In
  \bibinfo{booktitle}{\emph{Proceedings of the 3rd International Workshop on
  Cyber-Physical Systems for Smart Water Networks}}. \bibinfo{publisher}{ACM},
  \bibinfo{address}{New York, NY, USA}, \bibinfo{pages}{25-28}.
\newblock
\href{https://doi.org/10.1145/3055366.3055375}{doi:\nolinkurl{10.1145/3055366.3055375}}


\bibitem[Alcaraz and Lopez(2022)]%
        {alcaraz2022digitaltwinsecurity}
\bibfield{author}{\bibinfo{person}{Cristina Alcaraz} {and}
  \bibinfo{person}{Javier Lopez}.} \bibinfo{year}{2022}\natexlab{}.
\newblock \showarticletitle{{Digital Twin: A Comprehensive Survey of Security
  Threats}}.
\newblock \bibinfo{journal}{\emph{IEEE Communications Surveys \& Tutorials}}
  \bibinfo{volume}{24}, \bibinfo{number}{3} (\bibinfo{year}{2022}),
  \bibinfo{pages}{1475-1503}.
\newblock
\href{https://doi.org/10.1109/COMST.2022.3171465}{doi:\nolinkurl{10.1109/COMST.2022.3171465}}


\bibitem[Allen(2019)]%
        {allen2019regulatorysandboxes}
\bibfield{author}{\bibinfo{person}{Hilary~J. Allen}.}
  \bibinfo{year}{2019}\natexlab{}.
\newblock \showarticletitle{{Regulatory Sandboxes}}.
\newblock \bibinfo{journal}{\emph{George Washington Law Review}}
  \bibinfo{volume}{87}, \bibinfo{number}{3} (\bibinfo{year}{2019}),
  \bibinfo{pages}{579-645}.
\newblock
\urldef\tempurl%
\url{https://www.gwlr.org/wp-content/uploads/2019/06/87-Geo.-Wash.-L.-Rev.-579.pdf}
\showURL{%
\tempurl}


\bibitem[Alshiekh et~al\mbox{.}(2018)]%
        {alshiekh2018shielding}
\bibfield{author}{\bibinfo{person}{Mohammed Alshiekh},
  \bibinfo{person}{Roderick Bloem}, \bibinfo{person}{R{\"u}diger Ehlers},
  \bibinfo{person}{Bettina K{\"o}nighofer}, \bibinfo{person}{Scott Niekum},
  {and} \bibinfo{person}{Ufuk Topcu}.} \bibinfo{year}{2018}\natexlab{}.
\newblock \showarticletitle{Safe Reinforcement Learning via Shielding}. In
  \bibinfo{booktitle}{\emph{Proceedings of the Thirty-Second AAAI Conference on
  Artificial Intelligence}}. \bibinfo{publisher}{AAAI Press},
  \bibinfo{address}{New Orleans, Louisiana, USA}, \bibinfo{pages}{2669-2678}.
\newblock
\href{https://doi.org/10.1609/aaai.v32i1.11797}{doi:\nolinkurl{10.1609/aaai.v32i1.11797}}


\bibitem[{Amazon Web Services}(2026)]%
        {awsIoTTwinMakerDocs2026}
\bibfield{author}{\bibinfo{person}{{Amazon Web Services}}.}
  \bibinfo{year}{2026}\natexlab{}.
\newblock \bibinfo{title}{{AWS IoT TwinMaker Documentation}}.
\newblock
\urldef\tempurl%
\url{https://docs.aws.amazon.com/iot-twinmaker/}
\showURL{%
\tempurl}
\newblock
\shownote{Accessed 17 May 2026}.


\bibitem[Ames et~al\mbox{.}(2017)]%
        {ames2017cbf}
\bibfield{author}{\bibinfo{person}{Aaron~D. Ames}, \bibinfo{person}{Xiangru
  Xu}, \bibinfo{person}{Jessy~W. Grizzle}, {and} \bibinfo{person}{Paulo
  Tabuada}.} \bibinfo{year}{2017}\natexlab{}.
\newblock \showarticletitle{Control Barrier Function Based Quadratic Programs
  for Safety Critical Systems}.
\newblock \bibinfo{journal}{\emph{IEEE Trans. Automat. Control}}
  \bibinfo{volume}{62}, \bibinfo{number}{8} (\bibinfo{year}{2017}),
  \bibinfo{pages}{3861-3876}.
\newblock
\href{https://doi.org/10.1109/TAC.2016.2638961}{doi:\nolinkurl{10.1109/TAC.2016.2638961}}


\bibitem[Annpureddy et~al\mbox{.}(2011)]%
        {annpureddy2011staliro}
\bibfield{author}{\bibinfo{person}{Yashwanth Annpureddy}, \bibinfo{person}{Che
  Liu}, \bibinfo{person}{Georgios Fainekos}, {and} \bibinfo{person}{Sriram
  Sankaranarayanan}.} \bibinfo{year}{2011}\natexlab{}.
\newblock \showarticletitle{{S-TaLiRo}: A Tool for Temporal Logic Falsification
  for Hybrid Systems}. In \bibinfo{booktitle}{\emph{Proceedings of TACAS}}.
  \bibinfo{pages}{254-257}.
\newblock


\bibitem[{ASAM e.V.}(2026)]%
        {asamOpenScenarioDsl2026}
\bibfield{author}{\bibinfo{person}{{ASAM e.V.}}}
  \bibinfo{year}{2026}\natexlab{}.
\newblock \bibinfo{title}{{ASAM OpenSCENARIO DSL}}.
\newblock
\urldef\tempurl%
\url{https://www.asam.net/standards/detail/openscenario-dsl/}
\showURL{%
\tempurl}
\newblock
\shownote{Accessed 30 April 2026}.


\bibitem[{ASME}(2018)]%
        {asme2018vv40}
\bibfield{author}{\bibinfo{person}{{ASME}}.} \bibinfo{year}{2018}\natexlab{}.
\newblock \bibinfo{booktitle}{\emph{{ASME V\&V 40-2018}}}.
\newblock {American Society of Mechanical Engineers}.
\newblock
\urldef\tempurl%
\url{https://www.asme.org/codes-standards}
\showURL{%
\tempurl}
\newblock
\shownote{Assessing Credibility of Computational Modeling through Verification
  and Validation: Application to Medical Devices; official ASME standard
  metadata page, accessed 17 May 2026}.


\bibitem[{Autonomous Vehicle Systems Laboratory}(2026)]%
        {basiliskDocs2026}
\bibfield{author}{\bibinfo{person}{{Autonomous Vehicle Systems Laboratory}}.}
  \bibinfo{year}{2026}\natexlab{}.
\newblock \bibinfo{title}{{Basilisk: Astrodynamics Simulation Framework}}.
\newblock
\urldef\tempurl%
\url{https://avslab.github.io/basilisk/}
\showURL{%
\tempurl}
\newblock
\shownote{Accessed 17 May 2026}.


\bibitem[Banbury et~al\mbox{.}(2021)]%
        {banbury2021mlperftiny}
\bibfield{author}{\bibinfo{person}{Colby Banbury},
  \bibinfo{person}{Vijay~Janapa Reddi}, \bibinfo{person}{Peter Torelli},
  \bibinfo{person}{Jeremy Holleman}, \bibinfo{person}{Nat Jeffries},
  \bibinfo{person}{Csaba Kir{\'a}ly}, \bibinfo{person}{Pietro Montino},
  \bibinfo{person}{David Kanter}, \bibinfo{person}{Sebastian Ahmed},
  \bibinfo{person}{Danilo Pau}, \bibinfo{person}{Urmish Thakker},
  \bibinfo{person}{Antonio Torrini}, \bibinfo{person}{Peter Warden},
  \bibinfo{person}{Jay Cordaro}, \bibinfo{person}{Giuseppe~Di Guglielmo},
  \bibinfo{person}{Javier Duarte}, \bibinfo{person}{Stephen Jia},
  \bibinfo{person}{Honson Tran}, \bibinfo{person}{Nhan Tran},
  \bibinfo{person}{Niu Wenxu}, {and} \bibinfo{person}{Xu Xuesong}.}
  \bibinfo{year}{2021}\natexlab{}.
\newblock \bibinfo{title}{{MLPerf} Tiny Benchmark}.
\newblock \bibinfo{howpublished}{Proceedings of the Neural Information
  Processing Systems Track on Datasets and Benchmarks}.
\newblock
\urldef\tempurl%
\url{https://proceedings.neurips.cc/paper/2021/hash/da4fb5c6e93e74d3df8527599fa62642-Abstract.html}
\showURL{%
\tempurl}


\bibitem[Barricelli et~al\mbox{.}(2019)]%
        {barricelli2019digitaltwin}
\bibfield{author}{\bibinfo{person}{Barbara~Rita Barricelli},
  \bibinfo{person}{Elena Casiraghi}, {and} \bibinfo{person}{Daniela Fogli}.}
  \bibinfo{year}{2019}\natexlab{}.
\newblock \showarticletitle{{A Survey on Digital Twin: Definitions,
  Characteristics, Applications, and Design Implications}}.
\newblock \bibinfo{journal}{\emph{IEEE Access}}  \bibinfo{volume}{7}
  (\bibinfo{year}{2019}), \bibinfo{pages}{167653-167671}.
\newblock
\href{https://doi.org/10.1109/ACCESS.2019.2953499}{doi:\nolinkurl{10.1109/ACCESS.2019.2953499}}


\bibitem[Biesiadecki et~al\mbox{.}(2007)]%
        {biesiadecki2007mer}
\bibfield{author}{\bibinfo{person}{Jeffrey~J. Biesiadecki},
  \bibinfo{person}{P.~Chris Leger}, {and} \bibinfo{person}{Mark~W. Maimone}.}
  \bibinfo{year}{2007}\natexlab{}.
\newblock \showarticletitle{Tradeoffs Between Directed and Autonomous Driving
  on the Mars Exploration Rovers}.
\newblock \bibinfo{journal}{\emph{The International Journal of Robotics
  Research}} \bibinfo{volume}{26}, \bibinfo{number}{1} (\bibinfo{year}{2007}),
  \bibinfo{pages}{91-104}.
\newblock
\href{https://doi.org/10.1177/0278364907073777}{doi:\nolinkurl{10.1177/0278364907073777}}


\bibitem[Biggio et~al\mbox{.}(2012)]%
        {biggio2012poisoning}
\bibfield{author}{\bibinfo{person}{Battista Biggio}, \bibinfo{person}{Blaine
  Nelson}, {and} \bibinfo{person}{Pavel Laskov}.}
  \bibinfo{year}{2012}\natexlab{}.
\newblock \showarticletitle{Poisoning Attacks against Support Vector Machines}.
  In \bibinfo{booktitle}{\emph{Proceedings of the 29th International Conference
  on Machine Learning}} \emph{(\bibinfo{series}{ICML'12})}.
  \bibinfo{publisher}{Omnipress}, \bibinfo{address}{Edinburgh, Scotland},
  \bibinfo{pages}{1467-1474}.
\newblock
\urldef\tempurl%
\url{https://icml.cc/2012/papers/880.pdf}
\showURL{%
\tempurl}


\bibitem[Blattnig et~al\mbox{.}(2008)]%
        {blattnig2008credibility}
\bibfield{author}{\bibinfo{person}{Steve~R. Blattnig},
  \bibinfo{person}{Lawrence~L. Green}, \bibinfo{person}{James~M. Luckring},
  \bibinfo{person}{Joseph~H. Morrison}, \bibinfo{person}{Ram~K. Tripathi},
  {and} \bibinfo{person}{Thomas~A. Zang}.} \bibinfo{year}{2008}\natexlab{}.
\newblock \showarticletitle{{Towards a Credibility Assessment of Models and
  Simulations}}. In \bibinfo{booktitle}{\emph{10th AIAA Non-Deterministic
  Approaches Conference}}. \bibinfo{address}{Schaumburg, IL}.
\newblock
\urldef\tempurl%
\url{https://ntrs.nasa.gov/citations/20080015742}
\showURL{%
\tempurl}
\newblock
\shownote{NASA NTRS Document ID 20080015742}.


\bibitem[Bloomfield and Bishop(2010)]%
        {bloomfield2010safetycase}
\bibfield{author}{\bibinfo{person}{Robin Bloomfield} {and}
  \bibinfo{person}{Peter Bishop}.} \bibinfo{year}{2010}\natexlab{}.
\newblock \showarticletitle{Safety and Assurance Cases: Past, Present and
  Possible Future - an {A}delard Perspective}.
\newblock \bibinfo{journal}{\emph{Proceedings of the 18th Safety-Critical
  Systems Symposium}} (\bibinfo{year}{2010}), \bibinfo{pages}{51-67}.
\newblock
\href{https://doi.org/10.1007/978-1-84996-086-1_4}{doi:\nolinkurl{10.1007/978-1-84996-086-1_4}}


\bibitem[Bonawitz et~al\mbox{.}(2017)]%
        {bonawitz2017secureagg}
\bibfield{author}{\bibinfo{person}{Keith Bonawitz}, \bibinfo{person}{Vladimir
  Ivanov}, \bibinfo{person}{Ben Kreuter}, \bibinfo{person}{Antonio Marcedone},
  \bibinfo{person}{H.~Brendan McMahan}, \bibinfo{person}{Sarvar Patel},
  \bibinfo{person}{Daniel Ramage}, \bibinfo{person}{Aaron Segal}, {and}
  \bibinfo{person}{Karn Seth}.} \bibinfo{year}{2017}\natexlab{}.
\newblock \showarticletitle{Practical Secure Aggregation for Privacy-Preserving
  Machine Learning}. In \bibinfo{booktitle}{\emph{Proceedings of the ACM
  Conference on Computer and Communications Security (CCS)}}.
  \bibinfo{pages}{1175-1191}.
\newblock


\bibitem[Breck et~al\mbox{.}(2017)]%
        {breck2017mltest}
\bibfield{author}{\bibinfo{person}{Eric Breck}, \bibinfo{person}{Shanqing Cai},
  \bibinfo{person}{Eric Nielsen}, \bibinfo{person}{Michael Salib}, {and}
  \bibinfo{person}{D. Sculley}.} \bibinfo{year}{2017}\natexlab{}.
\newblock \showarticletitle{{The ML Test Score: A Rubric for ML Production
  Readiness and Technical Debt Reduction}}. In \bibinfo{booktitle}{\emph{2017
  IEEE International Conference on Big Data}}. \bibinfo{publisher}{IEEE},
  \bibinfo{address}{Piscataway, NJ, USA}, \bibinfo{pages}{1123-1132}.
\newblock
\href{https://doi.org/10.1109/BigData.2017.8258038}{doi:\nolinkurl{10.1109/BigData.2017.8258038}}


\bibitem[Brohan et~al\mbox{.}(2023)]%
        {zitkovich2023rt2}
\bibfield{author}{\bibinfo{person}{Anthony Brohan}, \bibinfo{person}{Noah
  Brown}, \bibinfo{person}{Justice Carbajal}, \bibinfo{person}{Yevgen
  Chebotar}, \bibinfo{person}{Xi Chen}, \bibinfo{person}{Krzysztof
  Choromanski}, \bibinfo{person}{Tianli Ding}, \bibinfo{person}{Danny Driess},
  \bibinfo{person}{Avinava Dubey}, \bibinfo{person}{Chelsea Finn},
  {et~al\mbox{.}}} \bibinfo{year}{2023}\natexlab{}.
\newblock \bibinfo{title}{{RT-2: Vision-Language-Action Models Transfer Web
  Knowledge to Robotic Control}}.
\newblock \bibinfo{howpublished}{arXiv preprint arXiv:2307.15818}.
\newblock
\urldef\tempurl%
\url{https://arxiv.org/abs/2307.15818}
\showURL{%
\tempurl}


\bibitem[Buscemi et~al\mbox{.}(2025)]%
        {buscemi2025sandboxconfigurator}
\bibfield{author}{\bibinfo{person}{Alessio Buscemi}, \bibinfo{person}{Thibault
  Simonetto}, \bibinfo{person}{Daniele Pagani}, \bibinfo{person}{German
  Castignani}, \bibinfo{person}{Maxime Cordy}, {and} \bibinfo{person}{Jordi
  Cabot}.} \bibinfo{year}{2025}\natexlab{}.
\newblock \bibinfo{title}{{Operationalising AI Regulatory Sandboxes:
  Activities, Requirements, and Technical Assessment under the EU AI Act}}.
\newblock \bibinfo{howpublished}{arXiv preprint arXiv:2509.25256}.
\newblock
\urldef\tempurl%
\url{https://arxiv.org/abs/2509.25256}
\showURL{%
\tempurl}


\bibitem[Cao et~al\mbox{.}(2019)]%
        {cao2019lidar}
\bibfield{author}{\bibinfo{person}{Yulong Cao}, \bibinfo{person}{Chaowei Xiao},
  \bibinfo{person}{Benjamin Cyr}, \bibinfo{person}{Yimeng Zhou},
  \bibinfo{person}{Won Park}, \bibinfo{person}{Sara Rampazzi},
  \bibinfo{person}{Qi~Alfred Chen}, \bibinfo{person}{Kevin Fu}, {and}
  \bibinfo{person}{Z.~Morley Mao}.} \bibinfo{year}{2019}\natexlab{}.
\newblock \showarticletitle{{Adversarial Sensor Attack on LiDAR-Based
  Perception in Autonomous Driving}}. In \bibinfo{booktitle}{\emph{Proceedings
  of the 2019 ACM SIGSAC Conference on Computer and Communications Security}}.
  \bibinfo{publisher}{ACM}, \bibinfo{address}{New York, NY, USA},
  \bibinfo{pages}{2267-2281}.
\newblock
\href{https://doi.org/10.1145/3319535.3339815}{doi:\nolinkurl{10.1145/3319535.3339815}}


\bibitem[C{\'a}rdenas et~al\mbox{.}(2008)]%
        {cardenas2008secure}
\bibfield{author}{\bibinfo{person}{Alvaro~A. C{\'a}rdenas},
  \bibinfo{person}{Saurabh Amin}, {and} \bibinfo{person}{Shankar Sastry}.}
  \bibinfo{year}{2008}\natexlab{}.
\newblock \showarticletitle{Secure Control: Towards Survivable Cyber-Physical
  Systems}. In \bibinfo{booktitle}{\emph{Proceedings of the 28th International
  Conference on Distributed Computing Systems Workshops}}.
  \bibinfo{pages}{495-500}.
\newblock


\bibitem[{CARLA Simulator Team}(2026)]%
        {carlaDocs2026}
\bibfield{author}{\bibinfo{person}{{CARLA Simulator Team}}.}
  \bibinfo{year}{2026}\natexlab{}.
\newblock \bibinfo{title}{{CARLA Autonomous Driving Simulator}}.
\newblock
\urldef\tempurl%
\url{https://carla.org/}
\showURL{%
\tempurl}
\newblock
\shownote{Accessed 10 May 2026}.


\bibitem[Chang et~al\mbox{.}(2024)]%
        {chang2024survey}
\bibfield{author}{\bibinfo{person}{Yupeng Chang}, \bibinfo{person}{Xu Wang},
  \bibinfo{person}{Jindong Wang}, \bibinfo{person}{Yuan Wu},
  \bibinfo{person}{Linyi Yang}, \bibinfo{person}{Kaijie Zhu},
  \bibinfo{person}{Hao Chen}, \bibinfo{person}{Xiaoyuan Yi},
  \bibinfo{person}{Cunxiang Wang}, \bibinfo{person}{Yidong Wang},
  \bibinfo{person}{Wei Ye}, \bibinfo{person}{Yue Zhang}, \bibinfo{person}{Yi
  Chang}, \bibinfo{person}{Philip~S. Yu}, \bibinfo{person}{Qiang Yang}, {and}
  \bibinfo{person}{Xing Xie}.} \bibinfo{year}{2024}\natexlab{}.
\newblock \showarticletitle{A Survey on Evaluation of Large Language Models}.
\newblock \bibinfo{journal}{\emph{ACM Transactions on Intelligent Systems and
  Technology}} \bibinfo{volume}{15}, \bibinfo{number}{3}
  (\bibinfo{year}{2024}).
\newblock
\href{https://doi.org/10.1145/3641289}{doi:\nolinkurl{10.1145/3641289}}


\bibitem[Cosseron et~al\mbox{.}(2024)]%
        {cosseron2024simulating}
\bibfield{author}{\bibinfo{person}{L{\'e}o Cosseron}, \bibinfo{person}{Louis
  Rilling}, \bibinfo{person}{Matthieu Simonin}, {and} \bibinfo{person}{Martin
  Quinson}.} \bibinfo{year}{2024}\natexlab{}.
\newblock \showarticletitle{Simulating the Network Environment of Sandboxes to
  Hide Virtual Machine Introspection Pauses}. In
  \bibinfo{booktitle}{\emph{Proceedings of the 17th European Workshop on
  Systems Security}}. \bibinfo{publisher}{ACM}, \bibinfo{address}{New York, NY,
  USA}, \bibinfo{pages}{1-7}.
\newblock
\href{https://doi.org/10.1145/3642974.3652280}{doi:\nolinkurl{10.1145/3642974.3652280}}


\bibitem[Crussell et~al\mbox{.}(2016)]%
        {crussell2016minimega}
\bibfield{author}{\bibinfo{person}{Jonathan Crussell}, \bibinfo{person}{Jeremy
  Erickson}, \bibinfo{person}{David Fritz}, {and} \bibinfo{person}{John
  Floren}.} \bibinfo{year}{2016}\natexlab{}.
\newblock \bibinfo{title}{{minimega v. 3.0}}.
\newblock
\href{https://doi.org/10.11578/dc.20171025.1714}{doi:\nolinkurl{10.11578/dc.20171025.1714}}


\bibitem[Das et~al\mbox{.}(2022)]%
        {das2022edgetwin}
\bibfield{author}{\bibinfo{person}{Sumit~K. Das},
  \bibinfo{person}{Mohammad~Helal Uddin}, {and} \bibinfo{person}{Sabur
  Baidya}.} \bibinfo{year}{2022}\natexlab{}.
\newblock \bibinfo{title}{Edge-Assisted Collaborative Digital Twin for
  Safety-Critical Robotics in Industrial {IoT}}.
\newblock \bibinfo{howpublished}{arXiv preprint arXiv:2209.12854}.
\newblock
\urldef\tempurl%
\url{https://arxiv.org/abs/2209.12854}
\showURL{%
\tempurl}


\bibitem[de~Vries et~al\mbox{.}(2020)]%
        {devries2020ecological}
\bibfield{author}{\bibinfo{person}{Harm de Vries}, \bibinfo{person}{Dzmitry
  Bahdanau}, {and} \bibinfo{person}{Christopher Manning}.}
  \bibinfo{year}{2020}\natexlab{}.
\newblock \bibinfo{title}{Towards Ecologically Valid Research on Language User
  Interfaces}.
\newblock \bibinfo{howpublished}{arXiv preprint arXiv:2007.14435}.
\newblock
\urldef\tempurl%
\url{https://arxiv.org/abs/2007.14435}
\showURL{%
\tempurl}


\bibitem[Deckenbrunnen et~al\mbox{.}(2026)]%
        {deckenbrunnen2026bathtubs}
\bibfield{author}{\bibinfo{person}{Tom Deckenbrunnen}, \bibinfo{person}{Alessio
  Buscemi}, \bibinfo{person}{Marco Almada}, \bibinfo{person}{Alfredo
  Capozucca}, {and} \bibinfo{person}{German Castignani}.}
  \bibinfo{year}{2026}\natexlab{}.
\newblock \showarticletitle{{Bathtubs, Boundaries, and Sandboxes: AI Regulatory
  Learning under Legal Uncertainty}}. In \bibinfo{booktitle}{\emph{Proceedings
  of the AAAI/ACM Conference on AI, Ethics, and Society (AIES '26)}},
  Vol.~\bibinfo{volume}{9}. \bibinfo{pages}{775--787}.
\newblock
\href{https://doi.org/10.1609/aies.v9i1.43182}{doi:\nolinkurl{10.1609/aies.v9i1.43182}}
\newblock
\shownote{Also available as arXiv:2601.04094}.


\bibitem[Denney and Pai(2018)]%
        {denney2018acas}
\bibfield{author}{\bibinfo{person}{Ewen Denney} {and} \bibinfo{person}{Ganesh
  Pai}.} \bibinfo{year}{2018}\natexlab{}.
\newblock \showarticletitle{Tool Support for Assurance Case Development}.
\newblock \bibinfo{journal}{\emph{Automated Software Engineering}}
  \bibinfo{volume}{25}, \bibinfo{number}{3} (\bibinfo{year}{2018}),
  \bibinfo{pages}{435-499}.
\newblock
\href{https://doi.org/10.1007/s10515-017-0230-5}{doi:\nolinkurl{10.1007/s10515-017-0230-5}}


\bibitem[Ding et~al\mbox{.}(2023)]%
        {ding2023survey}
\bibfield{author}{\bibinfo{person}{Wenhao Ding}, \bibinfo{person}{Chejian Xu},
  \bibinfo{person}{Mansur Arief}, \bibinfo{person}{Haohong Lin},
  \bibinfo{person}{Bo Li}, {and} \bibinfo{person}{Ding Zhao}.}
  \bibinfo{year}{2023}\natexlab{}.
\newblock \showarticletitle{A Survey on Safety-Critical Driving Scenario
  Generation - a Methodological Perspective}.
\newblock \bibinfo{journal}{\emph{IEEE Transactions on Intelligent
  Transportation Systems}} \bibinfo{volume}{24}, \bibinfo{number}{7}
  (\bibinfo{year}{2023}), \bibinfo{pages}{6971-6988}.
\newblock


\bibitem[Dodge et~al\mbox{.}(2019)]%
        {dodge2019showyourwork}
\bibfield{author}{\bibinfo{person}{Jesse Dodge}, \bibinfo{person}{Suchin
  Gururangan}, \bibinfo{person}{Dallas Card}, \bibinfo{person}{Roy Schwartz},
  {and} \bibinfo{person}{Noah~A. Smith}.} \bibinfo{year}{2019}\natexlab{}.
\newblock \showarticletitle{Show Your Work: Improved Reporting of Experimental
  Results}. In \bibinfo{booktitle}{\emph{Proceedings of the 2019 Conference on
  Empirical Methods in Natural Language Processing and the 9th International
  Joint Conference on Natural Language Processing (EMNLP-IJCNLP)}}.
  \bibinfo{publisher}{Association for Computational Linguistics},
  \bibinfo{address}{Hong Kong, China}, \bibinfo{pages}{2185-2194}.
\newblock
\href{https://doi.org/10.18653/v1/D19-1224}{doi:\nolinkurl{10.18653/v1/D19-1224}}


\bibitem[Donz{\'e}(2010)]%
        {donze2010breach}
\bibfield{author}{\bibinfo{person}{Alexandre Donz{\'e}}.}
  \bibinfo{year}{2010}\natexlab{}.
\newblock \showarticletitle{Breach, a Toolbox for Verification and Parameter
  Synthesis of Hybrid Systems}. In \bibinfo{booktitle}{\emph{Proceedings of
  CAV}}. \bibinfo{pages}{167-170}.
\newblock


\bibitem[Dosovitskiy et~al\mbox{.}(2017)]%
        {dosovitskiy2017carla}
\bibfield{author}{\bibinfo{person}{Alexey Dosovitskiy}, \bibinfo{person}{German
  Ros}, \bibinfo{person}{Felipe Codevilla}, \bibinfo{person}{Antonio Lopez},
  {and} \bibinfo{person}{Vladlen Koltun}.} \bibinfo{year}{2017}\natexlab{}.
\newblock \showarticletitle{{CARLA: An Open Urban Driving Simulator}}. In
  \bibinfo{booktitle}{\emph{Proceedings of the 1st Annual Conference on Robot
  Learning}} \emph{(\bibinfo{series}{Proceedings of Machine Learning Research},
  Vol.~\bibinfo{volume}{78})}, \bibfield{editor}{\bibinfo{person}{Sergey
  Levine}, \bibinfo{person}{Vincent Vanhoucke}, {and} \bibinfo{person}{Ken
  Goldberg}} (Eds.). \bibinfo{publisher}{PMLR}, \bibinfo{address}{Mountain
  View, CA, USA}, \bibinfo{pages}{1-16}.
\newblock
\urldef\tempurl%
\url{https://proceedings.mlr.press/v78/dosovitskiy17a.html}
\showURL{%
\tempurl}


\bibitem[Driess et~al\mbox{.}(2023)]%
        {driess2023palme}
\bibfield{author}{\bibinfo{person}{Danny Driess}, \bibinfo{person}{Fei Xia},
  \bibinfo{person}{Mehdi S.~M. Sajjadi}, \bibinfo{person}{Corey Lynch},
  \bibinfo{person}{Aakanksha Chowdhery}, \bibinfo{person}{Brian Ichter},
  \bibinfo{person}{Ayzaan Wahid}, \bibinfo{person}{Jonathan Tompson},
  \bibinfo{person}{Quan Vuong}, \bibinfo{person}{Tianhe Yu}, {et~al\mbox{.}}}
  \bibinfo{year}{2023}\natexlab{}.
\newblock \bibinfo{title}{{PaLM-E: An Embodied Multimodal Language Model}}.
\newblock \bibinfo{howpublished}{arXiv preprint arXiv:2303.03378}.
\newblock
\urldef\tempurl%
\url{https://arxiv.org/abs/2303.03378}
\showURL{%
\tempurl}


\bibitem[Du et~al\mbox{.}(2021)]%
        {du2021autotuned}
\bibfield{author}{\bibinfo{person}{Yuqing Du}, \bibinfo{person}{Olivia
  Watkins}, \bibinfo{person}{Trevor Darrell}, \bibinfo{person}{Pieter Abbeel},
  {and} \bibinfo{person}{Deepak Pathak}.} \bibinfo{year}{2021}\natexlab{}.
\newblock \showarticletitle{Auto-Tuned Sim-to-Real Transfer}. In
  \bibinfo{booktitle}{\emph{Proceedings of the IEEE International Conference on
  Robotics and Automation (ICRA)}}. \bibinfo{publisher}{IEEE},
  \bibinfo{address}{Piscataway, NJ, USA}, \bibinfo{pages}{1290-1296}.
\newblock
\href{https://doi.org/10.1109/ICRA48506.2021.9562091}{doi:\nolinkurl{10.1109/ICRA48506.2021.9562091}}


\bibitem[Duan et~al\mbox{.}(2022)]%
        {duan2022embodiedaisurvey}
\bibfield{author}{\bibinfo{person}{Jiafei Duan}, \bibinfo{person}{Samson Yu},
  \bibinfo{person}{Hui~Li Tan}, \bibinfo{person}{Hongyuan Zhu}, {and}
  \bibinfo{person}{Cheston Tan}.} \bibinfo{year}{2022}\natexlab{}.
\newblock \showarticletitle{{A Survey of Embodied AI: From Simulators to
  Research Tasks}}.
\newblock \bibinfo{journal}{\emph{IEEE Transactions on Emerging Topics in
  Computational Intelligence}} \bibinfo{volume}{6}, \bibinfo{number}{2}
  (\bibinfo{date}{April} \bibinfo{year}{2022}), \bibinfo{pages}{230-244}.
\newblock
\href{https://doi.org/10.1109/TETCI.2022.3141105}{doi:\nolinkurl{10.1109/TETCI.2022.3141105}}


\bibitem[{Eclipse Foundation}(2026)]%
        {eclipseDittoDocs2026}
\bibfield{author}{\bibinfo{person}{{Eclipse Foundation}}.}
  \bibinfo{year}{2026}\natexlab{}.
\newblock \bibinfo{title}{{Eclipse Ditto Documentation}}.
\newblock
\urldef\tempurl%
\url{https://eclipse.dev/ditto/}
\showURL{%
\tempurl}
\newblock
\shownote{Accessed 17 May 2026}.


\bibitem[Elmquist et~al\mbox{.}(2025)]%
        {elmquist2025simtoreal}
\bibfield{author}{\bibinfo{person}{Asher Elmquist}, \bibinfo{person}{Radu
  Serban}, {and} \bibinfo{person}{Dan Negrut}.}
  \bibinfo{year}{2025}\natexlab{}.
\newblock \showarticletitle{{A Methodology to Quantify
  Simulation-Versus-Reality Differences in Images for Autonomous Robots}}.
\newblock \bibinfo{journal}{\emph{IEEE Sensors Journal}} \bibinfo{volume}{25},
  \bibinfo{number}{4} (\bibinfo{year}{2025}), \bibinfo{pages}{6522-6533}.
\newblock
\href{https://doi.org/10.1109/JSEN.2024.3522050}{doi:\nolinkurl{10.1109/JSEN.2024.3522050}}


\bibitem[Erickson and Anderson(2022)]%
        {erickson2022soft}
\bibfield{author}{\bibinfo{person}{Jeremy~P Erickson} {and}
  \bibinfo{person}{James~H Anderson}.} \bibinfo{year}{2022}\natexlab{}.
\newblock \showarticletitle{Soft real-time scheduling}.
\newblock In \bibinfo{booktitle}{\emph{Handbook of Real-Time Computing}}.
  \bibinfo{publisher}{Springer}, \bibinfo{address}{Cham, Switzerland},
  \bibinfo{pages}{233-267}.
\newblock
\urldef\tempurl%
\url{https://link.springer.com/rwe/10.1007/978-981-287-251-7_4}
\showURL{%
\tempurl}


\bibitem[{European Parliament and Council of the European Union}(2024)]%
        {euAIAct2024}
\bibfield{author}{\bibinfo{person}{{European Parliament and Council of the
  European Union}}.} \bibinfo{year}{2024}\natexlab{}.
\newblock \bibinfo{title}{{Regulation (EU) 2024/1689 of the European Parliament
  and of the Council of 13 June 2024 laying down harmonised rules on artificial
  intelligence}}.
\newblock \bibinfo{howpublished}{{Official Journal of the European Union}}.
\newblock
\urldef\tempurl%
\url{https://eur-lex.europa.eu/eli/reg/2024/1689/oj/eng}
\showURL{%
\tempurl}


\bibitem[{European Telecommunications Standards Institute}(2024)]%
        {etsi3036452024}
\bibfield{author}{\bibinfo{person}{{European Telecommunications Standards
  Institute}}.} \bibinfo{year}{2024}\natexlab{}.
\newblock \bibinfo{title}{{ETSI EN 303 645 V3.1.3: Cyber Security for Consumer
  Internet of Things: Baseline Requirements}}.
\newblock
\urldef\tempurl%
\url{https://www.etsi.org/deliver/etsi_en/303600_303699/303645/03.01.03_60/en_303645v030103p.pdf}
\showURL{%
\tempurl}
\newblock
\shownote{Accessed 30 April 2026}.


\bibitem[Eykholt et~al\mbox{.}(2018)]%
        {eykholt2018physical}
\bibfield{author}{\bibinfo{person}{Kevin Eykholt}, \bibinfo{person}{Ivan
  Evtimov}, \bibinfo{person}{Earlence Fernandes}, \bibinfo{person}{Bo Li},
  \bibinfo{person}{Amir Rahmati}, \bibinfo{person}{Chaowei Xiao},
  \bibinfo{person}{Atul Prakash}, \bibinfo{person}{Tadayoshi Kohno}, {and}
  \bibinfo{person}{Dawn Song}.} \bibinfo{year}{2018}\natexlab{}.
\newblock \showarticletitle{{Robust Physical-World Attacks on Deep Learning
  Visual Classification}}. In \bibinfo{booktitle}{\emph{2018 IEEE/CVF
  Conference on Computer Vision and Pattern Recognition}}.
  \bibinfo{publisher}{IEEE}, \bibinfo{address}{Piscataway, NJ, USA},
  \bibinfo{pages}{1625-1634}.
\newblock
\href{https://doi.org/10.1109/CVPR.2018.00175}{doi:\nolinkurl{10.1109/CVPR.2018.00175}}


\bibitem[Fagan et~al\mbox{.}(2020)]%
        {fagan2020iotbaseline}
\bibfield{author}{\bibinfo{person}{Michael Fagan}, \bibinfo{person}{Katerina~N.
  Megas}, \bibinfo{person}{Karen Scarfone}, {and} \bibinfo{person}{Matthew
  Smith}.} \bibinfo{year}{2020}\natexlab{}.
\newblock \bibinfo{booktitle}{\emph{{IoT Device Cybersecurity Capability Core
  Baseline}}}.
\newblock \bibinfo{type}{{T}echnical {R}eport} NISTIR 8259A.
  \bibinfo{institution}{{National Institute of Standards and Technology}}.
\newblock
\href{https://doi.org/10.6028/NIST.IR.8259A}{doi:\nolinkurl{10.6028/NIST.IR.8259A}}


\bibitem[{Financial Conduct Authority}(2015)]%
        {fca2015regulatorysandbox}
\bibfield{author}{\bibinfo{person}{{Financial Conduct Authority}}.}
  \bibinfo{year}{2015}\natexlab{}.
\newblock \bibinfo{title}{{Regulatory Sandbox}}.
\newblock \bibinfo{howpublished}{{FCA Research Paper}}.
\newblock
\urldef\tempurl%
\url{https://www.fca.org.uk/publication/research/regulatory-sandbox.pdf}
\showURL{%
\tempurl}


\bibitem[{Financial Conduct Authority}(2017)]%
        {fca2017regulatorysandboxlessons}
\bibfield{author}{\bibinfo{person}{{Financial Conduct Authority}}.}
  \bibinfo{year}{2017}\natexlab{}.
\newblock \bibinfo{booktitle}{\emph{Regulatory Sandbox Lessons Learned
  Report}}.
\newblock \bibinfo{type}{{T}echnical {R}eport}.
  \bibinfo{institution}{{Financial Conduct Authority}}.
\newblock
\urldef\tempurl%
\url{https://www.fca.org.uk/publication/research-and-data/regulatory-sandbox-lessons-learned-report.pdf}
\showURL{%
\tempurl}
\newblock
\shownote{Accessed 17 May 2026}.


\bibitem[Furrer et~al\mbox{.}(2016)]%
        {furrer2016rotors}
\bibfield{author}{\bibinfo{person}{Fadri Furrer}, \bibinfo{person}{Michael
  Burri}, \bibinfo{person}{Markus Achtelik}, {and} \bibinfo{person}{Roland
  Siegwart}.} \bibinfo{year}{2016}\natexlab{}.
\newblock \showarticletitle{{RotorS} - A Modular {Gazebo} {MAV} Simulator
  Framework}.
\newblock In \bibinfo{booktitle}{\emph{Robot Operating System ({ROS}): The
  Complete Reference (Volume 1)}}, \bibfield{editor}{\bibinfo{person}{Anis
  Koubaa}} (Ed.). \bibinfo{publisher}{Springer}, \bibinfo{address}{Cham,
  Switzerland}, \bibinfo{pages}{595-625}.
\newblock
\href{https://doi.org/10.1007/978-3-319-26054-9_23}{doi:\nolinkurl{10.1007/978-3-319-26054-9_23}}


\bibitem[Gama et~al\mbox{.}(2014)]%
        {gama2014conceptdrift}
\bibfield{author}{\bibinfo{person}{Jo{\~a}o Gama}, \bibinfo{person}{Indr{\.{e}}
  {\v{Z}}liobait{\.{e}}}, \bibinfo{person}{Albert Bifet},
  \bibinfo{person}{Mykola Pechenizkiy}, {and} \bibinfo{person}{Abdelhamid
  Bouchachia}.} \bibinfo{year}{2014}\natexlab{}.
\newblock \showarticletitle{A Survey on Concept Drift Adaptation}.
\newblock \bibinfo{journal}{\emph{Comput. Surveys}} \bibinfo{volume}{46},
  \bibinfo{number}{4} (\bibinfo{year}{2014}), \bibinfo{pages}{44:1-44:37}.
\newblock


\bibitem[Garfinkel and Rosenblum(2003)]%
        {garfinkel2003vmi}
\bibfield{author}{\bibinfo{person}{Tal Garfinkel} {and} \bibinfo{person}{Mendel
  Rosenblum}.} \bibinfo{year}{2003}\natexlab{}.
\newblock \showarticletitle{{A Virtual Machine Introspection Based Architecture
  for Intrusion Detection}}. In \bibinfo{booktitle}{\emph{Proceedings of the
  Network and Distributed System Security Symposium}}.
  \bibinfo{publisher}{Internet Society}, \bibinfo{address}{Reston, VA, USA},
  \bibinfo{pages}{191-206}.
\newblock
\urldef\tempurl%
\url{https://www.ndss-symposium.org/ndss2003/virtual-machine-introspection-based-architecture-intrusion-detection/}
\showURL{%
\tempurl}


\bibitem[Gebru et~al\mbox{.}(2021)]%
        {gebru2021datasheets}
\bibfield{author}{\bibinfo{person}{Timnit Gebru}, \bibinfo{person}{Jamie
  Morgenstern}, \bibinfo{person}{Briana Vecchione},
  \bibinfo{person}{Jennifer~Wortman Vaughan}, \bibinfo{person}{Hanna Wallach},
  \bibinfo{person}{Hal Daum{\'e}}, {and} \bibinfo{person}{Kate Crawford}.}
  \bibinfo{year}{2021}\natexlab{}.
\newblock \showarticletitle{{Datasheets for Datasets}}.
\newblock \bibinfo{journal}{\emph{Commun. ACM}} \bibinfo{volume}{64},
  \bibinfo{number}{12} (\bibinfo{year}{2021}), \bibinfo{pages}{86-92}.
\newblock
\href{https://doi.org/10.1145/3458723}{doi:\nolinkurl{10.1145/3458723}}


\bibitem[Giraldo et~al\mbox{.}(2018)]%
        {giraldo2018physics}
\bibfield{author}{\bibinfo{person}{Jairo Giraldo}, \bibinfo{person}{David~I.
  Urbina}, \bibinfo{person}{Alvaro C{\'a}rdenas}, \bibinfo{person}{Junia
  Valente}, \bibinfo{person}{Mustafa~Amir Faisal}, \bibinfo{person}{Justin
  Ruths}, \bibinfo{person}{Nils~Ole Tippenhauer}, \bibinfo{person}{Henrik
  Sandberg}, {and} \bibinfo{person}{Richard Candell}.}
  \bibinfo{year}{2018}\natexlab{}.
\newblock \showarticletitle{A Survey of Physics-Based Attack Detection in
  Cyber-Physical Systems}.
\newblock \bibinfo{journal}{\emph{Comput. Surveys}} \bibinfo{volume}{51},
  \bibinfo{number}{4}, Article \bibinfo{articleno}{76} (\bibinfo{year}{2018}),
  \bibinfo{numpages}{36}~pages.
\newblock
\href{https://doi.org/10.1145/3203245}{doi:\nolinkurl{10.1145/3203245}}


\bibitem[Greshake et~al\mbox{.}(2023)]%
        {greshake2023indirectprompt}
\bibfield{author}{\bibinfo{person}{Kai Greshake}, \bibinfo{person}{Sahar
  Abdelnabi}, \bibinfo{person}{Shailesh Mishra}, \bibinfo{person}{Christoph
  Endres}, \bibinfo{person}{Thorsten Holz}, {and} \bibinfo{person}{Mario
  Fritz}.} \bibinfo{year}{2023}\natexlab{}.
\newblock \showarticletitle{{Not What You've Signed Up For: Compromising
  Real-World LLM-Integrated Applications with Indirect Prompt Injection}}. In
  \bibinfo{booktitle}{\emph{Proceedings of the 16th ACM Workshop on Artificial
  Intelligence and Security}}. \bibinfo{publisher}{ACM}, \bibinfo{address}{New
  York, NY, USA}, \bibinfo{pages}{79-90}.
\newblock
\href{https://doi.org/10.1145/3605764.3623985}{doi:\nolinkurl{10.1145/3605764.3623985}}


\bibitem[Hahn and Fasano(2021)]%
        {hahn2021sceptre}
\bibfield{author}{\bibinfo{person}{Adam~S. Hahn} {and}
  \bibinfo{person}{Robert~E. Fasano}.} \bibinfo{year}{2021}\natexlab{}.
\newblock \bibinfo{booktitle}{\emph{{OT Emulation Data Broker (SCEPTRE
  Capability)}}}.
\newblock \bibinfo{type}{{T}echnical {R}eport}. \bibinfo{institution}{{Sandia
  National Laboratories}}.
\newblock
\urldef\tempurl%
\url{https://www.sandia.gov/emulytics/}
\showURL{%
\tempurl}


\bibitem[{Health and Safety Executive}(2026)]%
        {hseAI2024}
\bibfield{author}{\bibinfo{person}{{Health and Safety Executive}}.}
  \bibinfo{year}{2026}\natexlab{}.
\newblock \bibinfo{title}{{HSE}'s Regulatory Approach to Artificial
  Intelligence ({AI})}.
\newblock \bibinfo{howpublished}{\url{https://www.hse.gov.uk/news/hse-ai.htm}}.
\newblock
\newblock
\shownote{Accessed 17 May 2026}.


\bibitem[{HELICS Project}(2026)]%
        {helicsDocs2026}
\bibfield{author}{\bibinfo{person}{{HELICS Project}}.}
  \bibinfo{year}{2026}\natexlab{}.
\newblock \bibinfo{title}{{HELICS Co-Simulation Framework}}.
\newblock
\urldef\tempurl%
\url{https://helics.org/}
\showURL{%
\tempurl}
\newblock
\shownote{Accessed 10 May 2026}.


\bibitem[Ichter et~al\mbox{.}(2023)]%
        {ichter2023saycan}
\bibfield{author}{\bibinfo{person}{Brian Ichter}, \bibinfo{person}{Anthony
  Brohan}, \bibinfo{person}{Yevgen Chebotar}, \bibinfo{person}{Chelsea Finn},
  \bibinfo{person}{Karol Hausman}, \bibinfo{person}{Alexander Herzog},
  \bibinfo{person}{Daniel Ho}, \bibinfo{person}{Julian Ibarz},
  \bibinfo{person}{Alex Irpan}, \bibinfo{person}{Eric Jang}, {et~al\mbox{.}}}
  \bibinfo{year}{2023}\natexlab{}.
\newblock \showarticletitle{{Do As I Can, Not As I Say: Grounding Language in
  Robotic Affordances}}. In \bibinfo{booktitle}{\emph{Proceedings of the 6th
  Conference on Robot Learning}} \emph{(\bibinfo{series}{Proceedings of Machine
  Learning Research}, Vol.~\bibinfo{volume}{205})}. \bibinfo{publisher}{PMLR},
  \bibinfo{address}{Auckland, New Zealand}, \bibinfo{pages}{287-318}.
\newblock
\urldef\tempurl%
\url{https://proceedings.mlr.press/v205/ichter23a.html}
\showURL{%
\tempurl}


\bibitem[{International Electrotechnical Commission}(2010)]%
        {iec61508}
\bibfield{author}{\bibinfo{person}{{International Electrotechnical
  Commission}}.} \bibinfo{year}{2010}\natexlab{}.
\newblock \bibinfo{booktitle}{\emph{{IEC} 61508: Functional Safety of
  Electrical/Electronic/Programmable Electronic Safety-Related Systems}
  (\bibinfo{edition}{2nd} ed.)}.
\newblock
\urldef\tempurl%
\url{https://www.iec.ch/functionalsafety/standards}
\showURL{%
\tempurl}
\newblock
\shownote{Parts 1-7}.


\bibitem[{International Electrotechnical Commission}(2026)]%
        {iec62443}
\bibfield{author}{\bibinfo{person}{{International Electrotechnical
  Commission}}.} \bibinfo{year}{2026}\natexlab{}.
\newblock \bibinfo{title}{{ISA/IEC 62443 series: Industrial communication
  networks - Network and system security}}.
\newblock
\urldef\tempurl%
\url{https://www.iec.ch/cyber-security-sector}
\showURL{%
\tempurl}
\newblock
\shownote{Accessed 30 April 2026}.


\bibitem[{International Organization for Standardization}(2018)]%
        {iso26262}
\bibfield{author}{\bibinfo{person}{{International Organization for
  Standardization}}.} \bibinfo{year}{2018}\natexlab{}.
\newblock \bibinfo{booktitle}{\emph{{ISO} 26262: Road Vehicles - Functional
  Safety} (\bibinfo{edition}{2nd} ed.)}.
\newblock
\urldef\tempurl%
\url{https://www.iso.org/standard/68383.html}
\showURL{%
\tempurl}
\newblock
\shownote{Parts 1-12}.


\bibitem[{International Organization for Standardization}(2022)]%
        {iso21448}
\bibfield{author}{\bibinfo{person}{{International Organization for
  Standardization}}.} \bibinfo{year}{2022}\natexlab{}.
\newblock \bibinfo{booktitle}{\emph{{ISO} 21448: Road Vehicles - Safety of the
  Intended Functionality ({SOTIF})}}.
\newblock
\urldef\tempurl%
\url{https://www.iso.org/standard/77490.html}
\showURL{%
\tempurl}


\bibitem[{International Organization for Standardization}(2023a)]%
        {iso23894}
\bibfield{author}{\bibinfo{person}{{International Organization for
  Standardization}}.} \bibinfo{year}{2023}\natexlab{a}.
\newblock \bibinfo{title}{{ISO/IEC 23894:2023 Information technology -
  Artificial intelligence - Guidance on risk management}}.
\newblock
\urldef\tempurl%
\url{https://www.iso.org/standard/77304.html}
\showURL{%
\tempurl}


\bibitem[{International Organization for Standardization}(2023b)]%
        {iso42001}
\bibfield{author}{\bibinfo{person}{{International Organization for
  Standardization}}.} \bibinfo{year}{2023}\natexlab{b}.
\newblock \bibinfo{title}{{ISO/IEC 42001:2023 Information technology -
  Artificial intelligence - Management system}}.
\newblock
\urldef\tempurl%
\url{https://www.iso.org/standard/42001}
\showURL{%
\tempurl}


\bibitem[{International Organization for Standardization} and {SAE
  International}(2021)]%
        {isosae21434}
\bibfield{author}{\bibinfo{person}{{International Organization for
  Standardization}} {and} \bibinfo{person}{{SAE International}}.}
  \bibinfo{year}{2021}\natexlab{}.
\newblock \bibinfo{booktitle}{\emph{{ISO/SAE} 21434: Road Vehicles -
  Cybersecurity Engineering}}.
\newblock
\urldef\tempurl%
\url{https://www.iso.org/standard/70918.html}
\showURL{%
\tempurl}


\bibitem[Jacobs and Wallach(2021)]%
        {jacobs2021measurement}
\bibfield{author}{\bibinfo{person}{Abigail~Z. Jacobs} {and}
  \bibinfo{person}{Hanna Wallach}.} \bibinfo{year}{2021}\natexlab{}.
\newblock \showarticletitle{Measurement and Fairness}. In
  \bibinfo{booktitle}{\emph{Proceedings of the 2021 ACM Conference on Fairness,
  Accountability, and Transparency (FAccT)}}. \bibinfo{publisher}{ACM},
  \bibinfo{address}{New York, NY, USA}, \bibinfo{pages}{375-385}.
\newblock
\href{https://doi.org/10.1145/3442188.3445901}{doi:\nolinkurl{10.1145/3442188.3445901}}


\bibitem[Jones et~al\mbox{.}(2020)]%
        {jones2020digitaltwin}
\bibfield{author}{\bibinfo{person}{David Jones}, \bibinfo{person}{Chris
  Snider}, \bibinfo{person}{Aydin Nassehi}, \bibinfo{person}{Jason Yon}, {and}
  \bibinfo{person}{Ben Hicks}.} \bibinfo{year}{2020}\natexlab{}.
\newblock \showarticletitle{{Characterising the Digital Twin: A Systematic
  Literature Review}}.
\newblock \bibinfo{journal}{\emph{CIRP Journal of Manufacturing Science and
  Technology}}  \bibinfo{volume}{29} (\bibinfo{year}{2020}),
  \bibinfo{pages}{36-52}.
\newblock
\href{https://doi.org/10.1016/j.cirpj.2020.02.002}{doi:\nolinkurl{10.1016/j.cirpj.2020.02.002}}


\bibitem[Kairouz et~al\mbox{.}(2021)]%
        {kairouz2021federated}
\bibfield{author}{\bibinfo{person}{Peter Kairouz}, \bibinfo{person}{H.~Brendan
  McMahan}, \bibinfo{person}{Brendan Avent}, \bibinfo{person}{Aur{\'e}lien
  Bellet}, \bibinfo{person}{Mehdi Bennis}, \bibinfo{person}{Arjun~Nitin
  Bhagoji}, \bibinfo{person}{Kallista Bonawitz}, \bibinfo{person}{Zachary
  Charles}, \bibinfo{person}{Graham Cormode}, \bibinfo{person}{Rachel
  Cummings}, {et~al\mbox{.}}} \bibinfo{year}{2021}\natexlab{}.
\newblock \showarticletitle{Advances and Open Problems in Federated Learning}.
\newblock \bibinfo{journal}{\emph{Foundations and Trends in Machine Learning}}
  \bibinfo{volume}{14}, \bibinfo{number}{1-2} (\bibinfo{year}{2021}),
  \bibinfo{pages}{1-210}.
\newblock


\bibitem[Kalra and Paddock(2016)]%
        {kalra2016driving}
\bibfield{author}{\bibinfo{person}{Nidhi Kalra} {and} \bibinfo{person}{Susan~M.
  Paddock}.} \bibinfo{year}{2016}\natexlab{}.
\newblock \showarticletitle{Driving to Safety: How Many Miles of Driving Would
  It Take to Demonstrate Autonomous Vehicle Reliability?}
\newblock \bibinfo{journal}{\emph{Transportation Research Part A: Policy and
  Practice}}  \bibinfo{volume}{94} (\bibinfo{year}{2016}),
  \bibinfo{pages}{182-193}.
\newblock
\href{https://doi.org/10.1016/j.tra.2016.09.010}{doi:\nolinkurl{10.1016/j.tra.2016.09.010}}


\bibitem[Kaur et~al\mbox{.}(2021)]%
        {kaur2021simulators}
\bibfield{author}{\bibinfo{person}{Prabhjot Kaur}, \bibinfo{person}{Samira
  Taghavi}, \bibinfo{person}{Zhaofeng Tian}, {and} \bibinfo{person}{Weisong
  Shi}.} \bibinfo{year}{2021}\natexlab{}.
\newblock \showarticletitle{{A Survey on Simulators for Testing Self-Driving
  Cars}}. In \bibinfo{booktitle}{\emph{2021 Fourth International Conference on
  Connected and Autonomous Driving (MetroCAD)}}. \bibinfo{publisher}{IEEE},
  \bibinfo{address}{Piscataway, NJ, USA}, \bibinfo{pages}{62-70}.
\newblock
\href{https://doi.org/10.1109/MetroCAD51599.2021.00018}{doi:\nolinkurl{10.1109/MetroCAD51599.2021.00018}}


\bibitem[Kelly and Weaver(2004)]%
        {kelly2004gsn}
\bibfield{author}{\bibinfo{person}{Tim Kelly} {and} \bibinfo{person}{Rob
  Weaver}.} \bibinfo{year}{2004}\natexlab{}.
\newblock \showarticletitle{The Goal Structuring Notation: A Safety Argument
  Notation}. In \bibinfo{booktitle}{\emph{Proceedings of the Dependable Systems
  and Networks Workshop on Assurance Cases}}. \bibinfo{publisher}{IEEE Computer
  Society}, \bibinfo{address}{Florence, Italy}, \bibinfo{pages}{1-6}.
\newblock
\urldef\tempurl%
\url{https://www-users.cs.york.ac.uk/~tpk/dsn2004.pdf}
\showURL{%
\tempurl}


\bibitem[Kenneally et~al\mbox{.}(2020)]%
        {kenneally2020basilisk}
\bibfield{author}{\bibinfo{person}{Patrick~W. Kenneally},
  \bibinfo{person}{Scott Piggott}, {and} \bibinfo{person}{Hanspeter Schaub}.}
  \bibinfo{year}{2020}\natexlab{}.
\newblock \showarticletitle{Basilisk: A Flexible, Scalable and Modular
  Astrodynamics Simulation Framework}.
\newblock \bibinfo{journal}{\emph{Journal of Aerospace Information Systems}}
  \bibinfo{volume}{17}, \bibinfo{number}{9} (\bibinfo{year}{2020}),
  \bibinfo{pages}{496-507}.
\newblock
\href{https://doi.org/10.2514/1.I010762}{doi:\nolinkurl{10.2514/1.I010762}}


\bibitem[Kim et~al\mbox{.}(2024)]%
        {kim2024openvla}
\bibfield{author}{\bibinfo{person}{Moo~Jin Kim}, \bibinfo{person}{Karl
  Pertsch}, \bibinfo{person}{Siddharth Karamcheti}, \bibinfo{person}{Ted Xiao},
  \bibinfo{person}{Ashwin Balakrishna}, \bibinfo{person}{Suraj Nair},
  \bibinfo{person}{Rafael Rafailov}, \bibinfo{person}{Ethan Foster},
  \bibinfo{person}{Grace Lam}, \bibinfo{person}{Pannag Sanketi},
  \bibinfo{person}{Quan Vuong}, \bibinfo{person}{Thomas Kollar},
  \bibinfo{person}{Benjamin Burchfiel}, \bibinfo{person}{Russ Tedrake},
  \bibinfo{person}{Dorsa Sadigh}, \bibinfo{person}{Sergey Levine},
  \bibinfo{person}{Percy Liang}, {and} \bibinfo{person}{Chelsea Finn}.}
  \bibinfo{year}{2024}\natexlab{}.
\newblock \showarticletitle{{OpenVLA}: An Open-Source Vision-Language-Action
  Model}.
\newblock \bibinfo{journal}{\emph{arXiv preprint arXiv:2406.09246}}
  (\bibinfo{year}{2024}).
\newblock
\urldef\tempurl%
\url{https://arxiv.org/abs/2406.09246}
\showURL{%
\tempurl}


\bibitem[Kitchenham and Charters(2007)]%
        {kitchenham2007slr}
\bibfield{author}{\bibinfo{person}{Barbara Kitchenham} {and}
  \bibinfo{person}{Stuart Charters}.} \bibinfo{year}{2007}\natexlab{}.
\newblock \bibinfo{booktitle}{\emph{{Guidelines for performing Systematic
  Literature Reviews in Software Engineering}}}.
\newblock \bibinfo{type}{{T}echnical {R}eport} EBSE 2007-001.
  \bibinfo{institution}{Keele University and Durham University Joint Report}.
\newblock
\urldef\tempurl%
\url{https://ebse.webspace.durham.ac.uk/ebse-bibliography/guidelines-for-performing-systematic-literature-reviews-in-software-engineering/}
\showURL{%
\tempurl}


\bibitem[Koenig and Howard(2004)]%
        {koenig2004gazebo}
\bibfield{author}{\bibinfo{person}{Nathan Koenig} {and} \bibinfo{person}{Andrew
  Howard}.} \bibinfo{year}{2004}\natexlab{}.
\newblock \showarticletitle{{Design and use paradigms for Gazebo, an
  open-source multi-robot simulator}}. In \bibinfo{booktitle}{\emph{2004
  IEEE/RSJ International Conference on Intelligent Robots and Systems (IROS)
  (IEEE Cat. No. 04CH37566)}}, Vol.~\bibinfo{volume}{3}.
  \bibinfo{publisher}{IEEE}, \bibinfo{address}{Piscataway, NJ, USA},
  \bibinfo{pages}{2149-2154}.
\newblock
\href{https://doi.org/10.1109/IROS.2004.1389727}{doi:\nolinkurl{10.1109/IROS.2004.1389727}}


\bibitem[Kolve et~al\mbox{.}(2017)]%
        {kolve2017ai2thor}
\bibfield{author}{\bibinfo{person}{Eric Kolve}, \bibinfo{person}{Roozbeh
  Mottaghi}, \bibinfo{person}{Winson Han}, \bibinfo{person}{Eli VanderBilt},
  \bibinfo{person}{Luca Weihs}, \bibinfo{person}{Alvaro Herrasti},
  \bibinfo{person}{Matt Deitke}, \bibinfo{person}{Kiana Ehsani},
  \bibinfo{person}{Daniel Gordon}, \bibinfo{person}{Yuke Zhu},
  \bibinfo{person}{Aniruddha Kembhavi}, \bibinfo{person}{Abhinav Gupta}, {and}
  \bibinfo{person}{Ali Farhadi}.} \bibinfo{year}{2017}\natexlab{}.
\newblock \bibinfo{title}{{AI2-THOR}: An Interactive {3D} Environment for
  Visual {AI}}.
\newblock \bibinfo{howpublished}{arXiv preprint arXiv:1712.05474}.
\newblock
\urldef\tempurl%
\url{https://arxiv.org/abs/1712.05474}
\showURL{%
\tempurl}


\bibitem[Koopman and Wagner(2017)]%
        {koopman2017avsafety}
\bibfield{author}{\bibinfo{person}{Philip Koopman} {and}
  \bibinfo{person}{Michael Wagner}.} \bibinfo{year}{2017}\natexlab{}.
\newblock \showarticletitle{Autonomous Vehicle Safety: An Interdisciplinary
  Challenge}.
\newblock \bibinfo{journal}{\emph{IEEE Intelligent Transportation Systems
  Magazine}} \bibinfo{volume}{9}, \bibinfo{number}{1} (\bibinfo{year}{2017}),
  \bibinfo{pages}{90-96}.
\newblock
\href{https://doi.org/10.1109/MITS.2016.2583491}{doi:\nolinkurl{10.1109/MITS.2016.2583491}}


\bibitem[Lee(2008)]%
        {lee2008cps}
\bibfield{author}{\bibinfo{person}{Edward~A. Lee}.}
  \bibinfo{year}{2008}\natexlab{}.
\newblock \showarticletitle{{Cyber Physical Systems: Design Challenges}}. In
  \bibinfo{booktitle}{\emph{2008 11th IEEE International Symposium on Object
  and Component-Oriented Real-Time Distributed Computing}}.
  \bibinfo{publisher}{IEEE}, \bibinfo{address}{Piscataway, NJ, USA},
  \bibinfo{pages}{363-369}.
\newblock
\href{https://doi.org/10.1109/ISORC.2008.25}{doi:\nolinkurl{10.1109/ISORC.2008.25}}


\bibitem[Leveson(2011)]%
        {leveson2011safetycase}
\bibfield{author}{\bibinfo{person}{Nancy~G. Leveson}.}
  \bibinfo{year}{2011}\natexlab{}.
\newblock \bibinfo{booktitle}{\emph{Engineering a Safer World: {S}ystems
  Thinking Applied to Safety}}.
\newblock \bibinfo{publisher}{MIT Press}, \bibinfo{address}{Cambridge, MA}.
\newblock


\bibitem[Li et~al\mbox{.}(2020)]%
        {li2020avfuzzer}
\bibfield{author}{\bibinfo{person}{Guanpeng Li}, \bibinfo{person}{Yiran Li},
  \bibinfo{person}{Saurabh Jha}, \bibinfo{person}{Timothy Tsai},
  \bibinfo{person}{Michael Sullivan}, \bibinfo{person}{Siva Kumar~Sastry Hari},
  \bibinfo{person}{Zbigniew Kalbarczyk}, {and} \bibinfo{person}{Ravishankar
  Iyer}.} \bibinfo{year}{2020}\natexlab{}.
\newblock \showarticletitle{{AV-FUZZER}: Finding Safety Violations in
  Autonomous Driving Systems}. In \bibinfo{booktitle}{\emph{Proceedings of the
  31st IEEE International Symposium on Software Reliability Engineering
  (ISSRE)}}. \bibinfo{pages}{25-36}.
\newblock


\bibitem[Li et~al\mbox{.}(2022)]%
        {li2022hilbuilding}
\bibfield{author}{\bibinfo{person}{Guowen Li}, \bibinfo{person}{Zhiyao Yang},
  \bibinfo{person}{Yangyang Fu}, \bibinfo{person}{Lingyu Ren},
  \bibinfo{person}{Zheng O'Neill}, {and} \bibinfo{person}{Chirag Parikh}.}
  \bibinfo{year}{2022}\natexlab{}.
\newblock \bibinfo{title}{Development of a Hardware-in-the-Loop ({HIL}) Testbed
  for Cyber-Physical Security in Smart Buildings}.
\newblock \bibinfo{howpublished}{arXiv preprint arXiv:2210.11234}.
\newblock
\urldef\tempurl%
\url{https://arxiv.org/abs/2210.11234}
\showURL{%
\tempurl}


\bibitem[Li et~al\mbox{.}(2024)]%
        {li2024simpler}
\bibfield{author}{\bibinfo{person}{Xuanlin Li}, \bibinfo{person}{Kyle Hsu},
  \bibinfo{person}{Jiayuan Gu}, \bibinfo{person}{Karl Pertsch},
  \bibinfo{person}{Oier Mees}, \bibinfo{person}{Homer~Rich Walke},
  \bibinfo{person}{Chuyuan Fu}, \bibinfo{person}{Ishikaa Lunawat},
  \bibinfo{person}{Isabel Sieh}, \bibinfo{person}{Sean Kirmani},
  \bibinfo{person}{Sergey Levine}, \bibinfo{person}{Jiajun Wu},
  \bibinfo{person}{Chelsea Finn}, \bibinfo{person}{Hao Su},
  \bibinfo{person}{Quan Vuong}, {and} \bibinfo{person}{Ted Xiao}.}
  \bibinfo{year}{2024}\natexlab{}.
\newblock \showarticletitle{Evaluating Real-World Robot Manipulation Policies
  in Simulation}.
\newblock \bibinfo{journal}{\emph{arXiv preprint arXiv:2405.05941}}
  (\bibinfo{year}{2024}).
\newblock
\urldef\tempurl%
\url{https://arxiv.org/abs/2405.05941}
\showURL{%
\tempurl}


\bibitem[Liang et~al\mbox{.}(2023)]%
        {liang2023helm}
\bibfield{author}{\bibinfo{person}{Percy Liang}, \bibinfo{person}{Rishi
  Bommasani}, \bibinfo{person}{Tony Lee}, {et~al\mbox{.}}}
  \bibinfo{year}{2023}\natexlab{}.
\newblock \bibinfo{title}{Holistic Evaluation of Language Models}.
\newblock \bibinfo{howpublished}{Transactions on Machine Learning Research;
  arXiv:2211.09110}.
\newblock
\urldef\tempurl%
\url{https://arxiv.org/abs/2211.09110}
\showURL{%
\tempurl}


\bibitem[Liu et~al\mbox{.}(2024)]%
        {liu2024agentbench}
\bibfield{author}{\bibinfo{person}{Xiao Liu}, \bibinfo{person}{Hao Yu},
  \bibinfo{person}{Hanchen Zhang}, \bibinfo{person}{Yifan Xu},
  \bibinfo{person}{Xuanyu Lei}, \bibinfo{person}{Hanyu Lai},
  \bibinfo{person}{Yu Gu}, \bibinfo{person}{Hangliang Ding},
  \bibinfo{person}{Kaiwen Men}, \bibinfo{person}{Kejuan Yang},
  \bibinfo{person}{Shudan Zhang}, \bibinfo{person}{Xiang Deng},
  \bibinfo{person}{Aohan Zeng}, \bibinfo{person}{Zhengxiao Du},
  \bibinfo{person}{Chenhui Zhang}, \bibinfo{person}{Sheng Shen},
  \bibinfo{person}{Tianjun Zhang}, \bibinfo{person}{Yu Su},
  \bibinfo{person}{Huan Sun}, \bibinfo{person}{Minlie Huang},
  \bibinfo{person}{Yuxiao Dong}, {and} \bibinfo{person}{Jie Tang}.}
  \bibinfo{year}{2024}\natexlab{}.
\newblock \bibinfo{title}{{AgentBench: Evaluating LLMs as Agents}}.
\newblock \bibinfo{howpublished}{The Twelfth International Conference on
  Learning Representations}.
\newblock
\urldef\tempurl%
\url{https://openreview.net/forum?id=zAdUB0aCTQ}
\showURL{%
\tempurl}


\bibitem[Liu et~al\mbox{.}(2009)]%
        {liu2009fdi}
\bibfield{author}{\bibinfo{person}{Yao Liu}, \bibinfo{person}{Peng Ning}, {and}
  \bibinfo{person}{Michael~K. Reiter}.} \bibinfo{year}{2009}\natexlab{}.
\newblock \showarticletitle{False Data Injection Attacks against State
  Estimation in Electric Power Grids}. In \bibinfo{booktitle}{\emph{Proceedings
  of the 16th ACM Conference on Computer and Communications Security}}.
  \bibinfo{publisher}{ACM}, \bibinfo{address}{New York, NY, USA},
  \bibinfo{pages}{21-32}.
\newblock
\href{https://doi.org/10.1145/1653662.1653666}{doi:\nolinkurl{10.1145/1653662.1653666}}


\bibitem[Liu et~al\mbox{.}(2022)]%
        {liu2022behaviorhabitat}
\bibfield{author}{\bibinfo{person}{Ziang Liu}, \bibinfo{person}{Roberto
  Mart{\'\i}n-Mart{\'\i}n}, \bibinfo{person}{Fei Xia}, \bibinfo{person}{Jiajun
  Wu}, {and} \bibinfo{person}{Li Fei-Fei}.} \bibinfo{year}{2022}\natexlab{}.
\newblock \bibinfo{title}{{BEHAVIOR} in Habitat 2.0: Simulator-Independent
  Logical Task Description for Benchmarking Embodied {AI} Agents}.
\newblock \bibinfo{howpublished}{arXiv preprint arXiv:2206.06489}.
\newblock
\urldef\tempurl%
\url{https://arxiv.org/abs/2206.06489}
\showURL{%
\tempurl}


\bibitem[Lu et~al\mbox{.}(2025)]%
        {lu2025toolsandbox}
\bibfield{author}{\bibinfo{person}{Jiarui Lu}, \bibinfo{person}{Thomas
  Holleis}, \bibinfo{person}{Yizhe Zhang}, \bibinfo{person}{Bernhard Aumayer},
  \bibinfo{person}{Feng Nan}, \bibinfo{person}{Felix Bai},
  \bibinfo{person}{Shuang Ma}, \bibinfo{person}{Shen Ma},
  \bibinfo{person}{Mengyu Li}, \bibinfo{person}{Guoli Yin},
  \bibinfo{person}{Zirui Wang}, {and} \bibinfo{person}{Ruoming Pang}.}
  \bibinfo{year}{2025}\natexlab{}.
\newblock \showarticletitle{{ToolSandbox: A Stateful, Conversational,
  Interactive Evaluation Benchmark for LLM Tool Use Capabilities}}. In
  \bibinfo{booktitle}{\emph{Findings of the Association for Computational
  Linguistics: NAACL 2025}}. \bibinfo{publisher}{Association for Computational
  Linguistics}, \bibinfo{address}{Albuquerque, New Mexico},
  \bibinfo{pages}{1160-1183}.
\newblock
\href{https://doi.org/10.18653/v1/2025.findings-naacl.65}{doi:\nolinkurl{10.18653/v1/2025.findings-naacl.65}}


\bibitem[Lu et~al\mbox{.}(2019)]%
        {lu2019driftsurvey}
\bibfield{author}{\bibinfo{person}{Jie Lu}, \bibinfo{person}{Anjin Liu},
  \bibinfo{person}{Fan Dong}, \bibinfo{person}{Feng Gu},
  \bibinfo{person}{Jo{\~a}o Gama}, {and} \bibinfo{person}{Guangquan Zhang}.}
  \bibinfo{year}{2019}\natexlab{}.
\newblock \showarticletitle{Learning under Concept Drift: A Review}.
\newblock \bibinfo{journal}{\emph{IEEE Transactions on Knowledge and Data
  Engineering}} \bibinfo{volume}{31}, \bibinfo{number}{12}
  (\bibinfo{year}{2019}), \bibinfo{pages}{2346-2363}.
\newblock


\bibitem[Makoviychuk et~al\mbox{.}(2021)]%
        {makoviychuk2021isaacgym}
\bibfield{author}{\bibinfo{person}{Viktor Makoviychuk}, \bibinfo{person}{Lukasz
  Wawrzyniak}, \bibinfo{person}{Yunrong Guo}, \bibinfo{person}{Michelle Lu},
  \bibinfo{person}{Kier Storey}, \bibinfo{person}{Miles Macklin},
  \bibinfo{person}{David Hoeller}, \bibinfo{person}{Nikita Rudin},
  \bibinfo{person}{Arthur Allshire}, \bibinfo{person}{Ankur Handa}, {and}
  \bibinfo{person}{Gavriel State}.} \bibinfo{year}{2021}\natexlab{}.
\newblock \bibinfo{title}{Isaac Gym: High Performance {GPU}-Based Physics
  Simulation For Robot Learning}.
\newblock \bibinfo{howpublished}{arXiv preprint arXiv:2108.10470}.
\newblock
\urldef\tempurl%
\url{https://arxiv.org/abs/2108.10470}
\showURL{%
\tempurl}


\bibitem[Mathur and Tippenhauer(2016)]%
        {mathur2016swat}
\bibfield{author}{\bibinfo{person}{Aditya~P. Mathur} {and}
  \bibinfo{person}{Nils~Ole Tippenhauer}.} \bibinfo{year}{2016}\natexlab{}.
\newblock \showarticletitle{{SWaT}: A Water Treatment Testbed for Research and
  Training on {ICS} Security}. In \bibinfo{booktitle}{\emph{2016 International
  Workshop on Cyber-Physical Systems for Smart Water Networks (CySWater)}}.
  \bibinfo{publisher}{IEEE}, \bibinfo{address}{Piscataway, NJ, USA},
  \bibinfo{pages}{31-36}.
\newblock
\href{https://doi.org/10.1109/CySWater.2016.7469060}{doi:\nolinkurl{10.1109/CySWater.2016.7469060}}


\bibitem[{Meta AI}(2026)]%
        {habitatSimRepo2026}
\bibfield{author}{\bibinfo{person}{{Meta AI}}.}
  \bibinfo{year}{2026}\natexlab{}.
\newblock \bibinfo{title}{{Habitat-Sim GitHub Repository}}.
\newblock
\urldef\tempurl%
\url{https://github.com/facebookresearch/habitat-sim}
\showURL{%
\tempurl}
\newblock
\shownote{Accessed 10 May 2026}.


\bibitem[Michel(2004)]%
        {michel2004webots}
\bibfield{author}{\bibinfo{person}{Olivier Michel}.}
  \bibinfo{year}{2004}\natexlab{}.
\newblock \showarticletitle{Webots: Professional Mobile Robot Simulation}.
\newblock \bibinfo{journal}{\emph{International Journal of Advanced Robotic
  Systems}} \bibinfo{volume}{1}, \bibinfo{number}{1} (\bibinfo{year}{2004}),
  \bibinfo{pages}{39-42}.
\newblock
\href{https://doi.org/10.5772/5618}{doi:\nolinkurl{10.5772/5618}}


\bibitem[{Microsoft}(2026a)]%
        {airsimRepo2026}
\bibfield{author}{\bibinfo{person}{{Microsoft}}.}
  \bibinfo{year}{2026}\natexlab{a}.
\newblock \bibinfo{title}{{AirSim GitHub Repository}}.
\newblock
\urldef\tempurl%
\url{https://github.com/microsoft/AirSim}
\showURL{%
\tempurl}
\newblock
\shownote{Accessed 10 May 2026}.


\bibitem[{Microsoft}(2026b)]%
        {azureDigitalTwinsDocs2026}
\bibfield{author}{\bibinfo{person}{{Microsoft}}.}
  \bibinfo{year}{2026}\natexlab{b}.
\newblock \bibinfo{title}{{Azure Digital Twins Documentation}}.
\newblock
\urldef\tempurl%
\url{https://learn.microsoft.com/azure/digital-twins/}
\showURL{%
\tempurl}
\newblock
\shownote{Accessed 17 May 2026}.


\bibitem[{MIT Marine Autonomy Lab}(2026)]%
        {moosivpDocs2026}
\bibfield{author}{\bibinfo{person}{{MIT Marine Autonomy Lab}}.}
  \bibinfo{year}{2026}\natexlab{}.
\newblock \bibinfo{title}{{MOOS-IvP Autonomy Tools Documentation}}.
\newblock
\urldef\tempurl%
\url{https://oceanai.mit.edu/moos-ivp/pmwiki/pmwiki.php}
\showURL{%
\tempurl}
\newblock
\shownote{Accessed 17 May 2026}.


\bibitem[Mitchell et~al\mbox{.}(2019)]%
        {mitchell2019modelcards}
\bibfield{author}{\bibinfo{person}{Margaret Mitchell}, \bibinfo{person}{Simone
  Wu}, \bibinfo{person}{Andrew Zaldivar}, \bibinfo{person}{Parker Barnes},
  \bibinfo{person}{Lucy Vasserman}, \bibinfo{person}{Ben Hutchinson},
  \bibinfo{person}{Elena Spitzer}, \bibinfo{person}{Inioluwa~Deborah Raji},
  {and} \bibinfo{person}{Timnit Gebru}.} \bibinfo{year}{2019}\natexlab{}.
\newblock \showarticletitle{{Model Cards for Model Reporting}}. In
  \bibinfo{booktitle}{\emph{Proceedings of the Conference on Fairness,
  Accountability, and Transparency}}. \bibinfo{publisher}{ACM},
  \bibinfo{address}{New York, NY, USA}, \bibinfo{pages}{220-229}.
\newblock
\href{https://doi.org/10.1145/3287560.3287596}{doi:\nolinkurl{10.1145/3287560.3287596}}


\bibitem[{MITRE}(2026)]%
        {mitreFederalAISandbox2026}
\bibfield{author}{\bibinfo{person}{{MITRE}}.} \bibinfo{year}{2026}\natexlab{}.
\newblock \bibinfo{title}{{MITRE's Federal AI Sandbox}}.
\newblock
  \bibinfo{howpublished}{\url{https://www.mitre.org/news-insights/fact-sheet/mitres-federal-ai-sandbox}}.
\newblock
\newblock
\shownote{Accessed: 2026-05-13}.


\bibitem[{MITRE Corporation}(2023)]%
        {mitreATLAS}
\bibfield{author}{\bibinfo{person}{{MITRE Corporation}}.}
  \bibinfo{year}{2023}\natexlab{}.
\newblock \bibinfo{title}{{ATLAS}: Adversarial Threat Landscape for
  Artificial-Intelligence Systems}.
\newblock \bibinfo{howpublished}{\url{https://atlas.mitre.org}}.
\newblock


\bibitem[Mittal et~al\mbox{.}(2025)]%
        {mittal2025isaaclab}
\bibfield{author}{\bibinfo{person}{Mayank Mittal}, \bibinfo{person}{Pascal
  Roth}, \bibinfo{person}{James Tigue}, \bibinfo{person}{Antoine Richard},
  {et~al\mbox{.}}} \bibinfo{year}{2025}\natexlab{}.
\newblock \bibinfo{title}{Isaac Lab: {A} {GPU}-Accelerated Simulation Framework
  for Multi-Modal Robot Learning}.
\newblock \bibinfo{howpublished}{arXiv preprint arXiv:2511.04831}.
\newblock
\urldef\tempurl%
\url{https://arxiv.org/abs/2511.04831}
\showURL{%
\tempurl}


\bibitem[{Modelica Association Project FMI}(2023)]%
        {fmiStandard3}
\bibfield{author}{\bibinfo{person}{{Modelica Association Project FMI}}.}
  \bibinfo{year}{2023}\natexlab{}.
\newblock \bibinfo{title}{{Functional Mock-up Interface Specification 3.0}}.
\newblock
\urldef\tempurl%
\url{https://fmi-standard.org/docs/3.0/}
\showURL{%
\tempurl}
\newblock
\shownote{Accessed 30 April 2026}.


\bibitem[{Modelica Association Project SSP}(2019)]%
        {ssp2019spec}
\bibfield{author}{\bibinfo{person}{{Modelica Association Project SSP}}.}
  \bibinfo{year}{2019}\natexlab{}.
\newblock \bibinfo{title}{System Structure and Parameterization ({SSP}),
  Version 1.0}.
\newblock \bibinfo{howpublished}{Modelica Association}.
\newblock
\urldef\tempurl%
\url{https://ssp-standard.org/}
\showURL{%
\tempurl}


\bibitem[Mohammadi et~al\mbox{.}(2025)]%
        {mohammadi2025agentevalsurvey}
\bibfield{author}{\bibinfo{person}{Mahmoud Mohammadi}, \bibinfo{person}{Yipeng
  Li}, \bibinfo{person}{Jane Lo}, {and} \bibinfo{person}{Wendy Yip}.}
  \bibinfo{year}{2025}\natexlab{}.
\newblock \showarticletitle{{Evaluation and Benchmarking of LLM Agents: A
  Survey}}. In \bibinfo{booktitle}{\emph{Proceedings of the 31st ACM SIGKDD
  Conference on Knowledge Discovery and Data Mining V.2 (KDD '25)}}.
  \bibinfo{pages}{6129-6139}.
\newblock
\href{https://doi.org/10.1145/3711896.3736570}{doi:\nolinkurl{10.1145/3711896.3736570}}


\bibitem[Muratore et~al\mbox{.}(2022)]%
        {muratore2022randomized}
\bibfield{author}{\bibinfo{person}{Fabio Muratore}, \bibinfo{person}{Michael
  Gienger}, {and} \bibinfo{person}{Jan Peters}.}
  \bibinfo{year}{2022}\natexlab{}.
\newblock \showarticletitle{Robot Learning from Randomized Simulations: {A}
  Review}.
\newblock \bibinfo{journal}{\emph{Frontiers in Robotics and AI}}
  \bibinfo{volume}{9} (\bibinfo{year}{2022}), \bibinfo{pages}{799893}.
\newblock
\href{https://doi.org/10.3389/frobt.2022.799893}{doi:\nolinkurl{10.3389/frobt.2022.799893}}


\bibitem[Naghibijouybari et~al\mbox{.}(2018)]%
        {naghibijouybari2018gpu}
\bibfield{author}{\bibinfo{person}{Hoda Naghibijouybari},
  \bibinfo{person}{Ajaya Neupane}, \bibinfo{person}{Zhiyun Qian}, {and}
  \bibinfo{person}{Nael Abu-Ghazaleh}.} \bibinfo{year}{2018}\natexlab{}.
\newblock \showarticletitle{Rendered Insecure: {GPU} Side Channel Attacks Are
  Practical}. In \bibinfo{booktitle}{\emph{Proceedings of the 2018 ACM SIGSAC
  Conference on Computer and Communications Security}}.
  \bibinfo{publisher}{ACM}, \bibinfo{address}{New York, NY, USA},
  \bibinfo{pages}{2139-2153}.
\newblock
\href{https://doi.org/10.1145/3243734.3243831}{doi:\nolinkurl{10.1145/3243734.3243831}}


\bibitem[{National Aeronautics and Space Administration}(2024)]%
        {nasa2024std7009b}
\bibfield{author}{\bibinfo{person}{{National Aeronautics and Space
  Administration}}.} \bibinfo{year}{2024}\natexlab{}.
\newblock \bibinfo{booktitle}{\emph{{NASA-STD-7009B: Standard for Models and
  Simulations}}}.
\newblock {NASA Technical Standards System}.
\newblock
\urldef\tempurl%
\url{https://standards.nasa.gov/standard/nasa/nasa-std-7009}
\showURL{%
\tempurl}
\newblock
\shownote{Accessed 17 May 2026}.


\bibitem[{National Institute of Standards and Technology}(2020)]%
        {nist2020privacy}
\bibfield{author}{\bibinfo{person}{{National Institute of Standards and
  Technology}}.} \bibinfo{year}{2020}\natexlab{}.
\newblock \bibinfo{booktitle}{\emph{NIST Privacy Framework: A Tool for
  Improving Privacy Through Enterprise Risk Management, Version 1.0}}.
\newblock \bibinfo{type}{{T}echnical {R}eport} NIST CSWP 10.
  \bibinfo{institution}{National Institute of Standards and Technology}.
\newblock
\href{https://doi.org/10.6028/NIST.CSWP.10}{doi:\nolinkurl{10.6028/NIST.CSWP.10}}


\bibitem[{National Institute of Standards and Technology}(2024a)]%
        {nistARIA2024}
\bibfield{author}{\bibinfo{person}{{National Institute of Standards and
  Technology}}.} \bibinfo{year}{2024}\natexlab{a}.
\newblock \bibinfo{title}{{ARIA: Assessing Risks and Impacts of AI}}.
\newblock
\urldef\tempurl%
\url{https://ai-challenges.nist.gov/aria}
\showURL{%
\tempurl}
\newblock
\shownote{See also NIST AI 700-2, doi:10.6028/NIST.AI.700-2. Accessed 11 June
  2026}.


\bibitem[{National Institute of Standards and Technology}(2024b)]%
        {nistDioptra2024}
\bibfield{author}{\bibinfo{person}{{National Institute of Standards and
  Technology}}.} \bibinfo{year}{2024}\natexlab{b}.
\newblock \bibinfo{title}{{Dioptra Test Platform}}.
\newblock
\urldef\tempurl%
\url{https://pages.nist.gov/dioptra/}
\showURL{%
\tempurl}
\newblock
\shownote{Accessed 11 June 2026}.


\bibitem[{National Institute of Standards and Technology}(2026a)]%
        {nistTEVV2026}
\bibfield{author}{\bibinfo{person}{{National Institute of Standards and
  Technology}}.} \bibinfo{year}{2026}\natexlab{a}.
\newblock \bibinfo{title}{{AI Test, Evaluation, Validation and Verification
  (TEVV)}}.
\newblock
\urldef\tempurl%
\url{https://www.nist.gov/ai-test-evaluation-validation-and-verification-tevv}
\showURL{%
\tempurl}
\newblock
\shownote{Accessed 30 April 2026}.


\bibitem[{National Institute of Standards and Technology}(2026b)]%
        {nistGlossarySandbox}
\bibfield{author}{\bibinfo{person}{{National Institute of Standards and
  Technology}}.} \bibinfo{year}{2026}\natexlab{b}.
\newblock \bibinfo{title}{{Sandbox}}.
\newblock \bibinfo{howpublished}{{NIST Computer Security Resource Center
  Glossary}}.
\newblock
\urldef\tempurl%
\url{https://csrc.nist.gov/glossary/term/Sandbox}
\showURL{%
\tempurl}
\newblock
\shownote{Accessed 30 April 2026}.


\bibitem[{National Telecommunications and Information Administration}(2021)]%
        {ntia2021sbom}
\bibfield{author}{\bibinfo{person}{{National Telecommunications and Information
  Administration}}.} \bibinfo{year}{2021}\natexlab{}.
\newblock \bibinfo{booktitle}{\emph{The Minimum Elements For a Software Bill of
  Materials ({SBOM})}}.
\newblock \bibinfo{type}{{T}echnical {R}eport}. \bibinfo{institution}{{U.S.
  Department of Commerce}}.
\newblock
\urldef\tempurl%
\url{https://www.ntia.gov/report/2021/minimum-elements-software-bill-materials-sbom}
\showURL{%
\tempurl}
\newblock
\shownote{Accessed 20 May 2026}.


\bibitem[Nguyen et~al\mbox{.}(2017)]%
        {bompard2017realtime}
\bibfield{author}{\bibinfo{person}{Van~Hoa Nguyen}, \bibinfo{person}{Yvon
  Besanger}, \bibinfo{person}{Quoc~Tuan Tran}, \bibinfo{person}{Tung~Lam
  Nguyen}, \bibinfo{person}{Cederic Boudinet}, \bibinfo{person}{Ron Brandl},
  \bibinfo{person}{Frank Marten}, \bibinfo{person}{Achilleas Markou},
  \bibinfo{person}{Panos Kotsampopoulos}, \bibinfo{person}{Arjen~A. van~der
  Meer}, {et~al\mbox{.}}} \bibinfo{year}{2017}\natexlab{}.
\newblock \bibinfo{title}{Real-Time Simulation and Hardware-in-the-Loop
  Approaches for Integrating Renewable Energy Sources into Smart Grids:
  Challenges and Actions}.
\newblock \bibinfo{howpublished}{arXiv preprint arXiv:1710.02306}.
\newblock
\urldef\tempurl%
\url{https://arxiv.org/abs/1710.02306}
\showURL{%
\tempurl}


\bibitem[Nguyen et~al\mbox{.}(2019)]%
        {steinbrink2019cosim}
\bibfield{author}{\bibinfo{person}{Van~Hoa Nguyen}, \bibinfo{person}{Tung~Lam
  Nguyen}, \bibinfo{person}{Quoc~Tuan Tran}, \bibinfo{person}{Yvon Besanger},
  {and} \bibinfo{person}{Raphael Caire}.} \bibinfo{year}{2019}\natexlab{}.
\newblock \bibinfo{title}{Integration of {SCADA} Services in
  Cross-Infrastructure Holistic Tests of Cyber-Physical Energy Systems}.
\newblock \bibinfo{howpublished}{arXiv preprint arXiv:1905.06008}.
\newblock
\urldef\tempurl%
\url{https://arxiv.org/abs/1905.06008}
\showURL{%
\tempurl}


\bibitem[{NVIDIA}(2026a)]%
        {isaacLabRepo2026}
\bibfield{author}{\bibinfo{person}{{NVIDIA}}.}
  \bibinfo{year}{2026}\natexlab{a}.
\newblock \bibinfo{title}{{Isaac Lab GitHub Repository}}.
\newblock
\urldef\tempurl%
\url{https://github.com/isaac-sim/IsaacLab}
\showURL{%
\tempurl}
\newblock
\shownote{Accessed 10 May 2026}.


\bibitem[{NVIDIA}(2026b)]%
        {isaacSimDocs2026}
\bibfield{author}{\bibinfo{person}{{NVIDIA}}.}
  \bibinfo{year}{2026}\natexlab{b}.
\newblock \bibinfo{title}{{Isaac Sim Documentation}}.
\newblock
\urldef\tempurl%
\url{https://docs.isaacsim.omniverse.nvidia.com/}
\showURL{%
\tempurl}
\newblock
\shownote{Accessed 10 May 2026}.


\bibitem[Oberkampf and Roy(2010)]%
        {oberkampf2010verification}
\bibfield{author}{\bibinfo{person}{William~L. Oberkampf} {and}
  \bibinfo{person}{Christopher~J. Roy}.} \bibinfo{year}{2010}\natexlab{}.
\newblock \bibinfo{booktitle}{\emph{Verification and Validation in Scientific
  Computing}}.
\newblock \bibinfo{publisher}{Cambridge University Press},
  \bibinfo{address}{Cambridge, United Kingdom}.
\newblock
\showISBNx{9780521113601}
\href{https://doi.org/10.1017/CBO9780511760396}{doi:\nolinkurl{10.1017/CBO9780511760396}}


\bibitem[{Object Management Group}(2020)]%
        {omgSACM2020}
\bibfield{author}{\bibinfo{person}{{Object Management Group}}.}
  \bibinfo{year}{2020}\natexlab{}.
\newblock \bibinfo{booktitle}{\emph{Structured Assurance Case Metamodel
  ({SACM}), Version 2.1}}.
\newblock \bibinfo{type}{{T}echnical {R}eport} formal/2020-12-02.
  \bibinfo{institution}{Object Management Group}.
\newblock
\urldef\tempurl%
\url{https://www.omg.org/spec/SACM/}
\showURL{%
\tempurl}


\bibitem[Ofenloch et~al\mbox{.}(2024)]%
        {ofenloch2024mosaik3}
\bibfield{author}{\bibinfo{person}{Annika Ofenloch},
  \bibinfo{person}{Jan~S{\"o}ren Schwarz}, \bibinfo{person}{Deborah Tolk},
  \bibinfo{person}{Tobias Brandt}, \bibinfo{person}{Reef Eilers}, {and}
  \bibinfo{person}{Rebeca Ramirez}.} \bibinfo{year}{2024}\natexlab{}.
\newblock \bibinfo{title}{{MOSAIK} 3.0: Combining Time-Stepped and Discrete
  Event Simulation}.
\newblock \bibinfo{howpublished}{arXiv preprint arXiv:2410.16937}.
\newblock
\urldef\tempurl%
\url{https://arxiv.org/abs/2410.16937}
\showURL{%
\tempurl}


\bibitem[{OFFIS}(2026)]%
        {mosaikDocs2026}
\bibfield{author}{\bibinfo{person}{{OFFIS}}.} \bibinfo{year}{2026}\natexlab{}.
\newblock \bibinfo{title}{{mosaik Co-Simulation Framework}}.
\newblock
\urldef\tempurl%
\url{https://mosaik.offis.de/}
\showURL{%
\tempurl}
\newblock
\shownote{Accessed 10 May 2026}.


\bibitem[O'Neill et~al\mbox{.}(2024)]%
        {oneill2024openx}
\bibfield{author}{\bibinfo{person}{Abby O'Neill}, \bibinfo{person}{Abdul
  Rehman}, \bibinfo{person}{Abhiram Maddukuri}, \bibinfo{person}{Abhishek
  Gupta}, \bibinfo{person}{Abhishek Padalkar}, \bibinfo{person}{Abraham Lee},
  \bibinfo{person}{Acorn Pooley}, \bibinfo{person}{Agrim Gupta},
  \bibinfo{person}{Ajay Mandlekar}, \bibinfo{person}{Ajinkya Jain},
  {et~al\mbox{.}}} \bibinfo{year}{2024}\natexlab{}.
\newblock \showarticletitle{{Open X-Embodiment: Robotic Learning Datasets and
  RT-X Models}}. In \bibinfo{booktitle}{\emph{2024 IEEE International
  Conference on Robotics and Automation (ICRA)}}. \bibinfo{publisher}{IEEE},
  \bibinfo{address}{Piscataway, NJ, USA}, \bibinfo{pages}{6892-6903}.
\newblock
\href{https://doi.org/10.1109/ICRA57147.2024.10611477}{doi:\nolinkurl{10.1109/ICRA57147.2024.10611477}}


\bibitem[{Open Robotics}(2026a)]%
        {gazeboDocs2026}
\bibfield{author}{\bibinfo{person}{{Open Robotics}}.}
  \bibinfo{year}{2026}\natexlab{a}.
\newblock \bibinfo{title}{{Gazebo Sim Documentation}}.
\newblock
\urldef\tempurl%
\url{https://gazebosim.org/docs/latest/}
\showURL{%
\tempurl}
\newblock
\shownote{Accessed 17 May 2026}.


\bibitem[{Open Robotics}(2026b)]%
        {vrxRepo2026}
\bibfield{author}{\bibinfo{person}{{Open Robotics}}.}
  \bibinfo{year}{2026}\natexlab{b}.
\newblock \bibinfo{title}{{Virtual RobotX (VRX) Simulation Environment GitHub
  Repository}}.
\newblock
\urldef\tempurl%
\url{https://github.com/osrf/vrx}
\showURL{%
\tempurl}
\newblock
\shownote{Accessed 17 May 2026}.


\bibitem[Palmintier et~al\mbox{.}(2017)]%
        {palmintier2017helics}
\bibfield{author}{\bibinfo{person}{Bryan Palmintier}, \bibinfo{person}{Dheepak
  Krishnamurthy}, \bibinfo{person}{Philip Top}, \bibinfo{person}{Steve Smith},
  \bibinfo{person}{Jeff Daily}, {and} \bibinfo{person}{Jason Fuller}.}
  \bibinfo{year}{2017}\natexlab{}.
\newblock \showarticletitle{{Design of the HELICS High-Performance
  Transmission-Distribution-Communication-Market Co-Simulation Framework}}. In
  \bibinfo{booktitle}{\emph{2017 Workshop on Modeling and Simulation of
  Cyber-Physical Energy Systems}}. \bibinfo{publisher}{IEEE},
  \bibinfo{address}{Piscataway, NJ, USA}, \bibinfo{pages}{1-6}.
\newblock
\href{https://doi.org/10.1109/MSCPES.2017.8064542}{doi:\nolinkurl{10.1109/MSCPES.2017.8064542}}


\bibitem[Peng et~al\mbox{.}(2018)]%
        {peng2018simtoreal}
\bibfield{author}{\bibinfo{person}{Xue~Bin Peng}, \bibinfo{person}{Marcin
  Andrychowicz}, \bibinfo{person}{Wojciech Zaremba}, {and}
  \bibinfo{person}{Pieter Abbeel}.} \bibinfo{year}{2018}\natexlab{}.
\newblock \showarticletitle{{Sim-to-Real Transfer of Robotic Control with
  Dynamics Randomization}}. In \bibinfo{booktitle}{\emph{2018 IEEE
  International Conference on Robotics and Automation}}.
  \bibinfo{publisher}{IEEE}, \bibinfo{address}{Piscataway, NJ, USA},
  \bibinfo{pages}{3803-3810}.
\newblock
\href{https://doi.org/10.1109/ICRA.2018.8460528}{doi:\nolinkurl{10.1109/ICRA.2018.8460528}}


\bibitem[Phuong et~al\mbox{.}(2024)]%
        {phuong2024frontier}
\bibfield{author}{\bibinfo{person}{Mary Phuong}, \bibinfo{person}{Matthew
  Aitchison}, {et~al\mbox{.}}} \bibinfo{year}{2024}\natexlab{}.
\newblock \bibinfo{title}{Evaluating Frontier Models for Dangerous
  Capabilities}.
\newblock \bibinfo{howpublished}{arXiv preprint arXiv:2403.13793}.
\newblock
\urldef\tempurl%
\url{https://arxiv.org/abs/2403.13793}
\showURL{%
\tempurl}


\bibitem[Quigley et~al\mbox{.}(2009)]%
        {quigley2009ros}
\bibfield{author}{\bibinfo{person}{Morgan Quigley}, \bibinfo{person}{Ken
  Conley}, \bibinfo{person}{Brian Gerkey}, \bibinfo{person}{Josh Faust},
  \bibinfo{person}{Tully Foote}, \bibinfo{person}{Jeremy Leibs},
  \bibinfo{person}{Eric Berger}, \bibinfo{person}{Rob Wheeler}, {and}
  \bibinfo{person}{Andrew Ng}.} \bibinfo{year}{2009}\natexlab{}.
\newblock \bibinfo{title}{{ROS}: An Open-Source Robot Operating System}.
\newblock \bibinfo{howpublished}{ICRA Workshop on Open Source Software}.
\newblock
\urldef\tempurl%
\url{https://www.ros.org/}
\showURL{%
\tempurl}


\bibitem[Raji et~al\mbox{.}(2021)]%
        {raji2021everything}
\bibfield{author}{\bibinfo{person}{Inioluwa~Deborah Raji},
  \bibinfo{person}{Emily~M. Bender}, \bibinfo{person}{Amandalynne Paullada},
  \bibinfo{person}{Emily Denton}, {and} \bibinfo{person}{Alex Hanna}.}
  \bibinfo{year}{2021}\natexlab{}.
\newblock \showarticletitle{{AI} and the Everything in the Whole Wide World
  Benchmark}. In \bibinfo{booktitle}{\emph{Proceedings of the Neural
  Information Processing Systems Track on Datasets and Benchmarks (NeurIPS
  Datasets and Benchmarks)}}.
\newblock
\urldef\tempurl%
\url{https://arxiv.org/abs/2111.15366}
\showURL{%
\tempurl}


\bibitem[Raji et~al\mbox{.}(2020)]%
        {raji2020closing}
\bibfield{author}{\bibinfo{person}{Inioluwa~Deborah Raji},
  \bibinfo{person}{Andrew Smart}, \bibinfo{person}{Rebecca~N. White},
  \bibinfo{person}{Margaret Mitchell}, \bibinfo{person}{Timnit Gebru},
  \bibinfo{person}{Ben Hutchinson}, \bibinfo{person}{Jamila Smith-Loud},
  \bibinfo{person}{Daniel Theron}, {and} \bibinfo{person}{Parker Barnes}.}
  \bibinfo{year}{2020}\natexlab{}.
\newblock \showarticletitle{{Closing the AI Accountability Gap: Defining an
  End-to-End Framework for Internal Algorithmic Auditing}}. In
  \bibinfo{booktitle}{\emph{Proceedings of the 2020 Conference on Fairness,
  Accountability, and Transparency}}. \bibinfo{publisher}{ACM},
  \bibinfo{address}{New York, NY, USA}, \bibinfo{pages}{33-44}.
\newblock
\href{https://doi.org/10.1145/3351095.3372873}{doi:\nolinkurl{10.1145/3351095.3372873}}


\bibitem[Rajkumar et~al\mbox{.}(2010)]%
        {rajkumar2010cps}
\bibfield{author}{\bibinfo{person}{Ragunathan Rajkumar}, \bibinfo{person}{Insup
  Lee}, \bibinfo{person}{Lui Sha}, {and} \bibinfo{person}{John Stankovic}.}
  \bibinfo{year}{2010}\natexlab{}.
\newblock \showarticletitle{{Cyber-Physical Systems: The Next Computing
  Revolution}}. In \bibinfo{booktitle}{\emph{Proceedings of the 47th Design
  Automation Conference}}. \bibinfo{publisher}{ACM}, \bibinfo{address}{New
  York, NY, USA}, \bibinfo{pages}{731-736}.
\newblock
\href{https://doi.org/10.1145/1837274.1837461}{doi:\nolinkurl{10.1145/1837274.1837461}}


\bibitem[Ranchordas(2021)]%
        {ranchordas2021sandboxes}
\bibfield{author}{\bibinfo{person}{Sofia Ranchordas}.}
  \bibinfo{year}{2021}\natexlab{}.
\newblock \showarticletitle{Experimental Regulations and Regulatory Sandboxes -
  {L}aw without Order?}
\newblock \bibinfo{journal}{\emph{Law and Method}} (\bibinfo{year}{2021}),
  \bibinfo{pages}{1-27}.
\newblock


\bibitem[Ratiu et~al\mbox{.}(2024)]%
        {ratiu2024spi}
\bibfield{author}{\bibinfo{person}{Daniel Ratiu}, \bibinfo{person}{Tihomir
  Rohlinger}, \bibinfo{person}{Torben Stolte}, {and} \bibinfo{person}{Stefan
  Wagner}.} \bibinfo{year}{2024}\natexlab{}.
\newblock \bibinfo{title}{Towards an Argument Pattern for the Use of Safety
  Performance Indicators}.
\newblock \bibinfo{howpublished}{arXiv preprint arXiv:2410.00578}.
\newblock
\urldef\tempurl%
\url{https://arxiv.org/abs/2410.00578}
\showURL{%
\tempurl}


\bibitem[Recht et~al\mbox{.}(2019)]%
        {recht2019imagenet}
\bibfield{author}{\bibinfo{person}{Benjamin Recht}, \bibinfo{person}{Rebecca
  Roelofs}, \bibinfo{person}{Ludwig Schmidt}, {and} \bibinfo{person}{Vaishaal
  Shankar}.} \bibinfo{year}{2019}\natexlab{}.
\newblock \showarticletitle{Do {ImageNet} Classifiers Generalize to
  {ImageNet}?}. In \bibinfo{booktitle}{\emph{Proceedings of the 36th
  International Conference on Machine Learning}}
  \emph{(\bibinfo{series}{Proceedings of Machine Learning Research},
  Vol.~\bibinfo{volume}{97})}. \bibinfo{publisher}{PMLR},
  \bibinfo{address}{Long Beach, California, USA}, \bibinfo{pages}{5389-5400}.
\newblock
\urldef\tempurl%
\url{https://proceedings.mlr.press/v97/recht19a.html}
\showURL{%
\tempurl}


\bibitem[Reuel et~al\mbox{.}(2024a)]%
        {reuel2024openproblems}
\bibfield{author}{\bibinfo{person}{Anka Reuel}, \bibinfo{person}{Ben Bucknall},
  \bibinfo{person}{Stephen Casper}, \bibinfo{person}{Tim Fist},
  \bibinfo{person}{Lisa Soder}, \bibinfo{person}{Onni Aarne},
  \bibinfo{person}{Lewis Hammond}, \bibinfo{person}{Lujain Ibrahim},
  \bibinfo{person}{Alan Chan}, \bibinfo{person}{Peter Wills}, {et~al\mbox{.}}}
  \bibinfo{year}{2024}\natexlab{a}.
\newblock \bibinfo{title}{{Open Problems in Technical AI Governance}}.
\newblock \bibinfo{howpublished}{arXiv preprint arXiv:2407.14981}.
\newblock
\urldef\tempurl%
\url{https://arxiv.org/abs/2407.14981}
\showURL{%
\tempurl}


\bibitem[Reuel et~al\mbox{.}(2024b)]%
        {reuel2024betterbench}
\bibfield{author}{\bibinfo{person}{Anka Reuel}, \bibinfo{person}{Amelia Hardy},
  \bibinfo{person}{Chandler Smith}, \bibinfo{person}{Max Lamparth},
  \bibinfo{person}{Malcolm Hardy}, {and} \bibinfo{person}{Mykel~J.
  Kochenderfer}.} \bibinfo{year}{2024}\natexlab{b}.
\newblock \showarticletitle{{BetterBench: Assessing AI Benchmarks, Uncovering
  Issues, and Establishing Best Practices}}. In
  \bibinfo{booktitle}{\emph{Advances in Neural Information Processing Systems
  (NeurIPS)}}.
\newblock
\urldef\tempurl%
\url{https://arxiv.org/abs/2411.12990}
\showURL{%
\tempurl}


\bibitem[Rushby(2015)]%
        {rushby2015safetycase}
\bibfield{author}{\bibinfo{person}{John Rushby}.}
  \bibinfo{year}{2015}\natexlab{}.
\newblock \showarticletitle{The Interpretation and Evaluation of Assurance
  Cases}.
\newblock \bibinfo{journal}{\emph{SRI International Technical Report}}
  \bibinfo{number}{SRI-CSL-15-01} (\bibinfo{year}{2015}).
\newblock


\bibitem[{SAE International}(2021)]%
        {saej3016}
\bibfield{author}{\bibinfo{person}{{SAE International}}.}
  \bibinfo{year}{2021}\natexlab{}.
\newblock \bibinfo{booktitle}{\emph{{SAE} {J3016}: Taxonomy and Definitions for
  Terms Related to Driving Automation Systems for On-Road Motor Vehicles}}.
\newblock
\urldef\tempurl%
\url{https://www.sae.org/standards/content/j3016_202104/}
\showURL{%
\tempurl}


\bibitem[Salamun et~al\mbox{.}(2023)]%
        {salamun2023weakly}
\bibfield{author}{\bibinfo{person}{Karla Salamun}, \bibinfo{person}{Ivan
  Pavi{\'c}}, \bibinfo{person}{Hrvoje D{\v{z}}apo}, {and}
  \bibinfo{person}{Ivana {\v{C}}uljak}.} \bibinfo{year}{2023}\natexlab{}.
\newblock \showarticletitle{Weakly hard real-time model for control systems: A
  survey}.
\newblock \bibinfo{journal}{\emph{Sensors}} \bibinfo{volume}{23},
  \bibinfo{number}{10} (\bibinfo{year}{2023}), \bibinfo{pages}{4652}.
\newblock
\href{https://doi.org/10.3390/s23104652}{doi:\nolinkurl{10.3390/s23104652}}


\bibitem[Salimpour et~al\mbox{.}(2025)]%
        {salimpour2025simtoreal}
\bibfield{author}{\bibinfo{person}{Sahar Salimpour}, \bibinfo{person}{Jorge
  Pe{\~n}a-Queralta}, \bibinfo{person}{Diego Paez-Granados},
  \bibinfo{person}{Jukka Heikkonen}, {and} \bibinfo{person}{Tomi Westerlund}.}
  \bibinfo{year}{2025}\natexlab{}.
\newblock \bibinfo{title}{Sim-to-Real Transfer for Mobile Robots with
  Reinforcement Learning: From {NVIDIA} {Isaac Sim} to {Gazebo} and Real
  {ROS}~2 Robots}.
\newblock \bibinfo{howpublished}{arXiv preprint arXiv:2501.02902}.
\newblock
\urldef\tempurl%
\url{https://arxiv.org/abs/2501.02902}
\showURL{%
\tempurl}


\bibitem[Saltzer and Schroeder(1975)]%
        {saltzer1975protection}
\bibfield{author}{\bibinfo{person}{Jerome~H. Saltzer} {and}
  \bibinfo{person}{Michael~D. Schroeder}.} \bibinfo{year}{1975}\natexlab{}.
\newblock \showarticletitle{{The Protection of Information in Computer
  Systems}}.
\newblock \bibinfo{journal}{\emph{Proc. IEEE}} \bibinfo{volume}{63},
  \bibinfo{number}{9} (\bibinfo{year}{1975}), \bibinfo{pages}{1278-1308}.
\newblock
\href{https://doi.org/10.1109/PROC.1975.9939}{doi:\nolinkurl{10.1109/PROC.1975.9939}}


\bibitem[Savva et~al\mbox{.}(2019)]%
        {savva2019habitat}
\bibfield{author}{\bibinfo{person}{Manolis Savva}, \bibinfo{person}{Abhishek
  Kadian}, \bibinfo{person}{Oleksandr Maksymets}, \bibinfo{person}{Yili Zhao},
  \bibinfo{person}{Erik Wijmans}, \bibinfo{person}{Bhavana Jain},
  \bibinfo{person}{Julian Straub}, \bibinfo{person}{Jia Liu},
  \bibinfo{person}{Vladlen Koltun}, \bibinfo{person}{Jitendra Malik},
  \bibinfo{person}{Devi Parikh}, {and} \bibinfo{person}{Dhruv Batra}.}
  \bibinfo{year}{2019}\natexlab{}.
\newblock \showarticletitle{{Habitat: A Platform for Embodied AI Research}}. In
  \bibinfo{booktitle}{\emph{Proceedings of the IEEE/CVF International
  Conference on Computer Vision (ICCV)}}. \bibinfo{publisher}{IEEE},
  \bibinfo{address}{Piscataway, NJ, USA}, \bibinfo{pages}{9339-9347}.
\newblock
\href{https://doi.org/10.1109/ICCV.2019.00943}{doi:\nolinkurl{10.1109/ICCV.2019.00943}}


\bibitem[Schierman et~al\mbox{.}(2020)]%
        {schierman2020rta}
\bibfield{author}{\bibinfo{person}{John~D. Schierman},
  \bibinfo{person}{Michael~D. DeVore}, \bibinfo{person}{Nathan~D. Richards},
  {and} \bibinfo{person}{Matthew~A. Clark}.} \bibinfo{year}{2020}\natexlab{}.
\newblock \showarticletitle{Runtime Assurance for Autonomous Aerospace
  Systems}.
\newblock \bibinfo{journal}{\emph{Journal of Guidance, Control, and Dynamics}}
  \bibinfo{volume}{43}, \bibinfo{number}{12} (\bibinfo{year}{2020}),
  \bibinfo{pages}{2205-2217}.
\newblock
\href{https://doi.org/10.2514/1.G004862}{doi:\nolinkurl{10.2514/1.G004862}}


\bibitem[Scholtes et~al\mbox{.}(2021)]%
        {scholtes2021odd}
\bibfield{author}{\bibinfo{person}{Maike Scholtes}, \bibinfo{person}{Lukas
  Westhofen}, \bibinfo{person}{Lara~R. Turner}, \bibinfo{person}{Katrin Lotto},
  \bibinfo{person}{Michael Schuldes}, \bibinfo{person}{Hendrik Weber},
  \bibinfo{person}{Nicolas Wagener}, \bibinfo{person}{Christian Neurohr},
  \bibinfo{person}{Marcus Bollmann}, \bibinfo{person}{Fabian K{\"o}rtke},
  \bibinfo{person}{Johannes Hiller}, \bibinfo{person}{Michael Hoss},
  \bibinfo{person}{Julian Bock}, {and} \bibinfo{person}{Lutz Eckstein}.}
  \bibinfo{year}{2021}\natexlab{}.
\newblock \showarticletitle{{6}-Layer Model for a Structured Description and
  Categorization of Urban Traffic and Environment}.
\newblock \bibinfo{journal}{\emph{IEEE Access}}  \bibinfo{volume}{9}
  (\bibinfo{year}{2021}), \bibinfo{pages}{59131-59147}.
\newblock


\bibitem[Sch{\"u}tte et~al\mbox{.}(2011)]%
        {schutte2011mosaik}
\bibfield{author}{\bibinfo{person}{Steffen Sch{\"u}tte},
  \bibinfo{person}{Stefan Scherfke}, {and} \bibinfo{person}{Martin
  Tr{\"o}schel}.} \bibinfo{year}{2011}\natexlab{}.
\newblock \showarticletitle{{Mosaik: A framework for modular simulation of
  active components in Smart Grids}}. In \bibinfo{booktitle}{\emph{2011 IEEE
  First International Workshop on Smart Grid Modeling and Simulation (SGMS)}}.
  \bibinfo{publisher}{IEEE}, \bibinfo{address}{Piscataway, NJ, USA},
  \bibinfo{pages}{55-60}.
\newblock
\href{https://doi.org/10.1109/SGMS.2011.6089027}{doi:\nolinkurl{10.1109/SGMS.2011.6089027}}


\bibitem[Shah et~al\mbox{.}(2017)]%
        {shah2017airsim}
\bibfield{author}{\bibinfo{person}{Shital Shah}, \bibinfo{person}{Debadeepta
  Dey}, \bibinfo{person}{Chris Lovett}, {and} \bibinfo{person}{Ashish Kapoor}.}
  \bibinfo{year}{2017}\natexlab{}.
\newblock \showarticletitle{{AirSim}: High-Fidelity Visual and Physical
  Simulation for Autonomous Vehicles}. In \bibinfo{booktitle}{\emph{Field and
  Service Robotics: Results of the 11th International Conference}}.
  \bibinfo{publisher}{Springer}, \bibinfo{address}{Cham, Switzerland},
  \bibinfo{pages}{621-635}.
\newblock
\href{https://doi.org/10.1007/978-3-319-67361-5_40}{doi:\nolinkurl{10.1007/978-3-319-67361-5_40}}


\bibitem[Shevlane et~al\mbox{.}(2023)]%
        {shevlane2023extreme}
\bibfield{author}{\bibinfo{person}{Toby Shevlane}, \bibinfo{person}{Sebastian
  Farquhar}, \bibinfo{person}{Ben Garfinkel}, \bibinfo{person}{Mary Phuong},
  \bibinfo{person}{Jess Whittlestone}, {et~al\mbox{.}}}
  \bibinfo{year}{2023}\natexlab{}.
\newblock \bibinfo{title}{Model Evaluation for Extreme Risks}.
\newblock \bibinfo{howpublished}{arXiv preprint arXiv:2305.15324}.
\newblock
\urldef\tempurl%
\url{https://arxiv.org/abs/2305.15324}
\showURL{%
\tempurl}


\bibitem[Song et~al\mbox{.}(2021)]%
        {song2020flightmare}
\bibfield{author}{\bibinfo{person}{Yunlong Song}, \bibinfo{person}{Selim Naji},
  \bibinfo{person}{Elia Kaufmann}, \bibinfo{person}{Antonio Loquercio}, {and}
  \bibinfo{person}{Davide Scaramuzza}.} \bibinfo{year}{2021}\natexlab{}.
\newblock \showarticletitle{Flightmare: A Flexible Quadrotor Simulator}. In
  \bibinfo{booktitle}{\emph{Proceedings of the 2020 Conference on Robot
  Learning}} \emph{(\bibinfo{series}{Proceedings of Machine Learning Research},
  Vol.~\bibinfo{volume}{155})}. \bibinfo{publisher}{PMLR},
  \bibinfo{address}{Virtual}, \bibinfo{pages}{1147-1157}.
\newblock
\urldef\tempurl%
\url{https://proceedings.mlr.press/v155/song21a.html}
\showURL{%
\tempurl}


\bibitem[Souppaya et~al\mbox{.}(2017)]%
        {souppaya2017containers}
\bibfield{author}{\bibinfo{person}{Murugiah Souppaya}, \bibinfo{person}{John
  Morello}, {and} \bibinfo{person}{Karen Scarfone}.}
  \bibinfo{year}{2017}\natexlab{}.
\newblock \bibinfo{booktitle}{\emph{{Application Container Security Guide}}}.
\newblock \bibinfo{type}{{T}echnical {R}eport} NIST SP 800-190.
  \bibinfo{institution}{{National Institute of Standards and Technology}}.
\newblock
\href{https://doi.org/10.6028/NIST.SP.800-190}{doi:\nolinkurl{10.6028/NIST.SP.800-190}}


\bibitem[Souppaya et~al\mbox{.}(2022)]%
        {nist2022ssdf}
\bibfield{author}{\bibinfo{person}{Murugiah Souppaya}, \bibinfo{person}{Karen
  Scarfone}, {and} \bibinfo{person}{Donna Dodson}.}
  \bibinfo{year}{2022}\natexlab{}.
\newblock \bibinfo{booktitle}{\emph{Secure Software Development Framework
  ({SSDF}) Version 1.1: Recommendations for Mitigating the Risk of Software
  Vulnerabilities}}.
\newblock \bibinfo{type}{{T}echnical {R}eport} NIST SP 800-218.
  \bibinfo{institution}{National Institute of Standards and Technology}.
\newblock
\href{https://doi.org/10.6028/NIST.SP.800-218}{doi:\nolinkurl{10.6028/NIST.SP.800-218}}


\bibitem[Szot et~al\mbox{.}(2021)]%
        {szot2021habitat2}
\bibfield{author}{\bibinfo{person}{Andrew Szot}, \bibinfo{person}{Alexander
  Clegg}, \bibinfo{person}{Eric Undersander}, \bibinfo{person}{Erik Wijmans},
  \bibinfo{person}{Yili Zhao}, \bibinfo{person}{John Turner},
  \bibinfo{person}{Noah Maestre}, \bibinfo{person}{Mustafa Mukadam},
  \bibinfo{person}{Devendra~Singh Chaplot}, \bibinfo{person}{Oleksandr
  Maksymets}, \bibinfo{person}{Aaron Gokaslan}, \bibinfo{person}{Vladim{\'\i}r
  Vondru{\v{s}}}, \bibinfo{person}{Samber Dharur}, \bibinfo{person}{Franziska
  Meier}, \bibinfo{person}{Wojciech Galuba}, \bibinfo{person}{Angel Chang},
  \bibinfo{person}{Zsolt Kira}, \bibinfo{person}{Vladlen Koltun},
  \bibinfo{person}{Jitendra Malik}, \bibinfo{person}{Manolis Savva}, {and}
  \bibinfo{person}{Dhruv Batra}.} \bibinfo{year}{2021}\natexlab{}.
\newblock \showarticletitle{Habitat 2.0: Training Home Assistants to Rearrange
  their Habitat}. In \bibinfo{booktitle}{\emph{Advances in Neural Information
  Processing Systems (NeurIPS)}}, Vol.~\bibinfo{volume}{34}.
  \bibinfo{publisher}{Curran Associates, Inc.}, \bibinfo{address}{Red Hook, NY,
  USA}, \bibinfo{pages}{251-266}.
\newblock
\urldef\tempurl%
\url{https://proceedings.neurips.cc/paper/2021/hash/021bbc7ee20b71134d53e20206bd6feb-Abstract.html}
\showURL{%
\tempurl}


\bibitem[Tabassi(2023)]%
        {tabassi2023airmf}
\bibfield{author}{\bibinfo{person}{Elham Tabassi}.}
  \bibinfo{year}{2023}\natexlab{}.
\newblock \bibinfo{booktitle}{\emph{{Artificial Intelligence Risk Management
  Framework (AI RMF 1.0)}}}.
\newblock \bibinfo{type}{{T}echnical {R}eport} NIST AI 100-1.
  \bibinfo{institution}{{National Institute of Standards and Technology}}.
\newblock
\href{https://doi.org/10.6028/NIST.AI.100-1}{doi:\nolinkurl{10.6028/NIST.AI.100-1}}


\bibitem[Tang et~al\mbox{.}(2023)]%
        {tang2023adstesting}
\bibfield{author}{\bibinfo{person}{Shuncheng Tang}, \bibinfo{person}{Zhenya
  Zhang}, \bibinfo{person}{Yi Zhang}, \bibinfo{person}{Jixiang Zhou},
  \bibinfo{person}{Yan Guo}, \bibinfo{person}{Shuang Liu},
  \bibinfo{person}{Shengjian Guo}, \bibinfo{person}{Yan-Fu Li},
  \bibinfo{person}{Lei Ma}, \bibinfo{person}{Yinxing Xue}, {and}
  \bibinfo{person}{Yang Liu}.} \bibinfo{year}{2023}\natexlab{}.
\newblock \showarticletitle{{A Survey on Automated Driving System Testing:
  Landscapes and Trends}}.
\newblock \bibinfo{journal}{\emph{ACM Transactions on Software Engineering and
  Methodology}} \bibinfo{volume}{32}, \bibinfo{number}{5} (\bibinfo{date}{July}
  \bibinfo{year}{2023}), \bibinfo{pages}{1-62}.
\newblock
\href{https://doi.org/10.1145/3579642}{doi:\nolinkurl{10.1145/3579642}}


\bibitem[Tian et~al\mbox{.}(2018)]%
        {tian2018deeptest}
\bibfield{author}{\bibinfo{person}{Yuchi Tian}, \bibinfo{person}{Kexin Pei},
  \bibinfo{person}{Suman Jana}, {and} \bibinfo{person}{Baishakhi Ray}.}
  \bibinfo{year}{2018}\natexlab{}.
\newblock \showarticletitle{{DeepTest}: Automated Testing of
  Deep-Neural-Network-Driven Autonomous Cars}. In
  \bibinfo{booktitle}{\emph{Proceedings of the 40th International Conference on
  Software Engineering (ICSE)}}. \bibinfo{pages}{303-314}.
\newblock


\bibitem[Tlaie and Farrell(2025)]%
        {tlaie2025gpaievals}
\bibfield{author}{\bibinfo{person}{Alejandro Tlaie} {and}
  \bibinfo{person}{Jimmy Farrell}.} \bibinfo{year}{2025}\natexlab{}.
\newblock \bibinfo{title}{{Securing External Deeper-than-black-box GPAI
  Evaluations}}.
\newblock \bibinfo{howpublished}{arXiv preprint arXiv:2503.07496}.
\newblock
\urldef\tempurl%
\url{https://arxiv.org/abs/2503.07496}
\showURL{%
\tempurl}


\bibitem[Tobin et~al\mbox{.}(2017)]%
        {tobin2017domainrandomization}
\bibfield{author}{\bibinfo{person}{Josh Tobin}, \bibinfo{person}{Rachel Fong},
  \bibinfo{person}{Alex Ray}, \bibinfo{person}{Jonas Schneider},
  \bibinfo{person}{Wojciech Zaremba}, {and} \bibinfo{person}{Pieter Abbeel}.}
  \bibinfo{year}{2017}\natexlab{}.
\newblock \showarticletitle{{Domain Randomization for Transferring Deep Neural
  Networks from Simulation to the Real World}}. In
  \bibinfo{booktitle}{\emph{2017 IEEE/RSJ International Conference on
  Intelligent Robots and Systems}}. \bibinfo{publisher}{IEEE},
  \bibinfo{address}{Piscataway, NJ, USA}, \bibinfo{pages}{23-30}.
\newblock
\href{https://doi.org/10.1109/IROS.2017.8202133}{doi:\nolinkurl{10.1109/IROS.2017.8202133}}


\bibitem[Todorov et~al\mbox{.}(2012)]%
        {todorov2012mujoco}
\bibfield{author}{\bibinfo{person}{Emanuel Todorov}, \bibinfo{person}{Tom
  Erez}, {and} \bibinfo{person}{Yuval Tassa}.} \bibinfo{year}{2012}\natexlab{}.
\newblock \showarticletitle{{MuJoCo}: {A} Physics Engine for Model-Based
  Control}. In \bibinfo{booktitle}{\emph{Proceedings of the IEEE/RSJ
  International Conference on Intelligent Robots and Systems (IROS)}}.
  \bibinfo{publisher}{IEEE}, \bibinfo{address}{Piscataway, NJ, USA},
  \bibinfo{pages}{5026-5033}.
\newblock
\href{https://doi.org/10.1109/IROS.2012.6386109}{doi:\nolinkurl{10.1109/IROS.2012.6386109}}


\bibitem[Truby et~al\mbox{.}(2022)]%
        {truby2022sandbox}
\bibfield{author}{\bibinfo{person}{Jon Truby}, \bibinfo{person}{Rafael~Dean
  Brown}, \bibinfo{person}{Imad~Antoine Ibrahim}, {and} \bibinfo{person}{Oriol
  Caudevilla~Parellada}.} \bibinfo{year}{2022}\natexlab{}.
\newblock \showarticletitle{{A Sandbox Approach to Regulating High-Risk
  Artificial Intelligence Applications}}.
\newblock \bibinfo{journal}{\emph{European Journal of Risk Regulation}}
  \bibinfo{volume}{13}, \bibinfo{number}{2} (\bibinfo{year}{2022}),
  \bibinfo{pages}{270-294}.
\newblock
\href{https://doi.org/10.1017/err.2021.52}{doi:\nolinkurl{10.1017/err.2021.52}}


\bibitem[{UK AI Security Institute}(2026)]%
        {ukaisandbox2024}
\bibfield{author}{\bibinfo{person}{{UK AI Security Institute}}.}
  \bibinfo{year}{2026}\natexlab{}.
\newblock \bibinfo{title}{{Inspect}: Sandboxing}.
\newblock
  \bibinfo{howpublished}{\url{https://inspect.aisi.org.uk/sandboxing.html}}.
\newblock


\bibitem[{UK Parliament}(2024)]%
        {avact2024}
\bibfield{author}{\bibinfo{person}{{UK Parliament}}.}
  \bibinfo{year}{2024}\natexlab{}.
\newblock \bibinfo{title}{Automated Vehicles Act 2024}.
\newblock
  \bibinfo{howpublished}{\url{https://www.legislation.gov.uk/ukpga/2024/10/contents}}.
\newblock
\newblock
\shownote{UK Public General Act, c.~10}.


\bibitem[{UL Standards and Engagement}(2022)]%
        {ul4600}
\bibfield{author}{\bibinfo{person}{{UL Standards and Engagement}}.}
  \bibinfo{year}{2022}\natexlab{}.
\newblock \bibinfo{booktitle}{\emph{{ANSI/UL} 4600: Standard for Safety for the
  Evaluation of Autonomous Products} (\bibinfo{edition}{2nd} ed.)}.
\newblock
\urldef\tempurl%
\url{https://ulse.org/UL4600}
\showURL{%
\tempurl}


\bibitem[Vu et~al\mbox{.}(2023)]%
        {nguyen2023shiphil}
\bibfield{author}{\bibinfo{person}{Linh Vu}, \bibinfo{person}{Lam Nguyen},
  \bibinfo{person}{Mahmoud Abdelaal}, \bibinfo{person}{Tuyen Vu}, {and}
  \bibinfo{person}{Osama Mohammed}.} \bibinfo{year}{2023}\natexlab{}.
\newblock \bibinfo{title}{A Cyber-{HIL} for Investigating Control Systems in
  Ship Cyber Physical Systems under Communication Issues and Cyber Attacks}.
\newblock \bibinfo{howpublished}{arXiv preprint arXiv:2306.14017}.
\newblock
\urldef\tempurl%
\url{https://arxiv.org/abs/2306.14017}
\showURL{%
\tempurl}


\bibitem[Wahbe et~al\mbox{.}(1993)]%
        {wahbe1993sfi}
\bibfield{author}{\bibinfo{person}{Robert Wahbe}, \bibinfo{person}{Steven
  Lucco}, \bibinfo{person}{Thomas~E. Anderson}, {and} \bibinfo{person}{Susan~L.
  Graham}.} \bibinfo{year}{1993}\natexlab{}.
\newblock \showarticletitle{{Efficient Software-Based Fault Isolation}}. In
  \bibinfo{booktitle}{\emph{Proceedings of the Fourteenth ACM Symposium on
  Operating Systems Principles}}. \bibinfo{publisher}{ACM},
  \bibinfo{address}{New York, NY, USA}, \bibinfo{pages}{203-216}.
\newblock
\href{https://doi.org/10.1145/168619.168635}{doi:\nolinkurl{10.1145/168619.168635}}


\bibitem[Waseem et~al\mbox{.}(2026)]%
        {waseem2026governance}
\bibfield{author}{\bibinfo{person}{Muhammad Waseem}, \bibinfo{person}{Md~Aidul
  Islam}, \bibinfo{person}{Md~Nasir~Uddin Shuvo}, \bibinfo{person}{Md~Mahade
  Hasan}, \bibinfo{person}{Kai-Kristian Kemell}, \bibinfo{person}{Jussi Rasku},
  \bibinfo{person}{Mika Saari}, \bibinfo{person}{Vilma Saari},
  \bibinfo{person}{Roope Pajasmaa}, \bibinfo{person}{Markku Oivo}, {and}
  \bibinfo{person}{Pekka Abrahamsson}.} \bibinfo{year}{2026}\natexlab{}.
\newblock \bibinfo{title}{{Engineering a Governance-Aware AI Sandbox: Design,
  Implementation, and Lessons Learned}}.
\newblock \bibinfo{howpublished}{arXiv preprint arXiv:2603.03394}.
\newblock
\urldef\tempurl%
\url{https://arxiv.org/abs/2603.03394}
\showURL{%
\tempurl}


\bibitem[Wei et~al\mbox{.}(2019)]%
        {wei2019assurance}
\bibfield{author}{\bibinfo{person}{Ran Wei}, \bibinfo{person}{Tim~P. Kelly},
  \bibinfo{person}{Xiaotian Dai}, \bibinfo{person}{Shuai Zhao}, {and}
  \bibinfo{person}{Richard Hawkins}.} \bibinfo{year}{2019}\natexlab{}.
\newblock \showarticletitle{Model Based System Assurance Using the Structured
  Assurance Case Metamodel}.
\newblock \bibinfo{journal}{\emph{Journal of Systems and Software}}
  \bibinfo{volume}{154} (\bibinfo{year}{2019}), \bibinfo{pages}{211-233}.
\newblock
\href{https://doi.org/10.1016/j.jss.2019.05.013}{doi:\nolinkurl{10.1016/j.jss.2019.05.013}}


\bibitem[Xu et~al\mbox{.}(2025)]%
        {xu2025fpgatwin}
\bibfield{author}{\bibinfo{person}{Bin Xu}, \bibinfo{person}{Ayan Banerjee},
  {and} \bibinfo{person}{Sandeep K.~S. Gupta}.}
  \bibinfo{year}{2025}\natexlab{}.
\newblock \bibinfo{title}{Fast Online Digital Twinning on {FPGA} for Mission
  Critical Applications}.
\newblock \bibinfo{howpublished}{arXiv preprint arXiv:2512.17942}.
\newblock
\urldef\tempurl%
\url{https://arxiv.org/abs/2512.17942}
\showURL{%
\tempurl}


\bibitem[Yamin et~al\mbox{.}(2020)]%
        {yamin2020cyberranges}
\bibfield{author}{\bibinfo{person}{Muhammad~Mudassar Yamin},
  \bibinfo{person}{Basel Katt}, {and} \bibinfo{person}{Vasileios Gkioulos}.}
  \bibinfo{year}{2020}\natexlab{}.
\newblock \showarticletitle{{Cyber Ranges and Security Testbeds: Scenarios,
  Functions, Tools and Architecture}}.
\newblock \bibinfo{journal}{\emph{Computers \& Security}}  \bibinfo{volume}{88}
  (\bibinfo{date}{Jan.} \bibinfo{year}{2020}), \bibinfo{pages}{101636}.
\newblock
\href{https://doi.org/10.1016/j.cose.2019.101636}{doi:\nolinkurl{10.1016/j.cose.2019.101636}}


\bibitem[Zetzsche et~al\mbox{.}(2017)]%
        {zetzsche2017regulating}
\bibfield{author}{\bibinfo{person}{Dirk~A. Zetzsche}, \bibinfo{person}{Ross~P.
  Buckley}, \bibinfo{person}{Janos~Nathan Barberis}, {and}
  \bibinfo{person}{Douglas~W. Arner}.} \bibinfo{year}{2017}\natexlab{}.
\newblock \showarticletitle{{Regulating a Revolution: From Regulatory Sandboxes
  to Smart Regulation}}.
\newblock \bibinfo{journal}{\emph{Fordham Journal of Corporate \& Financial
  Law}} \bibinfo{volume}{23}, \bibinfo{number}{1} (\bibinfo{year}{2017}),
  \bibinfo{pages}{31-103}.
\newblock
\urldef\tempurl%
\url{https://ir.lawnet.fordham.edu/jcfl/vol23/iss1/2/}
\showURL{%
\tempurl}


\bibitem[Zhang et~al\mbox{.}(2013)]%
        {zhang2013timesync}
\bibfield{author}{\bibinfo{person}{Zhenghao Zhang}, \bibinfo{person}{Shuping
  Gong}, \bibinfo{person}{Aleksandar~D. Dimitrovski}, {and}
  \bibinfo{person}{Husheng Li}.} \bibinfo{year}{2013}\natexlab{}.
\newblock \showarticletitle{Time Synchronization Attack in Smart Grid: Impact
  and Analysis}.
\newblock \bibinfo{journal}{\emph{IEEE Transactions on Smart Grid}}
  \bibinfo{volume}{4}, \bibinfo{number}{1} (\bibinfo{year}{2013}),
  \bibinfo{pages}{87-98}.
\newblock
\href{https://doi.org/10.1109/TSG.2012.2227342}{doi:\nolinkurl{10.1109/TSG.2012.2227342}}


\bibitem[Zhou et~al\mbox{.}(2024)]%
        {zhou2024webarena}
\bibfield{author}{\bibinfo{person}{Shuyan Zhou}, \bibinfo{person}{Frank~F. Xu},
  \bibinfo{person}{Hao Zhu}, \bibinfo{person}{Xuhui Zhou},
  \bibinfo{person}{Robert Lo}, \bibinfo{person}{Abishek Sridhar},
  \bibinfo{person}{Xianyi Cheng}, \bibinfo{person}{Tianyue Ou},
  \bibinfo{person}{Yonatan Bisk}, \bibinfo{person}{Daniel Fried},
  \bibinfo{person}{Uri Alon}, {and} \bibinfo{person}{Graham Neubig}.}
  \bibinfo{year}{2024}\natexlab{}.
\newblock \bibinfo{title}{{WebArena: A Realistic Web Environment for Building
  Autonomous Agents}}.
\newblock \bibinfo{howpublished}{The Twelfth International Conference on
  Learning Representations}.
\newblock
\urldef\tempurl%
\url{https://openreview.net/forum?id=oKn9c6ytLx}
\showURL{%
\tempurl}


\bibitem[Zhou et~al\mbox{.}(2025)]%
        {zhou2025badvla}
\bibfield{author}{\bibinfo{person}{Xueyang Zhou}, \bibinfo{person}{Guiyao Tie},
  \bibinfo{person}{Guowen Zhang}, \bibinfo{person}{Hechang Wang},
  \bibinfo{person}{Pan Zhou}, {and} \bibinfo{person}{Lichao Sun}.}
  \bibinfo{year}{2025}\natexlab{}.
\newblock \showarticletitle{{BadVLA}: Towards Backdoor Attacks on
  Vision-Language-Action Models via Objective-Decoupled Optimization}.
\newblock \bibinfo{journal}{\emph{arXiv preprint arXiv:2505.16640}}
  (\bibinfo{year}{2025}).
\newblock
\urldef\tempurl%
\url{https://arxiv.org/abs/2505.16640}
\showURL{%
\tempurl}


\bibitem[Zhu et~al\mbox{.}(2025)]%
        {zhu2025abc}
\bibfield{author}{\bibinfo{person}{Yuxuan Zhu}, \bibinfo{person}{Tengjun Jin},
  \bibinfo{person}{Yada Pruksachatkun}, \bibinfo{person}{Andy Zhang},
  \bibinfo{person}{Shu Liu}, \bibinfo{person}{Sasha Cui},
  \bibinfo{person}{Sayash Kapoor}, \bibinfo{person}{Shayne Longpre},
  \bibinfo{person}{Kevin Meng}, \bibinfo{person}{Rebecca Weiss},
  {et~al\mbox{.}}} \bibinfo{year}{2025}\natexlab{}.
\newblock \bibinfo{title}{{Establishing Best Practices for Building Rigorous
  Agentic Benchmarks}}.
\newblock \bibinfo{howpublished}{arXiv preprint arXiv:2507.02825}.
\newblock
\urldef\tempurl%
\url{https://arxiv.org/abs/2507.02825}
\showURL{%
\tempurl}


\end{thebibliography}
\endgroup

\appendix
\setcounter{table}{0}
\renewcommand{\thetable}{A\arabic{table}}

\section{Heatmap Coding Protocol and Per-Family Backing}
\label{app:heatmap-backing}

Table~\ref{tab:heatmap-backing} backs the cells in Figure~\ref{fig:heatmap}. Column~2 states the local target \(D_i^{\mathrm{local}}\), and Column~3 states the deployment reference claim \(D_i^*\) against which the figure is coded. The remaining columns explain the family-level grades and their source basis. A grade describes evidence for that reference claim. A specific platform configuration requires separate assessment. Missing evidence must be distinguished from an inapplicable local dimension. The operational criteria and cited source rationales allow each family-level judgment to be re-examined.

\begingroup
\scriptsize
\setlength{\LTcapwidth}{\textwidth}
\setlength{\tabcolsep}{3pt}
\renewcommand{\arraystretch}{1.10}
\begin{longtable}{>{\raggedright\arraybackslash}p{0.12\textwidth}>{\raggedright\arraybackslash}p{0.125\textwidth}>{\raggedright\arraybackslash}p{0.155\textwidth}>{\raggedright\arraybackslash}p{0.21\textwidth}>{\raggedright\arraybackslash}p{0.21\textwidth}>{\raggedright\arraybackslash}p{0.075\textwidth}}
\caption{Per-family backing for the Figure~\ref{fig:heatmap} coding, organized by sandbox family and reference claim: the local claim $D_i^{\mathrm{local}}$, the deployment target $D_i^*$ (with required artifacts and caution), and the rationale for its \gS{} and \gW{}/\gA{} grades. Families are numbered F1 to F11 in row order, and the worked instantiations of Section~\ref{subsec:worked-instantiations} cite these IDs. Grades are claim-relative and reproduced verbatim from the figure.}
\label{tab:heatmap-backing}\\
\toprule
\rowcolor{tabhead}\textbf{Family (platforms)} & \textbf{Local claim \(D_i^{\mathrm{local}}\)} & \textbf{Deployment target \(D_i^*\), artifacts, caution} & \textbf{\gS{} dimensions and rationale} & \textbf{\gW{}/\gA{} dimensions and rationale} & \textbf{Sources}\\
\midrule
\endfirsthead
\multicolumn{6}{@{}l}{\small\textit{Table \ref{tab:heatmap-backing} continued.}}\\
\toprule
\rowcolor{tabhead}\textbf{Family (platforms)} & \textbf{Local claim \(D_i^{\mathrm{local}}\)} & \textbf{Deployment target \(D_i^*\), artifacts, caution} & \textbf{\gS{} dimensions and rationale} & \textbf{\gW{}/\gA{} dimensions and rationale} & \textbf{Sources}\\
\midrule
\endhead
\bottomrule
\endfoot
\textbf{F1}\;General robotics simulation (Gazebo, ROS) & Controlled algorithm and robot-behavior experiments & Robot deployment-safety assurance. \emph{Artifacts:} URDF/SDF/MJCF, physics settings, seeds, logs. Sensing/world fidelity is task-specific & \gS{} Controllability, observability, reproducibility, openness: scriptable SDF/URDF worlds and seeds, complete simulator ground truth with rosbag logging, deterministic pinned reruns, open code & \gW{} Timing, network, actuator/plant, sim-to-real: not real-time, no fieldbus emulation, idealized contact, documented reality gap & \cite{koenig2004gazebo,quigley2009ros,michel2004webots,muratore2022randomized,salimpour2025simtoreal} \\
\addlinespace
\textbf{F2}\;GPU-scale robot learning (Isaac Gym/Sim/Lab) & Scalable policy training, synthetic-data generation, large ablations & Safety assurance for deployed robot learning policies. \emph{Artifacts:} hardware config, randomization ranges, seeds, assets. Fast biased simulation remains biased & \gS{} Controllability, reproducibility, scalability: randomization ranges and seeds over thousands of GPU-parallel environments & \gA{} Network realism (no network model). \gW{} timing, actuator realism, HiL/SiL, sim-to-real: transfer remains unvalidated & \cite{makoviychuk2021isaacgym,mittal2025isaaclab,isaacSimDocs2026,tobin2017domainrandomization} \\
\addlinespace
\textbf{F3}\;Embodied / HRI benchmarks (Habitat, AI2-THOR, BEHAVIOR) & Reproducible virtual task comparison and ground-truth evaluation & Physical robot readiness and safety. \emph{Artifacts:} task specs, datasets, metrics, embodiment assumptions. Deployment embodiment requires separate validation & \gS{} Observability, reproducibility, scalability: full scene/object ground truth, fixed datasets and seeds, high throughput & \gA{} Network realism, HiL/SiL. \gW{} sensing/world fidelity, actuator realism, and sim-to-real: visual distribution shift, simplified contact, and deployment timing remain unresolved & \cite{savva2019habitat,szot2021habitat2,kolve2017ai2thor,liu2022behaviorhabitat,duan2022embodiedaisurvey} \\
\addlinespace
\textbf{F4}\;Autonomous-driving simulation (CARLA) & Scenario regression, closed-loop testing, sensor/weather variation & Road-safety assurance within a declared operating domain. \emph{Artifacts:} maps, OpenSCENARIO files, sensor configs, timing traces. Scenario coverage dominates interpretation & \gS{} Controllability, scenario portability: rich scenario scripting with OpenSCENARIO/OpenDRIVE support & \gW{} Timing, network, sim-to-real: not real-time, no native V2X. ODD coverage, traffic-behavior realism, and field correlation unestablished & \cite{dosovitskiy2017carla,asamOpenScenarioDsl2026,tang2023adstesting,kaur2021simulators} \\
\addlinespace
\textbf{F5}\;Aerial and maritime autonomy (AirSim, Flightmare, VRX, MOOS-IvP) & Dangerous-condition screening, autonomy-stack regression, mission-logic tests & Field robustness under target weather and water conditions. \emph{Artifacts:} vehicle dynamics, wind/wave configs, logs. Archived projects require exact version preservation & \gS{} Controllability: scriptable missions, vehicles, and environments. SiL/HiL paths in mature stacks & \gW{} Network realism, actuator realism, sim-to-real, openness: radio/datalink degradation under-modeled, simplified propulsion and hydrodynamics, several projects archived & \cite{shah2017airsim,song2020flightmare,furrer2016rotors,airsimRepo2026,vrxRepo2026,moosivpDocs2026} \\
\addlinespace
\textbf{F6}\;Space-system simulation (Basilisk, mission M\&S) & Mission-phase logic, orbital/terrain dynamics, fault rehearsal & Mission assurance under the target hardware and mission envelope. \emph{Artifacts:} ephemeris, terrain/orbit models, fault scripts, review records. Expert validation is mandatory & \gS{} Containment (with dynamics modeling and fault logic): simulation-only mission rehearsal contains real-world consequence by construction & \gW{} Scenario portability, network, sim-to-real, openness: bespoke mission models. Radiation and terrain uncertainty require domain VV\&A and hardware evidence & \cite{kenneally2020basilisk,basiliskDocs2026,nasa2024std7009b,biesiadecki2007mer} \\
\addlinespace
\textbf{F7}\;Smart-grid / infrastructure co-simulation (HELICS, mosaik, FMI) & Power-communication-control interaction studies & Assurance of target plant and control behavior. \emph{Artifacts:} federate configs, time-step logs, plant models. Coupled models require individual validation & \gS{} Network realism, HiL/SiL integration (interoperability, timing structure): explicit communication federates and demonstrated real-time HiL co-simulation & \gW{} Sim-to-real, openness: federates, solvers, and assets are heterogeneous with incomplete joint auditability. Federate validity and solver assumptions unverified & \cite{palmintier2017helics,helicsDocs2026,schutte2011mosaik,fmiStandard3,bompard2017realtime,steinbrink2019cosim} \\
\addlinespace
\textbf{F8}\;AIoT and digital twins (Ditto, Azure DT, AWS TwinMaker, edge twins) & Telemetry replay, monitoring, what-if analysis, audit trails & Real-time operational assurance and safe actuation. \emph{Artifacts:} twin binding, telemetry provenance, access controls. Twin drift must be measured & \gS{} Observability, governance: operational telemetry and replay, telemetry lineage, access logs, audit trails & \gW{} Scenario portability, sim-to-real, attack/fault injection: bespoke twins, drift, actuation safety unestablished, connectivity that is itself an attack surface & \cite{barricelli2019digitaltwin,jones2020digitaltwin,alcaraz2022digitaltwinsecurity,das2022edgetwin,eclipseDittoDocs2026,azureDigitalTwinsDocs2026,awsIoTTwinMakerDocs2026} \\
\addlinespace
\textbf{F9}\;CPS / ICS security ranges (SCEPTRE, minimega, SWaT/WADI testbeds) & Attack/fault exercises, detector evaluation, blue-team observability & Physical resilience under target-system cyber attacks. \emph{Artifacts:} threat model, attack scripts, PCAP, process traces. Physical consequence is decisive & \gS{} Controllability, observability, containment, network realism, HiL/SiL, attack/fault injection: the security-range core, with physical testbeds adding real actuation and process telemetry & \gW{} Sim-to-real, scalability: cross-plant generalization remains unestablished. Physical testbeds are hard to scale. Attacker realism is bounded by the campaign & \cite{yamin2020cyberranges,hahn2021sceptre,crussell2016minimega,mathur2016swat,ahmed2017wadi,giraldo2018physics} \\
\addlinespace
\textbf{F10}\;Edge-AI / tinyML benchmarks (MLPerf Tiny) & Device-level accuracy, latency, energy, memory, runtime comparison & System safety and AIoT resilience. \emph{Artifacts:} board config, benchmark rules, power traces. System hazards remain outside the inference protocol & \gS{} Reproducibility, timing fidelity: standard run rules and on-device latency/energy measurement & \gA{} Containment, network realism, actuator/plant realism, sim-to-real: a device-inference benchmark with no field-transfer protocol & \cite{banbury2021mlperftiny} \\
\addlinespace
\textbf{F11}\;Regulatory sandboxes (FCA lineage, EU AI Act sandboxes) & Supervised experimentation, documentation, regulator-provider learning & Technical deployment-safety assurance. \emph{Artifacts:} sandbox plan, risk records, test reports. Technical validation requires a paired apparatus & \gS{} Governance (with auditability and procedural containment): process legitimacy, supervision, and documentation & \gA{} Nine technical dimensions (sensing/world fidelity, scenario portability, timing, network, actuator, HiL/SiL, sim-to-real, attack, scalability): the instrument models no technical dynamics by itself & \cite{euAIAct2024,allen2019regulatorysandboxes,zetzsche2017regulating,fca2017regulatorysandboxlessons} \\
\end{longtable}
\endgroup

\end{document}